\documentclass[letterpaper]{article}
\usepackage{geometry}
\usepackage{xcolor}
\usepackage{amsmath}
\usepackage[some]{background}
\usepackage{lipsum}

\usepackage{amsmath}
\usepackage{amsfonts}
\usepackage{amssymb}
\usepackage{graphicx, rotating}
\usepackage{epstopdf}
\usepackage{epsfig}
\usepackage{latexsym}
\usepackage{graphicx}
\usepackage{dsfont}
\usepackage{color}
\usepackage{amsmath,amssymb}
\usepackage{mathrsfs}  
\usepackage{cancel}

\DeclareMathAlphabet{\mathpzc}{T1}{pzc}{m}{it}
\usepackage{cite}
\usepackage{bbold}
\usepackage{slashed}
\usepackage{hyperref}
\usepackage{verbatim}
\definecolor{rossos}{cmyk}{0,1,1,0.55}
\definecolor{bluscuro}{rgb}{0.15, 0.2, .85}
\definecolor{bluchiaro}{cmyk}{1,.3,0.,0.1}
\usepackage{cancel}
\def\hhref#1{\href{http://arxiv.org/abs/#1}{#1}} 
\definecolor{oucrimsonred}{rgb}{0.6, 0.0, 0.0}
\definecolor{persianblue}{rgb}{0.11, 0.22, 0.73}
\definecolor{forestgreen}{rgb}{0.13,0.35,0.13}
\definecolor{lightergreen}{rgb}{.144,.238,.144}
 \hypersetup{colorlinks, citecolor=oucrimsonred, linkcolor=black, urlcolor=oucrimsonred}

\newcommand{\be}{\begin{equation}}
\newcommand{\ee}{\end{equation}}
\newcommand{\bea}{\begin{eqnarray}}
\newcommand{\eea}{\end{eqnarray}}
\newcommand{\bc}{\begin{center}}
\newcommand{\ec}{\end{center}}
\definecolor{Gray}{gray}{0.95}

\usepackage{amsmath}
\usepackage{empheq}
\usepackage[theorems,skins]{tcolorbox}

\newtcolorbox{mymathbox}[1][]{colback=white, #1}

\usepackage{adjustbox}
\usepackage{array}
\usepackage{booktabs}
\usepackage{multirow}
\usepackage{cancel}

\newcolumntype{R}[2]{%
    >{\adjustbox{angle=#1,lap=\width-(#2)}\bgroup}%
    l%
    <{\egroup}%
}

\def\Re{{\rm Re\,}}
\def\Im{{\rm Im\,}}

\definecolor{titlepagecolor}{cmyk}{.76,0,.76,.45}

\DeclareFixedFont{\bigsf}{T1}{phv}{b}{n}{1.5cm}

\backgroundsetup{
scale=1,
angle=0,
opacity=1,
contents={\begin{tikzpicture}[remember picture,overlay]
 \path [fill=titlepagecolor] (-0.5\paperwidth,5) rectangle (0.5\paperwidth,10);  
\end{tikzpicture}}
}
\makeatletter                       
\def\printauthor{%
    {\large \@author}}              
\makeatother
\author{%
\vspace{30pt}
    Rodrigo Alonso       \texttt{rodrigo.alonso@cern.ch}\vspace{30pt} \\     Alfredo Urbano \       \texttt{aurbano@cern.ch}  \vspace{30pt} \\ Theoretical Physics Department \\ CERN \\ Geneva, Switzerland
    }
\begin{document}
\begin{titlepage}
\BgThispage
\newgeometry{left=1cm,right=4cm}
\vspace*{2cm}
\noindent
\textcolor{white}{\Huge ~~~~Wormholes and masses for Goldstone bosons}\vspace{1cm}
\begin{flushright}
\large{\textcolor{white}{CERN-TH-2017-135}}
\end{flushright}
\vspace*{2.5cm}\par
\noindent
\begin{minipage}{0.35\linewidth}
    \begin{flushright}
        \printauthor
    \end{flushright}
\end{minipage} \hspace{15pt}
\begin{minipage}{0.02\linewidth}
    \rule{1pt}{175pt}
\end{minipage} \hspace{-10pt}
\begin{minipage}{0.6\linewidth}
\vspace{20pt}
    \begin{abstract} 
There exist non-trivial stationary points of the Euclidean action for 
an axion particle minimally coupled to Einstein gravity, dubbed wormholes. They explicitly break  
the continuos global shift symmetry of the axion in a non-perturbative way, and generate an effective 
 potential that may compete with QCD depending on the 
value of the axion decay constant. In this paper, we explore both theoretical and phenomenological aspects of this issue.
On the theory side, we address the problem of stability of the wormhole solutions, and we show that the spectrum of the quadratic action features only positive eigenvalues.
On the phenomenological side, we discuss, beside the obvious application to the QCD axion, 
relevant consequences for models with ultralight  dark matter, black hole superradiance, and the relaxation of the electroweak scale.
 We conclude discussing  wormhole solutions for a generic coset and the potential they generate.
    \end{abstract}
\end{minipage}
\end{titlepage}
\restoregeometry


\pagebreak

\tableofcontents

\section{Introduction}\label{sec:Intro}

The  explicit
 breaking of global symmetries due to gravitational effects is a topic of  great relevance in physics, both for  theoretical and phenomenological reasons.
Its study might shed light on the reason why 
gravity is so different from the other  fundamental interactions, and what r\^ole did these special features play in the history of the Universe.

Let us focus for simplicity and definiteness on 
 the case of a $U(1)$ Peccei-Quinn (PQ) global symmetry that is spontaneously broken at some scale $f_a$ by the vacuum expectation value (VEV) of a 
complex scalar field $\Phi$. 
By introducing the polar parametrization $\Phi = \rho e^{\imath \phi/f_a}$, the VEV 
of the complex field is $\langle |\Phi|\rangle = f_a$, and the axion can be identified with the angular direction  
$\phi$ -- the Nambu-Goldstone boson arising from the 
spontaneous symmetry breaking.
Below the scale $f_a$, the axion enjoys the non-linearly realized global $U(1)$ symmetry
\begin{equation}\label{eq:ContinuosShift}
\phi \to \phi + \alpha f_a~,
\end{equation}
for arbitrary values of the parameter $\alpha$ describing  the 
shift in field space.
The conserved Noether current implied by the symmetry in eq.~(\ref{eq:ContinuosShift}) is $J^{\mu} = f_a \partial^{\mu}\phi$, and the 
associated charge $\mathcal{Q}$ generates the symmetry transformation. 
 For the purpose of the present discussion, it is important to notice that the subgroup of eq.~(\ref{eq:ContinuosShift}) given by the discrete shift 
\begin{equation}\label{eq:DiscreteShift}
\phi \to \phi + 2k\pi f_a~,~~~~~ k\in \mathbb{Z}~,
\end{equation}
represents a gauge symmetry of the system.
This symmetry is inherent to the mere definition of the axion field in terms of an angular dynamical variable.
Referring for definiteness to the $U(1)$ case discussed above, 
the discrete shift in eq.~(\ref{eq:DiscreteShift}) maps the complex field $\Phi$ into itself, and thus corresponds to a redundancy 
in the description of the physical degrees of freedom.

At the QCD scale $\Lambda_{\rm QCD}$, Yang-Mills 
instanton effects break the continuos shift in eq.~(\ref{eq:ContinuosShift}) into its discrete subgroup $\phi \to \phi + 2k\pi f_a$
generating the potential 
\begin{equation}\label{eq:QCDpotentialterm}
V_{\rm QCD}(\phi) = - \Lambda_{\rm QCD}^4\cos\left(\frac{\phi}{f_a}\right)~,
\end{equation}
minimized at $\phi = 0$ in agreement with the axion solution of the strong CP problem.

What does gravity do with global symmetries, like the axion shift symmetry in eq.~(\ref{eq:ContinuosShift})?

A number of `folk theorems' state the impossibility to have global symmetries in a consistent theory of quantum gravity. 
To support this hypothesis, there exists a general semi-classical argument based on black hole physics. 
 The reasoning is that if some amount of  global charge 
  is thrown into a black hole, then the subsequent thermal decay of the black hole into photons and gravitons  via Hawking radiation 
  destroys the global charge without any possibility to reconstruct it in the final state, thus defining  a 
  process by means of which the global charge is violated. 
  To this respect, one also understands the crucial difference between global and local symmetries. 
  Local  symmetries obey the Gauss's law, and any observer
outside the black hole can determine its charge. 
The electric charge of a Kerr-Newman black hole represents an explicit example of this fact. 
 On the contrary,  there is no Gauss's law associated with global symmetries, 
and when charged particles are thrown into the black hole, there is no way to track them from
the outside: The global charge appears to be deleted, in obvious contradiction to its conservation.

In addition to the aforementioned theorem related to black hole physics, 
 various arguments in perturbative string theory~\cite{Banks:1987cy} -- where global symmetries on the world-sheet 
become gauge symmetries in the target space -- and AdS/CFT~\cite{Witten:1998qj} -- where global symmetries on the boundary correspond to gauge symmetries in the bulk --  
seem to corroborate this common belief. 

Motivated by these arguments, 
many authors studied the consequences  of the explicit breaking induced by higher-dimensional operators suppressed by the Planck scale $M_{\rm Pl} = \sqrt{1/G_N}$
of the form~\cite{Kamionkowski:1992mf,Barr:1992qq}
\begin{equation}\label{eq:Kamion}
 V_{\rm gravity}(\Phi) = g_{2m + n}\frac{|\Phi|^{2m}\Phi^{n}}{M_{\rm Pl}^{2m + n - 4}} + h.c.~,
 \end{equation}
with mass-dimension $2m + n$, and generic coupling $g_{2m + n}$ that in principle is of order unity. 
Needless to say, the impact of these operators  is catastrophic.
To give an idea, in the QCD axion case  including dimension-5 symmetry breaking operator induced by Planck scale physics
would require a coupling of order $|g_5| \lesssim 10^{-55}$ in order to 
avoid dangerous CP violation effects for $f_a \sim 10^{12}$ GeV.
This is of course unacceptable, since it introduces a fine-tuning by far more severe than the one the axion claims to solve.

The higher-dimensional operators in eq.~(\ref{eq:Kamion}) therefore represent, if present, a threat for the axion solution of the strong CP problem.
To solve the issue, it is possible to tailor suitable extensions of the simple $U(1)$ model discussed at the beginning of this section.
For instance, if the axion global symmetry is an accidental symmetry descending from an exact (discrete) gauge symmetry, 
 then the problematic higher-dimensional operators 
 can be forbidden to very high order~\cite{Babu:2002ic,Holman:1992us,DiLuzio:2017tjx,Fukuda:2017ylt,Kim:1981bb,Georgi:1981pu,Harigaya:2013vja,Dias:2002gg} 
(a similar conclusion remains true considering the  axion global symmetry as 
an accidental symmetry descending from exact discrete global symmetries~\cite{Dias:2014osa,Kim:2015yna}).

However, before engineering possible solutions, it is important to pose the following question:
Does gravity really generate power-suppressed symmetry breaking operators like those in eq.~(\ref{eq:Kamion})?

Let us tackle the problem from the simplest perspective, and consider the action 
describing an axion field minimally coupled to Einstein gravity. 
In this setup, it is possible to show that the global symmetry in eq.~(\ref{eq:ContinuosShift}) remains intact at any finite order in a perturbative expansion in $G_N$, 
and no power-suppressed operators are generated. 
Albeit surprising, this is not in contrast with 
the arguments presented at the beginning of this section 
 about the breaking of global symmetries due to gravity. What is the origin of this conundrum? 
 First  of all, the reader should keep in mind that the  `folk theorems' mentioned above have to do with 
 black holes, that is with non-perturbative objects. 
 This fact suggests that the breaking of global symmetries 
 induced by gravity is to be found at the non-perturbative level, and the lack
 of perturbative effects is not, after all, surprising.
 At the same time, this may sound  discouraging -- in particular from a phenomenological perspective -- since 
 referring to non-perturbative gravitational effects seems to be nothing but a vague and unclear indication.
Is there a computable non-perturbative effect showing indisputably that gravity breaks explicitly global symmetries?

A careful analysis confirms the expectation sketched above, and gives a positive answer to the aforementioned question.
Gravity breaks the global symmetry in eq.~(\ref{eq:ContinuosShift}) down to 
its discrete version in eq.~(\ref{eq:DiscreteShift})
in a non-perturbative way~\cite{Kallosh:1995hi}. In more detail, 
Euclidean wormhole solutions~\cite{Giddings:1987cg}  swallow axionic charge, break the shift symmetry it generates, 
and create a potential for the axion 
that  may compete, depending on the value of the axion decay constant, with the QCD effect in eq.~(\ref{eq:QCDpotentialterm}).
These effects are, as promised, controllably calculable -- up to some extent that we shall discuss in detail -- 
in the context of Einstein gravity, thus relying surprisingly little on our ignorance of its ultraviolet (UV) completion.

This paper is structured as follows. 
In section~\ref{sec:EuclideanWormholeSolution} we review the Euclidean wormhole solutions~\cite{Giddings:1987cg} 
in the context of Einstein gravity. Section~\ref{sec:Stability} contains the computation for the spectrum of the quadratic action, and 
in section~\ref{sec:EffectivePotential} the effective potential for a generic $U(1)$ Goldstone boson
generated by Euclidean wormholes can be found.
This concludes the theoretical part of this work.
In section~\ref{sec:Pheno} we illustrate several phenomenological implications, 
including the QCD axion, the case of ultralight scalar dark matter, and the relaxation of the electroweak scale.
In section~\ref{sec:UV} possible UV completions are discussed, and their impact on the existence of wormhole solutions assessed.  
In section~\ref{sec:GenericCoset}, we derive wormhole solutions for a generic coset and look into $O(n+1)/O(n)$ as an specific example.
Conclusions can be found in~\ref{sec:Conclusions}.
Since part of this letter reviews existing material, for quick reference main results can be found in green boxes.

\section{The Euclidean wormhole solution}\label{sec:EuclideanWormholeSolution}

In this section we review the wormhole instanton solution first discussed in~\cite{Giddings:1987cg}. 
The computation we are about to discuss is not novel  and  can be found
in the literature (see, e.g.,~\cite{Kallosh:1995hi,Lee:1988ge,Hebecker:2016dsw,Montero:2015ofa,Bachlechner:2015qja}). 
However, we consider useful and important to re-elaborate its derivation both for completeness and to highlight the
points that will matter most for the rest of this work.

\subsection{The bulk action}\label{sec:BulkAction}

The starting point is the Euclidean action of a three-form $H$ coupled to Einstein gravity\footnote{
Unlike the case of Yang-Mills theories, the Euclidean action in pure gravity is unbounded from below. 
Na\"{\i}vely, this fact seems to prevent from a well-defined path integral formulation of gravity. 
We shall discuss further this issue in section~\ref{sec:Fluctuations}.}
\begin{equation}\label{eq:WormholeAction}
\mathcal{S}_{\rm E} = \int d^4x\sqrt{g}\left(
-\frac{M_{\rm Pl}^2}{16\pi}\,\mathcal{R} + \frac{\mathcal{F}}{2}H_{\mu\nu\rho}H^{\mu\nu\rho}
\right)~,
\end{equation}
where $M_{\rm Pl} = 1/\sqrt{G_N}\simeq 1.22\times 10^{19}$ GeV is the Planck mass, and $\mathcal{F}\equiv 1/(3!\,f_a^2)$, with $f_a$ the PQ scale.
The (dimensionless) pseudo-scalar axion field $\theta$ is related to the  three-form $H$ via the 
relation $H_{\mu\nu\rho} = f_a^2\epsilon_{\mu\nu\rho\sigma}(\partial^{\sigma}\theta)$.
The variation of $\mathcal{S}_{\rm E}$ w.r.t. the metric is
\begin{equation}
\left.\delta\mathcal{S}_{\rm E}\right|_{\delta g_{\mu\nu}} = \int d^4x\sqrt{g}\left[
\frac{M_{\rm Pl}^2}{16\pi}\left(
-\frac{1}{2}g^{\mu\nu}\mathcal{R} + \mathcal{R}^{\mu\nu}
\right) + \frac{\mathcal{F}}{4}g^{\mu\nu}H^2 - \frac{3\mathcal{F}}{2}H^{\mu}_{\,\,\alpha\beta}H^{\nu\alpha\beta}
\right]\delta g_{\mu\nu}~,
\end{equation}
and the corresponding Einstein equation reads
\begin{equation}\label{eq:Einstein}
\mathcal{G}^{\mu\nu}\equiv \mathcal{R}^{\mu\nu} - \frac{1}{2}g^{\mu\nu}\mathcal{R} = 
\frac{8\pi}{M_{\rm Pl}^2}T^{\mu\nu}_H~,~~~~~
T^{\mu\nu}_H \equiv
\mathcal{F}
\left(
3H^{\mu}_{\,\,\alpha\beta}H^{\nu\alpha\beta}  - \frac{1}{2}g^{\mu\nu}H^2
\right)~.
\end{equation}
By taking the trace of eq.~(\ref{eq:Einstein}) we find $-\mathcal{R} = 8\pi G_N\mathcal{F}H^2$.
The equation of motion takes the form
\begin{equation}
\nonumber
\left.\delta \mathcal{S}_{\rm E}\right|_{\delta B_{\nu\rho}}=\int d^4x\delta B_{\nu\rho} \left[-3\mathcal F^2 \partial_\mu\left(\sqrt{g} H^{\mu\nu\rho}\right)\right] = 0~,
\end{equation}
 where we write the three form as the field strength of an antisymmetric tensor $B_{\mu\nu}$, i.e. $H_{\mu\nu\rho}=\partial_{\mu}B_{\nu\rho}+\partial_{\rho}B_{\mu\nu}+\partial_{\nu}B_{\rho\mu}$.

In terms of the angular axion field $\theta$ the Euclidean action in eq.~(\ref{eq:WormholeAction}) reads 
\begin{equation}\label{eq:AxionAction}
\mathcal{S}_{\rm E} = \int d^4x\sqrt{g}\left[
-\frac{M_{\rm Pl}^2}{16\pi}\,\mathcal{R} + \frac{f_a^2}{2}(\partial_{\rho}\theta)(\partial^{\rho}\theta)
\right]~.
\end{equation}
The canonically normalized axion field is $\phi \equiv f_a \theta$.
Notice that the kinetic term in eq.~(\ref{eq:AxionAction}) has the `wrong' sign if compared to the 
Euclidean action of an ordinary scalar. At the technical level, this is related to the Levi-Civita contraction 
$\epsilon_{\mu\nu\rho\sigma}\epsilon^{\mu\nu\rho\lambda} = (-1)^t\,3!\,\delta_{\sigma}^{\lambda}$,
where $t$ is the number of negative eigenvalues of the metric: Continuing 
from Minkowski ($t = 1$) to Euclidean ($t = 0$) space using the dual description, the axion kinetic term flips its overall sign.\footnote{
Notice that $\epsilon$ is a genuine tensor 
under general coordinate transformations, and we have the relation
\begin{equation}\label{eq:Levi}
\epsilon_{i_1,\dots,i_n} = \sqrt{|g|}\varepsilon_{i_1,\dots,i_n}~,~~~~\epsilon^{i_1,\dots,i_n} = \frac{1}{\sqrt{|g|}}\varepsilon^{i_1,\dots,i_n}~,
\end{equation}
where the tensor density $\varepsilon$ of weight $-1$  is the usual total antisymmetric  
Levi-Civita symbol, 
defined to be  $\pm 1$ and $0$ in
{\it all}
frames. In eq.~(\ref{eq:Levi}), $\sqrt{|g|} = \sqrt{g}$ if $t=0$ (or more generally if $t$ is even), and $\sqrt{|g|} = \sqrt{-g}$ if $t=1$  (or more generally if $t$ is odd).

Finally, in full generality, we have 
\begin{equation}
\epsilon_{i_1,\dots,i_k,i_{k+1},\dots,i_n}\epsilon^{i_1,\dots,i_k,j_{k+1},\dots,j_n} = (-1)^t\,k!\,\delta_{i_{k+1},\dots,i_n}^{j_{k+1},\dots,j_n}~,~~
~~~~~{\rm with}~~~
\delta_{\nu_1,\dots,\nu_p}^{\mu_1,\dots,\mu_p} \equiv\left|
\begin{array}{ccc}
\delta_{\nu_1}^{\mu_1}  & \dots  & \delta_{\nu_1}^{\mu_p}  \\
 \vdots & \ddots  &  \vdots \\
\delta_{\nu_1}^{\mu_p}  & \dots   &   \delta_{\nu_p}^{\mu_p}
\end{array}
\right|~.
\end{equation}
} We refer to appendix~\ref{app:A} for a more detailed explanation and careful derivation.
The same property appears at the level of the energy-momentum tensor. From eq.~(\ref{eq:Einstein}), we find
\begin{equation}\label{eq:TAxion}
T_{\theta}^{\mu\nu} = f_a^2\left[
-(\partial^{\mu}\theta)(\partial^{\nu}\theta) + \frac{1}{2}g^{\mu\nu}(\partial^{\rho}\theta)(\partial_{\rho}\theta)
\right]~,
\end{equation}
where we used the contraction $\epsilon^{\alpha\beta\mu\nu}\epsilon_{\alpha\beta\lambda\eta} = (-1)^t 2\left(
\delta^{\mu}_{\lambda}\delta^{\nu}_{\eta} - \delta^{\mu}_{\eta}\delta^{\nu}_{\lambda}
\right)$ with again $t = 0$ in the Euclidean case. Note that this energy-momentum tensor has the opposite overall sign if compared with the one of an ordinary real scalar.  
\begin{figure}[!htb!]
\minipage{0.5\textwidth}
  \includegraphics[width=1.\linewidth]{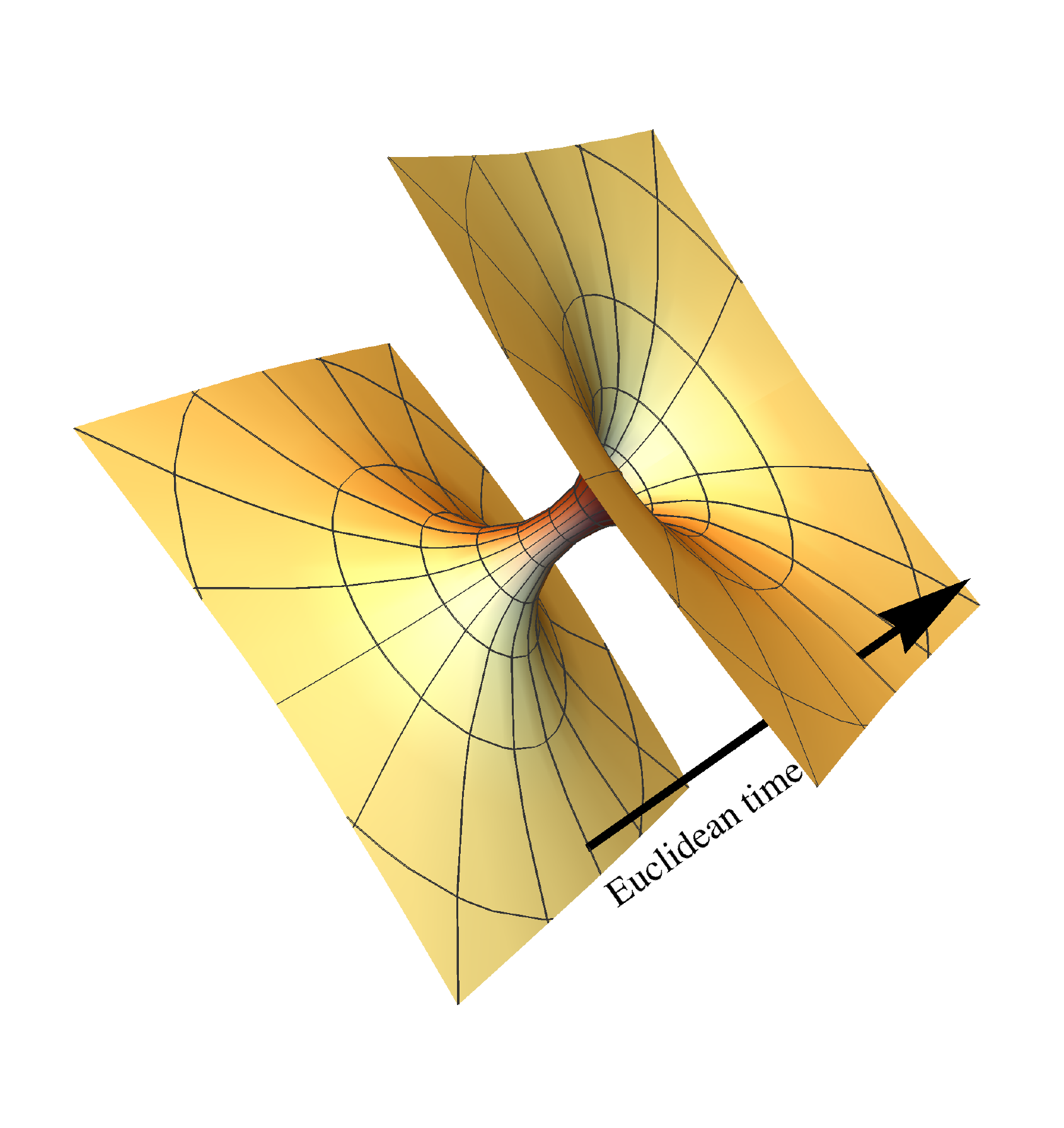}
\endminipage 
\minipage{0.5\textwidth}
  \includegraphics[width=1.\linewidth]{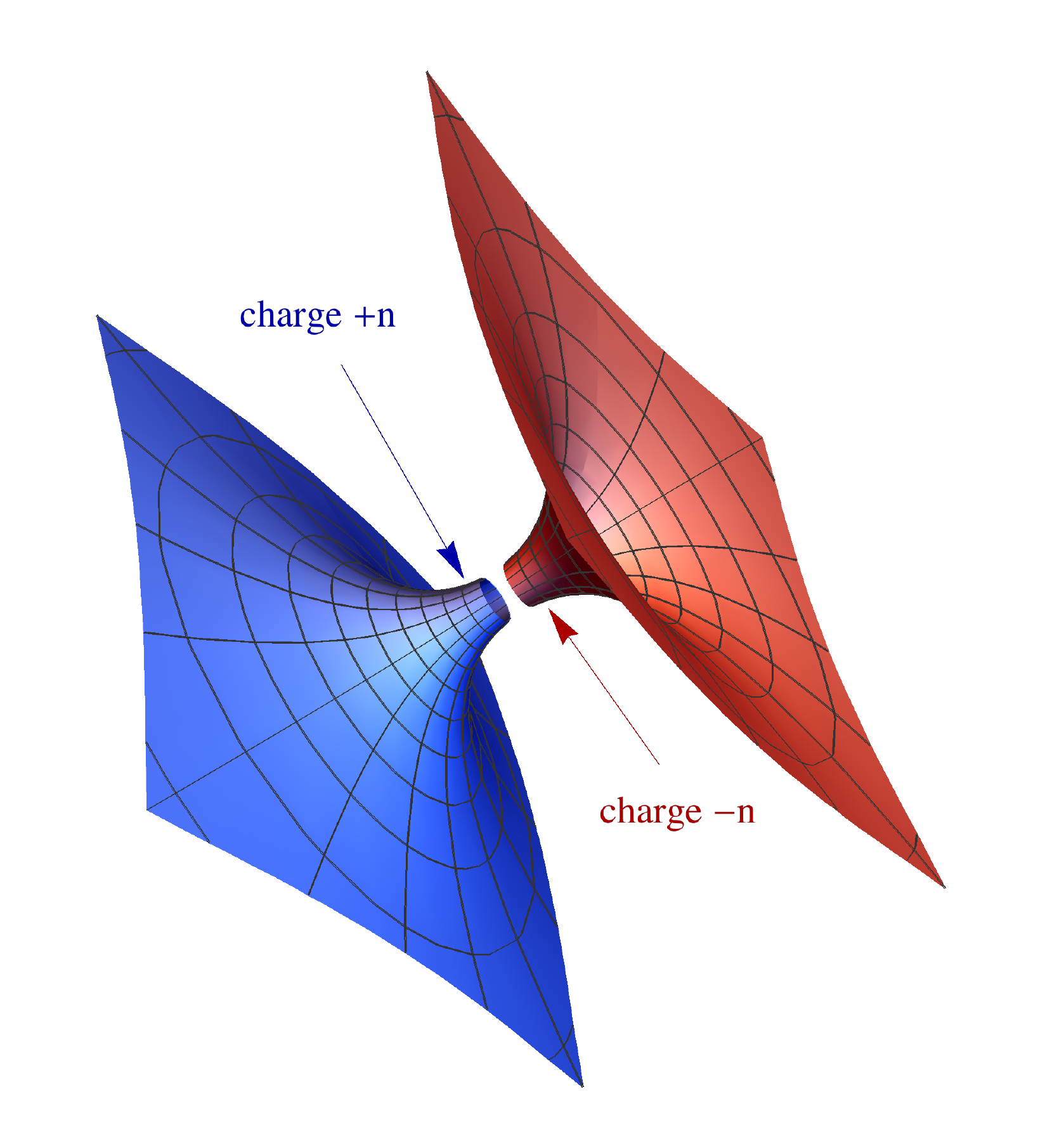}
\endminipage 
\vspace{-.2 cm}
\caption{\label{fig:Wormhole}\em 
The Euclidean wormhole solution connects two asymptotically flat regions (left panel).
The minimum size of the wormhole throat defines the characteristic length $L$ in eq.~(\ref{eq:Wormhole}).
The wormhole throat is characterized by the axion charge $n$ (see eq.~(\ref{eq:Charge})). 
This characteristic  allows for the 
possibility to interpret the wormhole as
the combination of an instanton with axion charge $+n$ and an anti-instanton with axion charge $-n$ (right panel).
}
\end{figure}
The flipped sign in eqs.~(\ref{eq:AxionAction},\ref{eq:TAxion}) plays a crucial r\^ole in what follows.
The mere existence of wormhole solutions -- as we shall see -- is a peculiar property of massless free pseudo-scalar (scalar) fields,
which admit a dual description in terms of a
two-index
antisymmetric
tensor (pseudo-tensor) whose field strength is the three-form $H$ in eq.~(\ref{eq:WormholeAction}). 
Indeed, even if we focus on the case of a pseudo-scalar axion field,  
it is important to remark that the construction we put forward in this section remains valid also for a genuine  scalar field.
 
We now look for a spherically symmetric solution with the following ansatz
\begin{equation}\label{eq:Ansatz}
ds^2 = \alpha^2(r)dr^2 + \beta^2(r)d\Omega^2_{D-1,\mathcal{K}}~,~~~~~~H_{\mu\nu\rho} = \mathcal{H}(r)\epsilon_{ijk}~,
\end{equation}
where $r$ is the time coordinate in Euclidean space, and 
the $D-1$ space has curvature $\mathcal{K}$. 
Without loss of generality, we write $d\Omega^2_{D-1,\mathcal{K}} = \hat{g}_{ij}d\varphi^i d\varphi^j$, 
with metric determinant $\sqrt{g} = \alpha\beta^3\sqrt{\hat{g}}$.
$\mathcal{K} =1$ corresponds to a closed geometry, 
and for $D=4$ $d\Omega^2_{D-1,\mathcal{K}=1}$ reduces to the ordinary three-sphere whose metric can be parametrized in terms of three angles as
$d\Omega^2_{3,1} = (d\psi^2,\sin^2\psi d\phi^2,\sin^2\psi\sin^2\phi d\varphi^2)$. 
On the contrary, $\mathcal{K} =-1$ describes an open geometry, 
with $d\Omega^2_{3,-1} = (d\psi^2,\sinh^2\psi d\phi^2,\sinh^2\psi\sin^2\phi d\varphi^2)$. 
In eq.~(\ref{eq:Ansatz}) the indices in $\epsilon_{ijk}$ run over the three-dimensional space $d\Omega^2_{3,\mathcal{K}}$, and 
they are raised and lowered by the induced metric $\hat{g}$.

The scalar curvature is
\begin{equation}
\mathcal{R} = \frac{6\left\{
\mathcal{K}\alpha^3 + \beta\beta^{\prime}\alpha^{\prime} - \alpha\left[
(\beta^{\prime})^2  + \beta\beta^{\prime\prime}
\right]
\right\}}{\alpha^3 \beta^2}~,
\end{equation}
and the Einstein equations are
\begin{eqnarray}
G_{rr} = -\frac{3\mathcal{K}\alpha^2}{\beta^2} + \frac{3(\beta^{\prime})^2}{\beta^2} &=& \frac{8\pi}{M_{\rm Pl}^2}\left( -3\mathcal{F}\alpha^2 \mathcal{H}^2\right)~,\label{eq:Einstein1} \\
G_{ij} = \hat{g}_{ij}\left[-\mathcal{K} - \frac{2\beta\beta^{\prime}\alpha^{\prime}}{\alpha^3} +\frac{(\beta^{\prime})^2}{\alpha^2} + \frac{2\beta\beta^{\prime\prime}}{\alpha^2} \right] &=& 
\hat{g}_{ij} \frac{8\pi}{M_{\rm Pl}^2}\left(3\mathcal{F}\beta^2\mathcal{H}^2\right)~.\label{eq:Einstein2}
\end{eqnarray}

Notice that the field ansatz in eq.~(\ref{eq:Ansatz}) automatically solves the equation of motion 
$\partial_{\mu}\left(
\sqrt{g}H_{\alpha\nu\rho}g^{\alpha\mu}
\right) = 0$. 
On the contrary, the Bianchi identity $\epsilon^{\mu\nu\rho\sigma}(\partial_{\rho}H_{\sigma\mu\nu})=0$ provides a non-trivial constraint. 
Using the field ansatz, and bearing in mind that $\epsilon_{ijk} = \beta^3\sqrt{\hat{g}}\varepsilon_{ijk}$ (see footnote $2$), we 
find $\partial_{r}(H\epsilon_{ijk})$ = $\partial_{r}(\mathcal{H}\beta^3\sqrt{\hat{g}}\varepsilon_{ijk})$ = 0. 
As a direct consequence, we have the $r$ dependence $\mathcal{H}(r) = \mathcal{H}_0/\beta^3(r)$, with $\mathcal{H}_0$ dimensionless constant.
To make contact with the original solution proposed in~\cite{Giddings:1987cg}, we now restrict the analysis 
to the three-sphere $S_3$, with $\mathcal{K} = 1$ and $\beta^2(r) = r^2$. The constant $\mathcal{H}_0$
can be fixed by normalizing the integral of the field strength over $S_3$. We have
\begin{equation}
 \int_{S_3} \mathcal{H}_0\,d\Omega_{3,1} = 2\pi^2\mathcal{H}_0 = n~, 
\end{equation}
where - as we shall see later - $n$ is an integer number.
We can now use eq.~(\ref{eq:Einstein1}). Solving for $\alpha$, we find
\begin{equation}\label{eq:Wormhole}
\alpha^2(r) = \frac{1}{1- L^4/r^4}~,~~~~~~~L\equiv \left(
\frac{n^2}{3\pi^3 M_{\rm Pl}^2 f_a^2}
\right)^{1/4}~.
\end{equation}
The metric in eqs.~(\ref{eq:Ansatz},\ref{eq:Wormhole}) describes the Giddings-Strominger wormhole
configuration. The wormhole connects two asymptotically flat Euclidean region, as shown in the left panel of fig.~\ref{fig:Wormhole}.
At any fixed Euclidean time $r$, the cross-section of the wormhole is a three-sphere.
The length scale $L$ in eq.~(\ref{eq:Wormhole}) defines the minimal radius of the wormhole throat.
Let us note that there is no 
singularity at $L=0$;  this is indeed nothing but a coordinate singularity, as we shall discuss later.

Using the relation $H_{\mu\nu\rho} = f_a^2\epsilon_{\mu\nu\rho\sigma}(\partial^{\sigma}\theta) =  \epsilon_{\mu\nu\rho\sigma}J^{\sigma}$
and the field ansatz in eq.~(\ref{eq:Ansatz}) we can compute the axion charge stored in a cross-section of the wormhole throat. We find
\begin{equation}\label{eq:Charge}
\mathcal{Q} \equiv \int_{S^3} f_a^2\left(\partial^r \theta\right)\beta^3(r) d\Omega_{3,1} = 
 \int_{S_3} \mathcal{H}_0\,d\Omega_{3,1} = 2\pi^2\mathcal{H}_0 = n~.
\end{equation}
The wormhole throat is therefore characterized by the axion charge $n$.
This characteristic  offers the 
possibility to consider the wormhole as
the combination of an instanton with axion charge $+n$ and an anti-instanton with axion charge $-n$ (fig.~\ref{fig:Wormhole}, right panel).
Furthermore, as we shall discuss at the end of this section, $n$ is quantized, and it only takes integer values.
As discussed in~\cite{Giddings:1987cg},  separating the wormhole in two halves has  an  important topological interpretation.
On the one hand, 
the wormhole interpolates between 
 two $\mathbb{R}^3$ geometries widely separated in time. On the other one, 
the instanton (anti-instanton) represents a topology change 
from $\mathbb{R}^3$ to $\mathbb{R}^3 \oplus S_3$ (from $\mathbb{R}^3 \oplus S_3$ to $\mathbb{R}^3$).
\begin{figure}[!htb!]
\centering
  \includegraphics[width=.7\linewidth]{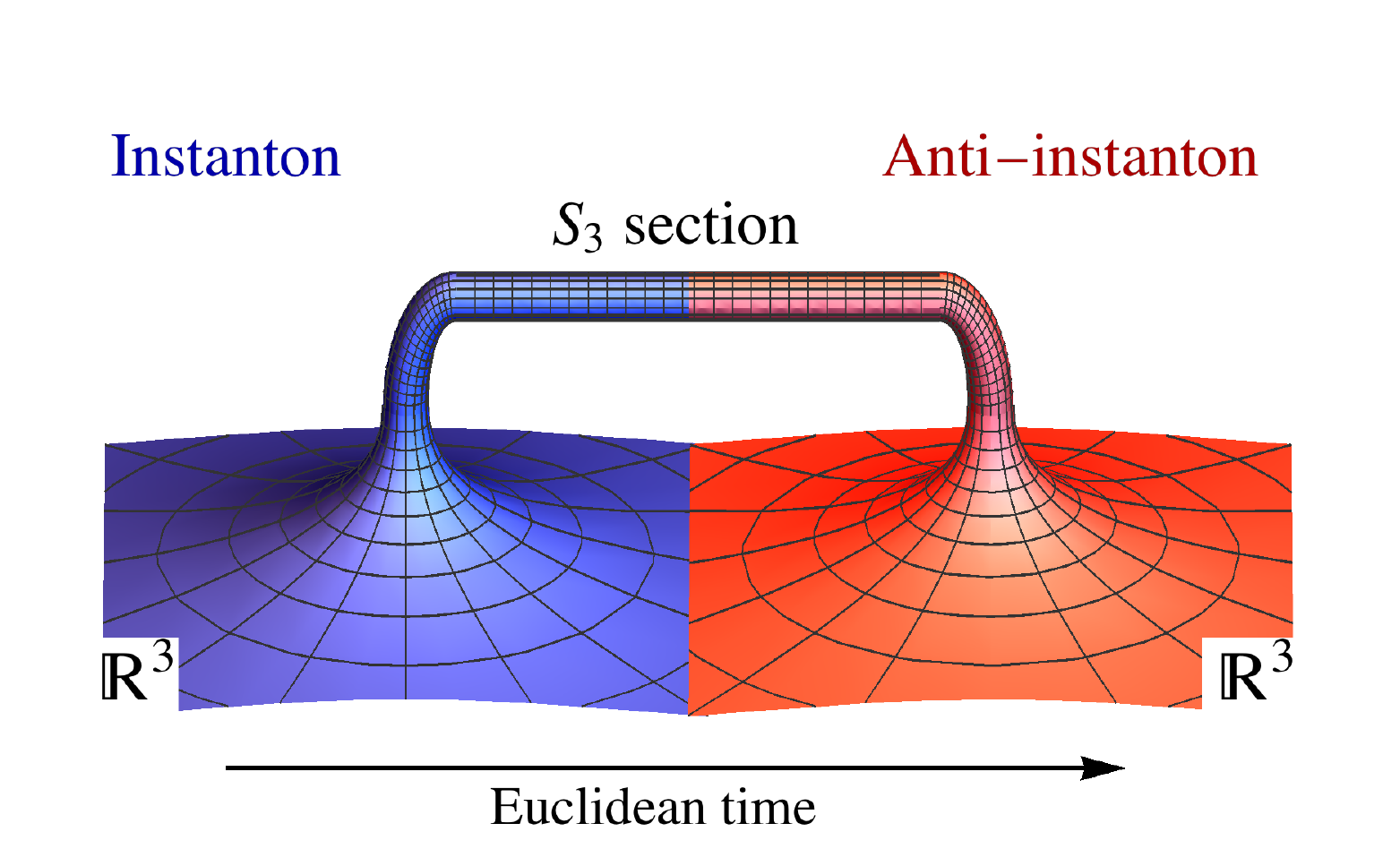}
\vspace{-.2 cm}
\caption{\label{fig:Wormhole2}\em 
Pictorial representation of an euclidean wormhole with endpoints connected
to the
same
asymptotically flat space.
The Euclidean time runs in the direction of the arrow. At each instant of time, two space dimensions have been suppressed.
The cross-section of the wormhole throat is a three-sphere $S_3$.
}
\end{figure}

Instead of considering the wormhole as connecting two separate asymptotic Euclidean spaces - a configuration that has an unclear physical  interpretation - it is possible to consider wormholes that connect back to the same space, as shown in fig.~\ref{fig:Wormhole2}.
By slicing the wormhole in half,  the instanton/anti-instanton  represents, as before,  a topology 
change 
from $\mathbb{R}^3$ to  $\mathbb{R}^3 \oplus S_3$ and viceversa. 

From eq.~(\ref{eq:WormholeAction}), we can compute the action of an instanton configuration with charge $n$.
Using $\mathcal{R} = - 8\pi G_N\mathcal{F}H^2$, 
we find\footnote{Notice that our result agrees with the original computation of~\cite{Giddings:1987cg}, and with~\cite{Kallosh:1995hi,Hebecker:2016dsw}. However, we report
a factor of $2$ discrepancy if compared with the result of~\cite{Montero:2015ofa,Bachlechner:2015qja}, where the action of the half-wormhole
 configuration is half of $\mathcal{S}_{\rm inst}$ in eq.~(\ref{eq:GIAction}).}
\begin{equation}\label{eq:GIAction}
\mathcal{S}_{\rm inst} = \int d^4x\sqrt{g}\mathcal{F}H^2 = \frac{n^2}{4\pi^4 f_a^2}\int_L^{\infty}  dr \int d\Omega_{3,1}
\frac{1}{r^3 \sqrt{1- \frac{L^4}{r^4}}} = 
\frac{\sqrt{3\pi} n M_{\rm Pl}}{8 f_a} = 
\frac{3\pi^2M_{\rm Pl}^2 L^2}{8}~.
\end{equation}
 This is the right moment to do some useful dimensional analysis, and  it is convenient to restore the appropriate powers of $\hbar$ (while keeping $c = 1$). 
 Equivalently, we can introduce  a unit of energy $\mathrm{E}$  and length $\mathrm{L}$, with $[\hbar] = \mathrm{E\,L}$.
 Natural units correspond to $\mathrm{E} = \mathrm{L}^{-1}$. It is straightforward to realize -- using the fact that a generic Lagrangian density 
 has dimension $[\mathcal{L}] = \mathrm{E}\,\mathrm{L}^{-3}$ -- that the canonically normalized axion field has dimension 
 $[a] = \mathrm{E}^{1/2}\mathrm{L}^{-1/2}$. Similarly, the dimension 
 of a coupling constant (like a gauge coupling) is $[\mathrm{g}] =  \mathrm{E}^{-1/2}\mathrm{L}^{-1/2}$.
 The axion decay constant $f_a$ is a VEV, and it shares the 
 same dimensionality of the electroweak VEV and the Planck scale: $[f_a] = [v] = [M_{\rm Pl}] =  \mathrm{E}^{1/2}\mathrm{L}^{-1/2}$.
 From eq.~(\ref{eq:Charge}) the dimensionality of the axion charge is $[n] = \mathrm{E}\,\mathrm{L}$, and 
 it corresponds to an inverse coupling squared. Finally, from eq.~(\ref{eq:Wormhole}) we see that $L$ is a genuine length scale.
 From this simple analysis we learn a number of interesting things.
 
First, we can exploit the analogy with the electroweak theory. The Fermi constant $G_F$ (or equivalently the electroweak VEV) 
is related to the ratio between the electroweak mass (defined here by the mass of the $W$
 boson, and corresponding to the physical threshold of energy $\mathrm{E}\sim M_W$ at which the 
 new degrees of freedom in the electroweak sector become dynamical) 
 and the electroweak coupling constant $\mathrm{g}_W$ by means of  
 $[v] = [G_F]^{-1/2} = [M_W]/[\mathrm{g}_W]$.
 Similarly, in gravity we expect the relation
 \begin{equation}\label{eq:StringMassScale}
 [M_{\rm Pl}] = [G_N]^{-1/2} = \frac{[M_S]}{[\mathrm{g}_S]}~,
 \end{equation}
 where we introduced a string mass scale $M_S$ and a string coupling $\mathrm{g}_S$.
 If the UV completion of general relativity is weakly coupled, we expect new degrees of freedom 
 to occur even below the Planck scale. If, on the contrary, the UV completion of general relativity is strongly coupled, $\mathrm{g}_S \gtrsim 1$,
we expect the new dynamics of quantum gravity to occur above $M_{\rm Pl}$.
We shall return on this point later, discussing the validity of the wormhole solutions (see section~\ref{sec:HigherCurv}).

Second, from eq.~(\ref{eq:GIAction}) we 
recover the correct dimensionality of an instanton action  since we have that $[\mathcal{S}_{\rm inst}] = 1/[\mathrm{g}]^2 =[\hbar]$.
On general ground, this also explains why instantons are non-perturbative.  If $\mathrm{g} \to 0$, 
$\exp(-\mathcal{S}_{\rm inst}/\hbar)$ goes to zero faster then any power of $\mathrm{g}$, and instanton effects cannot be captured at any finite order in perturbation theory.

Finally, notice that the wormholes with action given in eq.~(\ref{eq:GIAction}) realize  the Weak Gravity Conjecture (WGC) proposed in~\cite{ArkaniHamed:2006dz}.
The generalization of the WGC to the case of axions states that  for an axion with decay constant $f_a$ 
there exists an instanton with action $\mathcal{S}_{\rm inst} \lesssim M_{\rm Pl}/f_a$. 
In the context of Einstein gravity, this is precisely the wormhole with $n=1$ and $\mathcal{S}_{\rm inst} \simeq 0.38\,M_{\rm Pl}/f_a$. 

Let us now 
explore in more detail our solution. 
The wormhole metric
\begin{equation}\label{eq:WormholeMetric}
ds^2 = \left(\frac{1}{1- L^4/r^4}\right)dr^2 + r^2d\Omega_{3,1}
\end{equation}
 is conformally equivalent 
to a flat metric on $S_3 \times \mathbb{R}$. The coordinate transformation 
\begin{equation}
\left(\frac{d\tau}{dr}\right)^2 = \frac{r^2}{r^4 - L^4}~~~~~~\Longrightarrow~~~~~~\tau(r) = \frac{1}{2}\ln\left[
\frac{r^2 + \sqrt{r^4 - L^4}}{L^2}
\right]~,
\end{equation}
brings the metric in eq.~(\ref{eq:WormholeMetric}) into the form 
\begin{equation}\label{eq:ConformalFactor}
ds^2 = a^2(\tau)\left(
d\tau^2 + d\Omega_{3,1}
\right)~,~~~~a^2(\tau) = L^2\cosh(2\tau)~.
\end{equation}
Notice that $r=L$ is mapped to $\tau=0$ and in this coordinate system is plain to see that there is no singularity and
the wormhole connects the two asymptotically flat regions at $\tau=\pm\infty$.
With respect to this metric, the Einstein equations and the equation of motion for the axion field, which can be obtained from taking $\alpha=\beta=a$ in eqs.~(\ref{eq:Einstein1},\ref{eq:Einstein2}), are 
\begin{equation}\label{eq:ConformalEinstein}
-3+3\left(\frac{a^{\prime}}{a}\right)^2 = -\frac{f_a^2\kappa}{2}(\theta^{\prime})^2~,~~~~
2\frac{a^{\prime\prime}}{a} - \left(
\frac{a^{\prime}}{a}
\right)^2 - 1 = \frac{f_a^2\kappa}{2}(\theta^{\prime})^2~,~~~~\theta^{\prime\prime} + (2a^{\prime}/a)\theta^{\prime} = 0~,
\end{equation}
where we defined $\kappa \equiv 8\pi/M_{\rm Pl}^2$.
From the first Einstein equation, we find 
\begin{equation}\label{eq:AxionWormhole}
f_a^2\kappa(\theta^{\prime})^2 = \frac{6}{\cosh^2(2\tau)}~.
\end{equation}

\begin{figure}[!htb!]
\minipage{0.5\textwidth}
  \includegraphics[width=.9\linewidth]{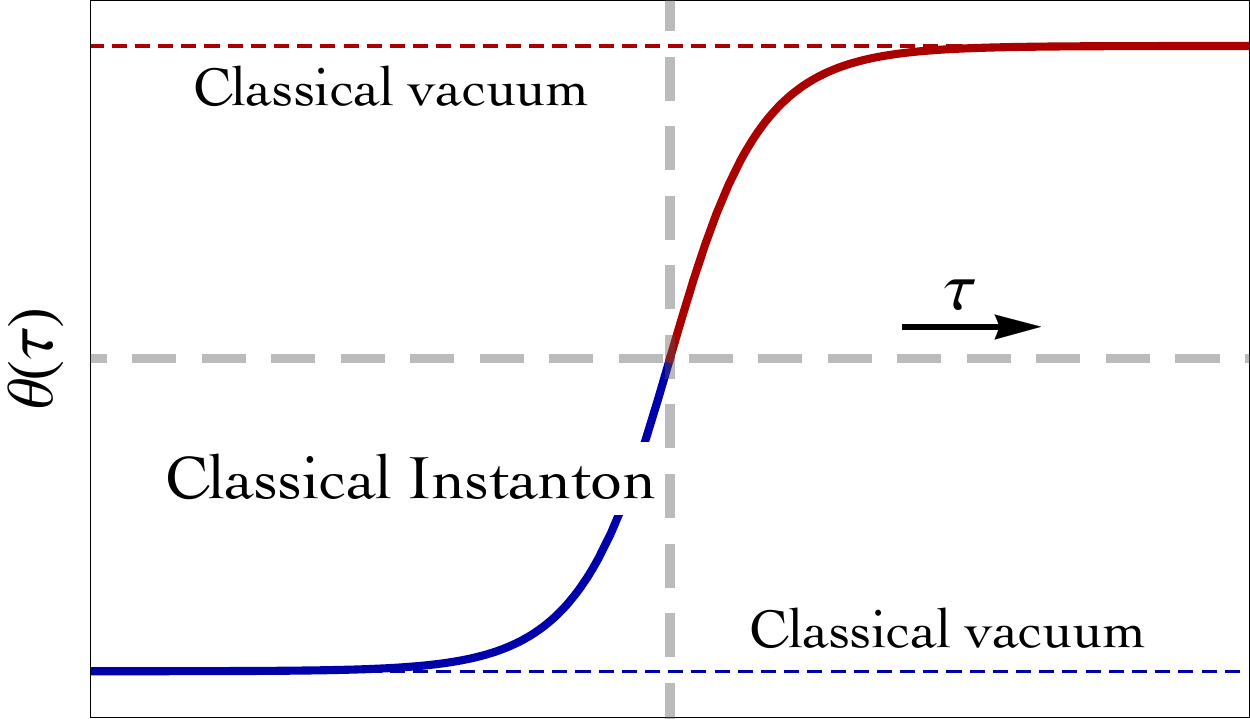}
\endminipage 
\minipage{0.5\textwidth}
  \includegraphics[width=.9\linewidth]{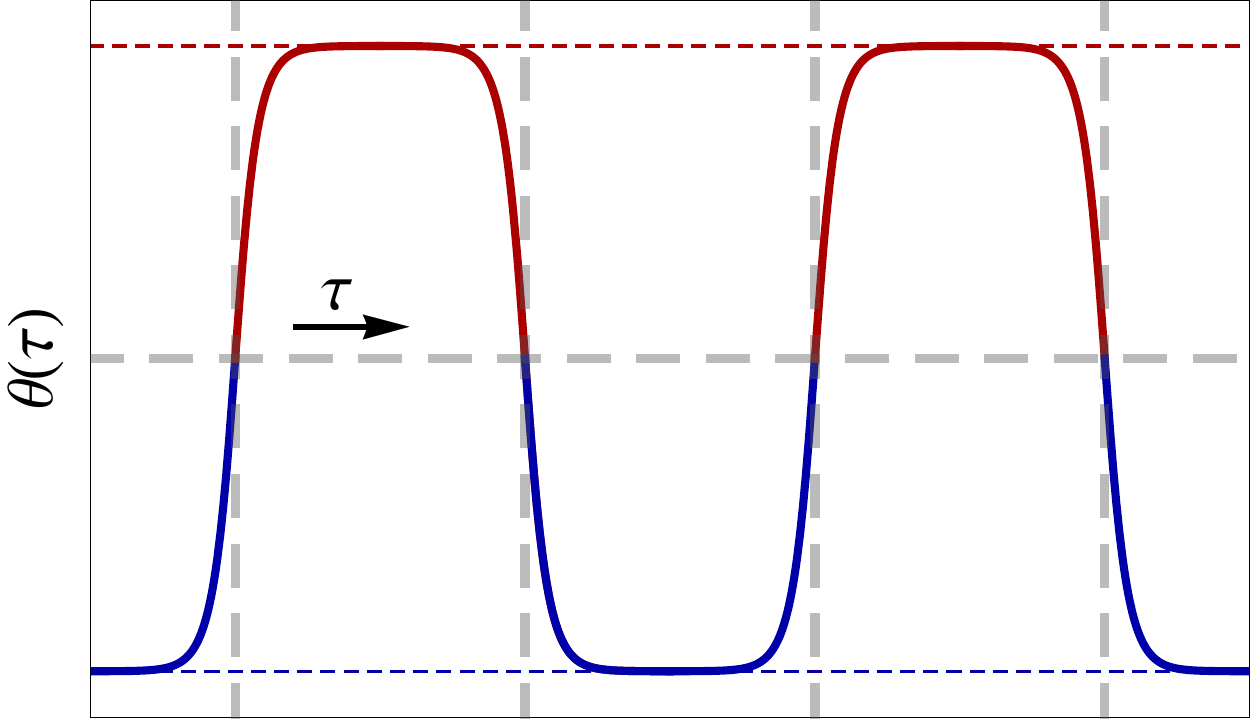}
\endminipage 
\caption{\label{fig:InstantonSolution}\em 
Axion field profile corresponding to the wormhole solution in eq.~(\ref{eq:Instanton}). The blue and red colors 
describe the instanton and anti-instanton part of the solution (see fig.~\ref{fig:Wormhole}).
A string of widely separated instantons and anti-instantons (right panel) provides an approximate 
stationary point of the Euclidean action (which becomes an exact solution of the
equation of motion only asymptotically, in the limit of infinite separation).
}
\end{figure}
Integrating over $\tau$, we find
\begin{equation}\label{eq:Instanton}
f_a^2\kappa\theta(\tau) = \sqrt{6}\arctan\left[\tanh(\tau)\right]~.
\end{equation}
In the left panel of fig.~\ref{fig:InstantonSolution} we show the instanton solution in eq.~(\ref{eq:Instanton}).
Asymptotically, the axion field configuration approaches the constant 
values $f_a^2\kappa\theta(\tau)/\sqrt{6} \overset{\tau \to \pm \infty}{\rightarrow} \pm \pi/4$,
and features  a sharp transition in between. Notice that the asymptotic limit is exponentially fast,  
$\tanh(\tau)\overset{\tau \to \pm \infty}{\rightarrow}  \pm \left(
1 - e^{-2\tau}
\right)$. Since instantons are  well-localized objects,
there are also approximate solutions consisting of strings of widely separated instantons and anti-instantons centered at 
$\tau_1,\dots, \tau_m$.
Notice that the possibility to center the instanton-anti-instanton pair 
at an arbitrary position in $\tau$ is due 
to the fact that the wormhole solution in eqs.~(\ref{eq:ConformalFactor},\ref{eq:AxionWormhole}) possesses a `time'-translation symmetry.
Indeed, the most general solution of the field equations  is $a^2(\tau) = L^2\cosh[2(\tau-\tau_0)]$, $f_a^2\kappa(\theta^{\prime})^2 = 6/\cosh^2[2(\tau-\tau_0)]$, and 
it depends on the arbitrary constant $\tau_0$.  
As we shall see in section~\ref{sec:Fluctuations}, the presence of this `time'-translation symmetry will imply the existence of a zero eigenvalue 
in the spectrum of the quadratic action describing fluctuations around the wormhole solution.
For illustrative purposes,  we show a multi-instanton solution in the right panel of fig.~\ref{fig:InstantonSolution}.
Since the $m$ instantons are widely separated, the classical action is just $m\mathcal{S}_{\rm inst}$.

Finally, we close this section with the following remark. 
As already anticipated, 
the axion charge in eq.~(\ref{eq:Charge}) turns out to be quantized.
This property can be understood as follows.
We consider the action for the axion field in the background metric of eq.~(\ref{eq:ConformalFactor}), 
and, after introducing 
the new variable $d\tau/d\sigma \equiv a^2$  and considering radially symmetric field configurations $\theta(\tau(\sigma))$, 
we find
\begin{equation}\label{eq:AxionAction2}
\mathcal{S}_{\theta} = 2\pi^2\int d\sigma\frac{f_a^2}{2}\left(
\frac{d\theta}{d\sigma}
\right)^2~.
\end{equation}
With respect to the new variable, the equation of motion of the axion field is simply given by $d^2\theta/d\sigma^2 = 0$. 
Equivalently, using eqs.~(\ref{eq:ConformalFactor},\ref{eq:AxionWormhole}),  we find $d\theta/d\sigma = {\rm const} = n/2\pi^2 f_a^2$.
We can now expand eq.~(\ref{eq:AxionAction2}) around the instanton solution, $\theta \to \theta + \delta\theta$. We find, using the equation of motion,
the following change in the action
\begin{equation}\label{eq:Quantization}
\delta\mathcal{S}_{\theta} = 2\pi^2f_a^2\int d\sigma\frac{d}{d\sigma}\left(
\delta\theta \frac{d\theta}{d\sigma}
\right) = 2\pi^2f_a^2 \delta\theta \left.\frac{d\theta}{d\sigma}\right|_{\rm boundary} = n\delta\theta~.
\end{equation}
In the presence of wormholes, the action is no longer invariant under the generic shift  $\theta \to \theta + \delta\theta$.
We now restrict to the discrete transformations $\delta\theta = 2k\pi$, with $k\in \mathbb{Z}$. Since $\theta$ is an angular variable, we know that $\theta \to \theta + 2k\pi$
corresponds to the same physical point, as discussed in section~\ref{sec:Intro}. This is a gauge redundancy, and, as such, it cannot be broken.
The simple observation is that 
quantum mechanical transition amplitudes 
involve $e^{\imath \mathcal{S}_{\theta}}$, and the phase exponential must be gauge invariant. 
This means that the only change $\delta\mathcal{S}_{\theta}$ 
that can be tolerated corresponds to an integer multiple of $2\pi$. 
We therefore find the 
quantization condition $n \in \mathbb{Z}$.

The quantization of the wormhole charge $\mathcal{Q}$ -- and the 
subsequent invariance of the action under the transformation $\theta \to \theta + 2k\pi$, with $k\in \mathbb{Z}$  -- is a crucial point, 
since it is related to the fact  that in the presence of wormholes the continuos shift symmetry
 in eq.~(\ref{eq:ContinuosShift}) is broken down 
 to its discrete subgroup in eq.~(\ref{eq:DiscreteShift}).
 We shall elaborate more on this point  in section~\ref{sec:EffectivePotential}.
 
 The quantization of the wormhole charge $\mathcal{Q}$ is tightly related to the periodicity of the axion field, and it has a 
 profound mathematical explanation. 
 It is indeed possible to show that $n$ can only be either an integer or zero.
The point is that if the flux of the three-form $H$ through $S_3$ is non-zero, then in general 
a unique two-form potential $B$ -- an antisymmetric tensor whose filed strength is $H$ according 
to $H_{\mu\nu\rho} = \partial_{\mu}B_{\nu\rho} + \partial_{\nu}B_{\rho\mu} + \partial_{\rho}B_{\mu\nu}$, see appendix~\ref{app:A}  --  covering 
the whole sphere $S_3$ does not exist. Only if $n$ is an integer the ambiguity can be bypassed 
by defining two potentials $B_{1,2}$, covering the upper and
lower halves of  $S_3$, and related to each other by a gauge transformation in an overlap
region which is topologically $S_2$~\cite{Ortin:2015hya,Witten:1983ar,Callan:1991at}. 
The argument is very similar to the Dirac quantization of the electric charge in the presence of a 
magnetic monopole. We discuss in more detail this point in appendix~\ref{app:AxionChargeQuantization}.
The analogy is indeed  deeper, and the Dirac quantization condition corresponds to the quantization of the axion charge 
over a spatial slice $S_3$~\cite{Rohm:1985jv,Allen:1991wc}.
 
\subsection{The Gibbons-Hawking-York boundary term \label{sec:GHY}}
 
 In the presence of  a manifold with boundary, the Einstein-Hilbert action in eq.~(\ref{eq:WormholeAction}) 
 must be supplemented by a boundary term so that the variational principle is well-defined~\cite{York:1971hw,York:1972sj,Gibbons:1976ue}.
 To make this point clear, let us only focus on the term involving the Ricci scalar
 for the metric ansatz in eq.~(\ref{eq:ConformalFactor}) with $\mathcal{R} = 6(a - a^{\prime\prime})/a^3$
\begin{equation}\label{eq:VariationalProblem}
\mathcal{S} = -\frac{3M_{\rm Pl}^2}{8\pi}\int d\Omega_{3,1} d\tau \left(a^2 - a a^{\prime\prime}\right)  \\
= -\frac{3M_{\rm Pl}^2}{8\pi}\int d\Omega_{3,1} d\tau
\left[
a^2 + (a^{\prime})^2 - \frac{d(a a^{\prime})}{d\tau}
\right]~.
\end{equation}
The presence of the second derivative prevents the correct formulation of the variational principle.\footnote{As a prototype, consider the action 
$S = \int_{t_1}^{t_2}dt L(q,\dot{q}) = \frac{1}{2}\int_{t_1}^{t_2}dt\,\dot{q}^2$ with variation
\begin{equation}
\delta S = \int_{t_1}^{t_2} dt\left(
\frac{\partial L}{\partial q} - \frac{d}{dt}\frac{\partial L}{\partial \dot{q}}\right)\delta q + 
\left.
\frac{\partial L}{\partial \dot{q}}\delta q
\right|_{t_1}^{t_2} = -\int_{t_1}^{t_2} dt\,\ddot{q}\delta q + \left.
\dot{q}\delta q
\right|_{t_1}^{t_2}~.
\end{equation}
In order to extract the equation of motion by means of the variational principle, it is necessary to 
specify boundary data prescribing the values of the variation at the boundaries $\delta q(t_2)$, $\delta q(t_1)$.  

Let us now suppose to add a total derivative term
\begin{equation}
\tilde{L} = \frac{1}{2}\dot{q}^2 - \frac{d}{dt}\left(
\frac{1}{2}q\dot{q}
\right) = -\frac{1}{2}q\ddot{q}~~~~~\Longrightarrow~~~~~\delta \tilde{L} = -\ddot{q}\delta q + \frac{1}{2}\frac{d}{dt}\left(
\dot{q}\delta q - q\delta\dot{q}
\right)~.
\end{equation}
The equation of motion remains unchanged, since a total derivative does not affect the dynamics. However, in order to apply the variational principle, 
four boundary data are needed, namely $\delta q(t_2)$, $\delta q(t_1)$, $\delta \dot{q}(t_2)$, $\delta \dot{q}(t_1)$.
This is inconsistent since fixing position and velocity violates the uncertainty principle.  It is therefore necessary to add a boundary term which  
cancels the extra total derivative in $\tilde{L}$. 
This example mimics the typical situation in general relativity -- and, in particular, the case of eq.~(\ref{eq:VariationalProblem}) -- with the r\^ole of $\tilde{L}$ played by the Einstein-Hilbert action.
In fact, the Ricci scalar has second derivative of the metric, and its variation generates a total derivative that is precisely 
canceled by the addition of the Gibbons-Hawking-York boundary term.
}
It is therefore necessary to add the Gibbons-Hawking-York boundary term, whose general form is given by 
\begin{equation}
\mathcal{S}_{\rm GHY} = -\frac{M_{\rm Pl}^2}{8\pi}\int_{\partial M} d^3\vec{x}\sqrt{h}\left(
K - K_0
\right)~,
\end{equation}
where $h$ is the determinant of the induced metric on the boundary manifold $\partial M$, $K$ is the trace of the extrinsic curvature tensor, and $K_0$ is the trace of the 
extrinsic curvature tensor of the same  boundary embedded in
flat spacetime. The explicit computation  
of $\mathcal{S}_{\rm GHY}$ is straightforward. 
The trace of the extrinsic curvature tensor is $K = \nabla_{\mu}n^{\mu} = \partial_{\mu}n^{\mu} + \Gamma^{\mu}_{\mu\nu}n^{\nu}$, 
where $n^{\mu}$ is the unit normal to $\partial M$. 
The boundary we are interested in are hypersurfaces of constant 
$\tau$, and the corresponding unit normal vector is $n^{\tau} = 1/a$. 
By direct computation, we find $K = 3a^{\prime}/a^2$. As far as $K_0$ is concerned, we use $a = e^{\tau}$ for the flat metric, and we find $K_0 = 3/a$.
All in all, the Gibbons-Hawking-York boundary term reduces to 
\begin{equation}\label{eq:GHY}
\mathcal{S}_{\rm GHY} = -\frac{3M_{\rm Pl}^2}{8\pi}\int d\Omega_{3,1}\left(
a a^{\prime} - a^2
\right)~,
\end{equation}
and it cancels the total derivative in eq.~(\ref{eq:VariationalProblem}).

The Gibbons-Hawking-York boundary term must be included for a correct formulation of the gravitational action in the presence of a manifold with boundary, and this is the case for the half-wormhole configuration. It is therefore important to evaluate the topological contribution of eq.~(\ref{eq:GHY}) to the bulk action in eq.~(\ref{eq:GIAction}).
Using the explicit solution in eq.~(\ref{eq:ConformalFactor}), we find
\begin{equation}
\mathcal{S}_{\rm GHY} = -\frac{3M_{\rm Pl}^2\pi}{4}\left|
aa^{\prime} - a^2
\right|_{\tau = 0}^{\tau = \infty} =-\frac{3M_{\rm Pl}^2\pi L^2}{4} =  -\left(
\frac{2}{\pi}
\right)\frac{\sqrt{3\pi} n M_{\rm Pl}}{8 f_a}~,
\end{equation}
where only the boundary at the wormhole throat gives a non-vanishing contribution.
Taking into account the boundary term, the action of an instanton configuration with charge $n$ is
\begin{mymathbox}[ams gather, title=Half-wormhole action, colframe=titlepagecolor]
\mathcal{S}_{\rm inst} =
\frac{\sqrt{3\pi} n M_{\rm Pl}}{8 f_a}\left(
1 - \frac{2}{\pi}
\right)~.\label{eq:Fullaction}
\end{mymathbox}

\section{Quadratic action and the absence of negative modes}\label{sec:Stability}

Instantons are responsible for 
one of the most important effect in Quantum Mechanics, that is the tunneling through a potential barrier. 
\begin{figure}[!htb!]
\minipage{0.5\textwidth}
  \includegraphics[width=.9\linewidth]{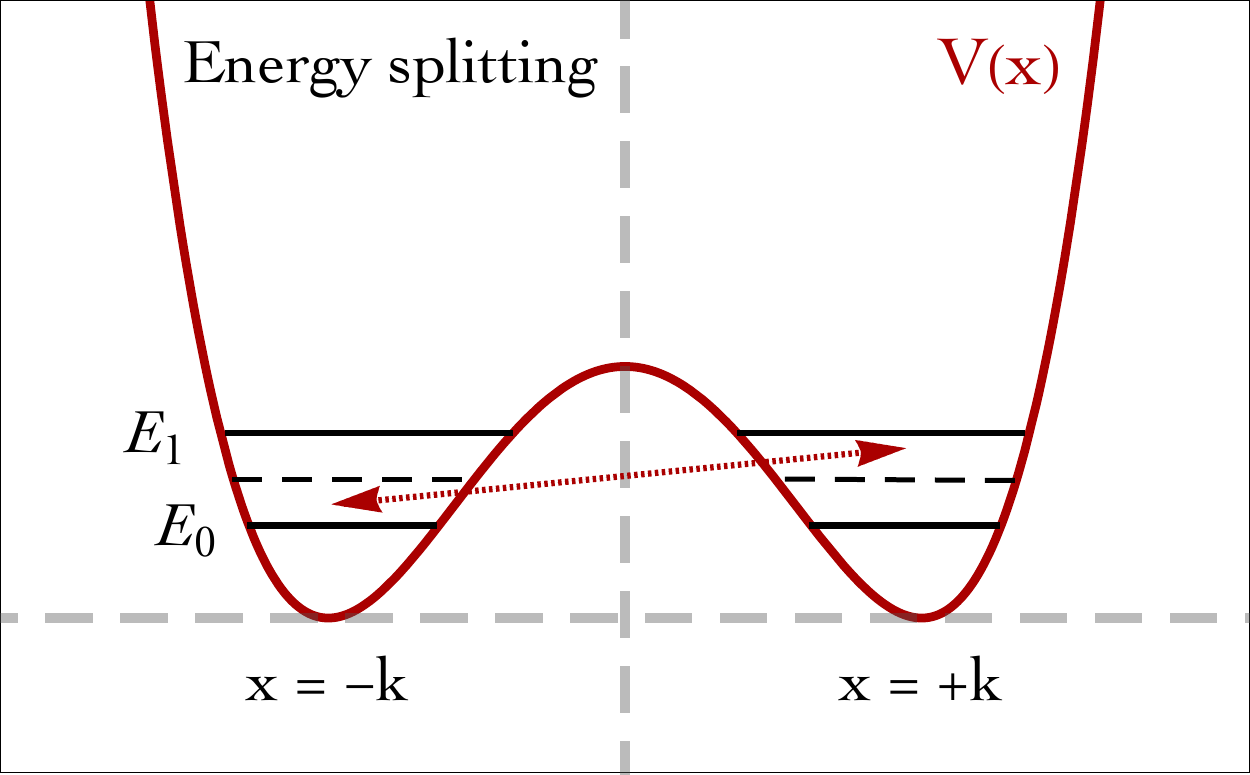}
\endminipage 
\minipage{0.5\textwidth}
  \includegraphics[width=.9\linewidth]{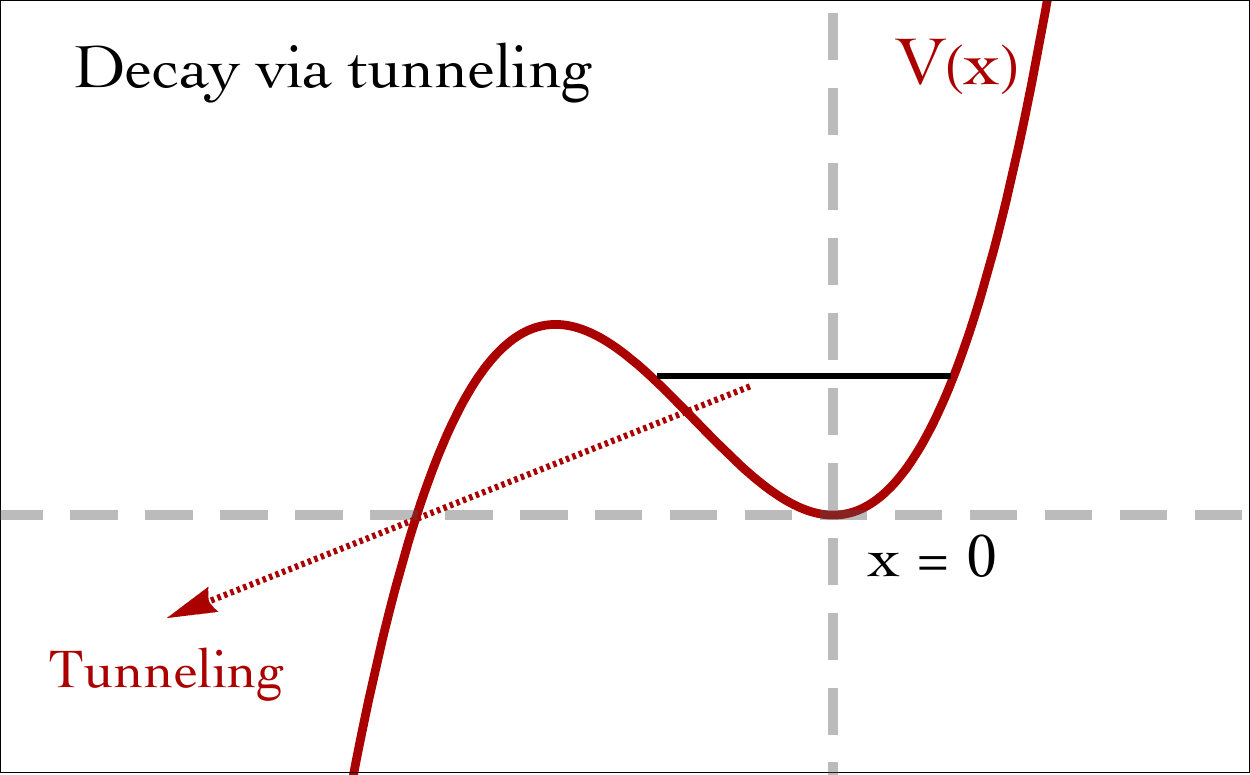}
\endminipage 
\caption{\label{fig:Instanton}\em 
Tunneling through a potential barrier in Quantum Mechanics mediated by non-perturbative instanton solutions.
}
\end{figure}
The importance of such effect lies in the fact that it changes the vacuum of the system, 
and there exist two paradigmatic situations in which this effect is particularly manifest~\cite{ColemanBook,Vainshtein:1981wh}.

In the left panel of fig.~\ref{fig:Instanton} we show a one-dimensional quantum system featuring a 
potential with a double well, of the form $V(x) =\lambda(x^2 - k^2)^2$. In perturbation theory, 
the ground state has two degenerate minima at $x = \pm k$, which are equivalent
because of the reflection symmetry of the system (dashed black lines). 
Instantons lift the degeneracy, and the energy difference between the true ground state and the first excited state (solid black lines -- respectively, the anti-symmetric and symmetric combination of the two vacua at $x = \pm k$) is controlled by the
 instanton action, according to the qualitative scaling 
$E_1(\mathrm{g}) - E_0(\mathrm{g}) \sim e^{-{\rm const}/\mathrm{g}^2}$. 

In the right panel of fig.~\ref{fig:Instanton} we show a potential with a metastable vacuum, of the form $V(x) = m\omega^2 x^2 + \lambda x^3$.
In perturbation theory -- assuming small $\lambda$ -- one can compute the energy of the 
ground state of the system at, in principle, any order (solid black line). 
However, perturbation theory can not describe the  tunneling through the barrier, and the instability of the vacuum.
Again, the tunneling is captured by instanton effects, that in this case introduce a small imaginary part in the ground state energy,
$E_0(\mathrm{g}) = \Re E_0(\mathrm{g}) + \imath\Im E_0(\mathrm{g})$, with $\Im E_0(\mathrm{g}) \sim e^{-{\rm const}/\mathrm{g}^2}$,
which in turn becomes a decay probability in the time evolution $e^{\imath E_0(\mathrm{g})t}$. 
In this case the instanton is called {\it bounce}.

In order to understand the true nature of the instanton, it is necessary to inspect the action at the quadratic order.
Let us clarify this point with an explicit example.  In Euclidean time, 
the  quantum probability amplitude  from  the initial state $|i\rangle$ at time $-T/2$ to the final state 
$| f \rangle$ at time $T/2$ -- that corresponds, for instance, to the transition amplitude between
 the two vacua  in the quantum systems discussed in fig.~\ref{fig:Instanton}  -- is given by
\begin{equation}\label{eq:FluctuationDet}
\langle f| e^{-T H} | i  \rangle = \sum_{n}e^{- TE_n} \langle f| n\rangle  \langle n | i  \rangle   \approx
\frac{\mathcal{N}}{\sqrt{{\rm det}^{\prime}\left(
\frac{\delta^2 \mathcal{S}_{\rm E}}{\delta x^2}
\right)}}e^{-\mathcal{S}_{\rm E}[x_{\rm cl}(\tau)]}~.
\end{equation}
where $\mathcal{N}$ is a normalization constant, $\mathcal{S}_{\rm E}[x_{\rm cl}(\tau)]$ is the Euclidean action computed for the classical 
instanton solution  $x_{\rm cl}(\tau)$, and the prefactor encodes the
quantum
fluctuation around the
classical trajectory. The latter, as a result of a Gaussian integration, takes the shape of a determinant with a prime denoting the
  subtraction of  zero eigenvalues. If we take $T$ to be very large, the lowest energy state dominates the sum,
   and we can extract the ground state energy from the computation of the path integral.
  The spectrum of the differential operator $\delta^2 \mathcal{S}_{\rm E}/\delta x^2$ is therefore crucial.
  In the presence of one (or an odd number of) negative eigenvalue(s), there is an imaginary contribution to the vacuum energy, and 
 the instanton plays the r\^ole of a bounce. If the spectrum of the quadratic action is positive (or if the number of negative eigenvalues is even) 
 the vacuum energy is real, and the instanton describes the transition between degenerate vacua, lifting their degeneracy as discussed in the case of the double well.

The analogy with one-dimensional systems in Quantum Mechanics suits particularly well our case since 
the wormhole solutions 
discussed in section~\ref{sec:EuclideanWormholeSolution} are characterized by one relevant dimension, that is the
time coordinate in the Euclidean 4D space.

\subsection{Bulk action and boundary terms}\label{sec:Fluctuations}

In this section we study, motivated by the previous discussion, 
the spectrum of the quadratic action describing perturbations around the Euclidean wormhole solution. 
To this end, we shall exploit the formalism developed 
in the context of  cosmological perturbation theory for the analysis 
of fluctuations of a scalar field $\varphi_0$ coupled to gravity in a Friedmann-Robertson-Walker metric~\cite{Garriga:1997wz}. 
This study requires the canonical formalism in order to 
correctly get rid of the non-dynamical degrees of freedom, as in any other gauge theory.
The canonical formulation, although non-explicitly, preserves gauge invariance.

In the following, we start considering metric perturbations in Minkowski space around the background line element $ds^2 = a^2(\eta)(- d\eta^2 + \gamma_{ij}dx^i dx^j)$. 
Including fluctuations, in full generality we write
$ds^2 = a^2(\eta)\left[
-(1+2A)d\eta^2 + 2B_i dx^id\eta  +(\gamma_{ij} + h_{ij})dx^i dx^j
\right]$. 
This perturbed line element can be further analyzed by means of the scalar-vector-tensor decomposition,  
which separates the fluctuations into components according to their transformations under spatial rotations.
 In particular, the three-vector $B_i$ splits into the gradient of a scalar, and a divergence-free vector $B_i = \partial_i B + \hat{B}_i$ while the symmetric tensor $h_{ij}$ can 
 be decomposed into a scalar part, a divergence-free vector, and a trace-free transverse tensor 
 $h_{ij} = -2\psi \gamma_{ij} + 2\partial_{\langle i}\partial_{j\rangle}E + 2\partial_{(i}\hat{E}_{j)} + t_{ij}$.\footnote{
 We use the following short-hand notation: $\partial_{\langle i}\partial_{j\rangle}E \equiv (\partial_i\partial_j - \frac{1}{3}\gamma_{ij}\nabla^2)E$, and
 $\partial_{(i}\hat{E}_{j)} \equiv (\partial_i\hat{E}_j + \partial_j \hat{E}_i)/2$.}
The advantage of this decomposition is that  scalar, vector and tensor Einstein 
equations are decoupled at the
 linear order and can therefore be treated separately.
 Vector perturbations are pure gauge modes, and they carry no dynamics~\cite{Garriga:1997wz}. As far as tensor perturbations are concerned, they do not couple to scalar modes, 
 and we can therefore rely on standard computations of gravitational waves in instanton background~\cite{Hertog:1999kg}.
 We found a positive spectrum, 
 and we summarize our computation in appendix~\ref{app:Perturbations}.
 
 We focus here on scalar perturbations. 
The perturbed line element is therefore
\begin{equation}\label{eq:Perturbations}
ds^2=a^2(\eta)\left\{-(1+A) d\eta^2+2\partial_i B dx^id\eta+[\gamma_{ij}(1-2\psi)+2\partial_{\langle i}\partial_{j\rangle}E]dx^idx^j\right\}~,
\end{equation}
where we simultaneously expand the scalar field, assumed to be function of $\eta$, as $\varphi_0+\delta\varphi$.

The second order Lagrangian in terms of the perturbations in eq.~(\ref{eq:Perturbations}) can be found in e.g.~\cite{Garriga:1997wz}, 
but it has, at this level, non-dynamical degrees of freedom and gauge redundancy.
First $A$ and $B$ are not dynamical, that is they do not have a `kinetic term'; they behave like $A_0$
in a `conventional' gauge theory. In addition, there is a two-parameter gauge transformation, 
corresponding to diffeomorphism acting on scalar perturbations, given by $\delta x^\mu=(\lambda^0, \partial^i \lambda)$, and we have
\begin{equation}\label{eq:GaugeonPer}
\delta\psi=-\frac{\dot a}{a}\lambda^0~,~~~~~~~\delta B=\dot \lambda-\lambda^0~,~~~~~~~ 
\delta E=\lambda~,~~~~~~~  \delta(\delta \varphi)=\dot\varphi_0 \lambda^0~,
\end{equation}
where we indicate with $\dot{}$ the derivative w.r.t. the time variable $\eta$. Out
of the 5 perturbations $A,B,E,\psi,\delta\varphi$, 2 are non dynamical, and 2 can be removed via a gauge transformation which leaves us with 1 d.o.f.
 All in all, we have the following quadratic action in Hamiltonian form $p\dot q-\mathscr H$ (see~\cite{Garriga:1997wz} for details of the derivation)
\begin{eqnarray}\label{eq:FullQuadratic}
&&\delta \mathcal{S}^{(2)} = \\ && \int d\eta d\vec{x}\left\{
\Pi_\Psi \dot \Psi -\frac{2a^2\sqrt{\gamma}}{\kappa^2 \dot\varphi_0 ^2}\left[
\left(\Delta + 3\mathcal{K}\right)\Psi + \frac{\kappa \dot{a}/a}{2a^2\sqrt{\gamma}\Pi_\Psi}
\right]^2 - \frac{a^2 \sqrt{\gamma}}{\kappa}\Psi\left(\Delta + 3\mathcal{K}\right)\Psi
+\frac{1}{4a^2\sqrt{\gamma}}\Pi_\Psi\frac{\kappa\mathcal{K}}{(\Delta + 3\mathcal{K})}\Pi_\Psi
\right\}~,\nonumber
\end{eqnarray}
where the dynamical variable $\Psi$
and its conjugate momentum $\Pi_\Psi$ in terms of the metric perturbations in eq.~(\ref{eq:Perturbations}) are
$\Psi \equiv \psi + (\dot{a}/a)\delta\varphi/\dot\varphi_0$ and 
$\Pi_\Psi  \equiv \Pi_{\psi} - 2a^2\sqrt{\gamma}(\Delta + 3\mathcal{K})\delta\varphi/\kappa\dot{\varphi}_0$, 
where $\Pi_{\psi}$ is the conjugate momentum of $\psi$ and $\Delta$ is the Laplacian on the three-sphere associated with $\gamma_{ij}$.
The background solutions satisfy
\begin{equation}\label{eq:MinkowskiEOM}
\left(\frac{\dot a}{a}\right)^2+\mathcal K -\frac\kappa6\dot\varphi_0 ^2=0~,~~~~~~~~~~~~~\ddot\varphi_0+2\frac{\dot a\dot\varphi_0}{a} =0~.
\end{equation}
In the following we focus on the spatially  homogeneous  $O(4)$ invariant modes, and we 
formally set $\Delta = 0$.\footnote{In a closed Universe, $\mathcal{K}=1$, 
the line element on $S_3$ is $dl^2 = \gamma_{ij}dx^i dx^j = d\psi^2 + \sin^2\psi(d\phi^2 + \sin^2\phi d\varphi^2)$. 
The scalar spherical harmonics $Q^{(n)}_{lm}(\psi,\phi,\varphi)$ are 
eigenfunctions of the Laplacian operator on $S_3$ 
with eigenvalue equation $\Delta Q^{(n)}_{lm} = -(n^2 - 1)Q^{(n)}_{lm}$, $n\in \mathbb{N}^+$. Because of the 
degeneracy in $l$ and $m$, the most general solution of the eigenvalue 
equation is $Q^{(n)}(\psi,\phi,\varphi) = \sum_{l=0}^{n-1}\sum_{m = -l}^{+l}A_{lm}^{(n)}Q^{(n)}_{lm}(\psi,\phi,\varphi)$ with 
$A_{lm}^{(n)}$ arbitrary constants. Explicitly, we have $Q^{(n)}_{lm}(\psi,\phi,\varphi) = \Pi_l^n(\psi)Y_{lm}(\phi,\varphi)$ where 
$Y_{lm}(\phi,\varphi)$ are the usual spherical harmonics on $S_2$ and $\Pi_l^n(\psi)$ are the Fock harmonics. The eigenfunction corresponding to $n = 1$ 
with zero eigenvalue defines the spatially homogeneous modes. 
In a open Universe, $\mathcal{K}= -1$, the Laplacian is again zero for the spatially homogeneous modes, 
and takes the values $-p^2 - 1$, with $p^2 > 0$, 
for the continuum of square integrable modes.} 
As discussed in~\cite{Gratton:2000fj,Gratton:1999ya}, the homogeneous modes are 
those that are potentially responsible for the presence of negative eigenvalues in the spectrum of the quadratic action, and 
they need to be carefully analyzed. 
Indeed, in~\cite{Rubakov:1996cn} the authors claimed the existence of a negative mode among the homogenous fluctuations 
around the Giddings-Strominger wormholes. 
If true, this result would imply that wormholes are bounce solutions mediating unstable vacuum decay via tunneling transition.
Motivated by the necessity to further explore and check this result, we now turn to discuss our own analysis. 
For completeness, in appendix~\ref{app:Perturbations} we discuss the spectrum of inhomogeneous scalar perturbations.

Restricting to the homogeneous modes, eq.~(\ref{eq:FullQuadratic}) becomes 
\begin{equation}
\left.\delta \mathcal{S}^{(2)}\right|_{\Delta = 0} = \int d\eta d^3\vec{x}\left[ \Pi_\Psi \dot \Psi-\left(\begin{array}{cc}\Psi&\Pi_\Psi \end{array} \right)\left(\begin{array}{cc}
\frac{3\mathcal{K}a^2\sqrt{\gamma}(6\mathcal K+\kappa \dot \varphi_0^2)}{\kappa \dot\varphi_0 ^2}& \frac{3\mathcal K  \dot a}{\kappa a \dot\varphi_0 ^2}\\
\frac{3\mathcal K  \dot a}{\kappa a \dot\varphi_0 ^2}& \frac{6\dot a^2-\kappa a^2 \dot\varphi_0 ^2}{12 a^4\sqrt{\gamma}\dot\varphi_0 ^2}
\end{array}
\right)\left(\begin{array}{c}\Psi\\\Pi_\Psi \end{array} \right)\right]~.\label{S2Psi}
\end{equation}

The first difficulty one faces is the conformal factor problem~\cite{Gibbons:1978ac}.\footnote{The natural 
definition of path integral in pure gravity involves the Euclidean partition function
\begin{equation}
\mathcal{Z}=\int_{\mathcal{M}}[\mathfrak{D}g]e^{-\mathcal{S}_{\rm E}[g]}~,~~~~~~
\mathcal{S}_{\rm E}[g] = -\frac{M_{\rm Pl}^2}{16\pi}\int_{\mathcal{M}}d^4x\sqrt{g}\mathcal{R}
-\frac{M_{\rm Pl}^2}{8\pi}\int_{\partial\mathcal{M}}d^3\vec{x}\sqrt{h}(K - K_0)~,
\end{equation}
where the sum runs over all metrics on a four-dimensional manifold $\mathcal{M}$ with boundary $\partial\mathcal{M}$. 
The Euclidean path integral is ill-defined due to the fact that the scalar curvature can be made arbitrarily negative. 
The physics behind is related to the fact that the gravitational potential energy is negative because gravity is attractive. 
A concrete way to look at the problem is to perform  the conformal transformation $g_{\mu\nu}\to \tilde{g}_{\mu\nu} = \Omega^2 g_{\mu\nu}$ which 
brings the Euclidean action into the form
\begin{equation}
\mathcal{S}_{\rm E}[g,\Omega] = -\frac{M_{\rm Pl}^2}{16\pi}\int_{\mathcal{M}}d^4x\sqrt{g}
\left(
\Omega^2\mathcal{R} +6g^{\mu\nu}\nabla_{\mu}\Omega\nabla_{\nu}\Omega\right) - \frac{M_{\rm Pl}^2}{8\pi}\int_{\partial\mathcal{M}}d^3\vec{x}\sqrt{h}\Omega^2(K-K_0)~.
\end{equation}
The Euclidean action can be made arbitrarily negative by choosing a rapidly varying conformal factor. 
In turn,  this implies that the path integral diverges since one has to integrate over all possible $\Omega$.
An alternative formulation of the problem, closer in spirit to the content of this section, consists in performing an expansion of the Euclidean action 
around a fixed on-shell metric $g$ obeying the classical field equations $R_{\mu\nu}[g]= 0$.  At the quadratic order in the fluctuation 
$h_{\mu\nu}$ we have~\cite{Mazur:1989by}
\begin{equation}\label{eq:Mottola}
\delta\mathcal{S}_{\rm E}^{(2)} = \frac{M_{\rm Pl}^2}{32\pi}\int_{\mathcal{M}}d^4x\sqrt{g}\left[
h^{{\rm T}\mu\nu}\left(
2\nabla^{\sigma}\nabla_{\mu}h^{\rm T}_{\sigma\nu} - \nabla^2h^{\rm T}_{\mu\nu} - \nabla_{\mu}\nabla_{\nu}h
\right) + \frac{3}{8}h\nabla^2h
\right]~,
\end{equation}
where we decomposed the fluctuation into its trace and a trace-free part, $h_{\mu\nu} = \frac{1}{4}hg_{\mu\nu} + h_{\mu\nu}^{\rm T}$.
The quadratic action in eq.~(\ref{eq:Mottola}) reveals that the trace part of the perturbation is unbounded from below, since 
the eigenvalue equation for the Laplace-Beltrami operator, $-\nabla^2u=\lambda u$, admits only positive eigenvalues, $\lambda \geqslant 0$,  
in a compact Riemannian manifold (a result known as the Lichnerowicz-Obata theorem).
}
The conformal factor 
problem can be spotted in the Hamiltonian of 
eq.~(\ref{S2Psi}) by using the equation of motion $6\dot{a}^2-\kappa a^2\dot\varphi_0^2 = -6\mathcal K a^2$.
 As a consequence, we find a `wrong sign' kinetic term since we will be looking at the closed  case $\mathcal K=1$. The literature often treats this problem by just turning $\Psi$ into the imaginary axis 
but this, other than pretty arbitrary, 
will lead to augmented confusion when convoluted with the rotation to Euclidean. Instead the most sensible way in our humble opinion is to perform a canonical (symplectic) transformation and study the resulting system as done in~\cite{Gratton:2000fj}. This will correspond roughly to swapping $\Psi$ and $\Pi_\Psi$ and then performing another 
transformation, explicitly and in a single step
\begin{equation}\label{eq:CanonicalTransformation}
\Pi_\Psi =-\frac{3a^2\sqrt{\gamma}}{\dot\varphi_0}q-\frac{\dot\varphi_0 \dot a}{\kappa a}p~,~~~~~~~~~
\Psi =-\frac{\kappa a}{2\dot\varphi_0 \dot a}q+\frac{\dot\varphi_0}{6a^2\sqrt{\gamma}}p~,
\end{equation}
which is symplectic since
\begin{equation}
J^{\rm T}
\left(
\begin{array}{cc}
 0 & 1    \\
 -1 & 0     
\end{array}
\right)J = 
\left(
\begin{array}{cc}
 0 & 1    \\
 -1 & 0     
\end{array}
\right)
~~~~~~{\rm with}~~~~~\left(\begin{array}{c}\Psi\\\Pi_\Psi
\end{array}\right)=J \left(\begin{array}{c}p\\q
\end{array}\right)
~.
\end{equation}
and leads to the action
\begin{equation}
\left.\delta \mathcal{S}^{(2)}\right|_{\Delta = 0} =\int d\eta d^3\vec{x} \left[p\dot q- \left(\begin{array}{cc}q&p \end{array} \right)\left(\begin{array}{cc}
0& \frac{3\mathcal K-\kappa\dot\varphi_0^2}{3\dot a/a}\\
\frac{3\mathcal K-\kappa\dot\varphi_0^2}{3\dot a/a}& \frac{2\mathcal K^2}{\kappa^2a^2\sqrt{\gamma}}
\end{array}
\right)\left(\begin{array}{c}q\\ p \end{array} \right)\right]~,\label{S2qR}
\end{equation}
where we discarded boundary terms since the equation above is the starting point for a canonical formulation, as discussed in sec.~\ref{sec:GHY}
and in accordance with \cite{Gratton:1999ya}.

The canonical transformation above might appear to have no other virtue than to simplify the Hamiltonian, 
so it is worth pausing and looking at its connection with the original variables and physical interpretation.
Indeed in ref.~\cite{Garriga:1997wz} a similar canonical transformation is used such that the resulting coordinate is proportional to Bardeen's potential.
In our case we have,
after substituting in the canonical transformation:
\begin{equation}\label{qOurs}
q=\frac{1}{\kappa}\left\{ 
\dot\varphi_0\left[ \mathcal K(B-\dot E)-\frac{\dot a}{a}\psi\right]+\left(\mathcal K-\frac{\dot a^2}{a^2}\right)\delta \varphi  \right\}~,
\end{equation}
which, under a gauge transformation as in eq.~(\ref{eq:GaugeonPer}) stays invariant. 
The gauge invariance of the variable speaks to the success of the proccess of removing redundancies and is a valuable check on the procedure
here employed.

Notice that we can write this action, since there is no `potential' in the Hamiltonian for $q$, as
\begin{equation}\label{DeqSm}
\left.\delta \mathcal{S}^{(2)}\right|_{\Delta = 0} =\int d\eta d^3\vec{x}\frac{2\mathcal K^2}{\kappa^2a^2\sqrt{\gamma}}p^2~,
\end{equation}
with $p=\partial\mathscr H/\partial q$.
It would seem that the discussion of the sign of eigenvalues ends here: they are $\geq 0$ from construction. 
The only non-trivial step to 
corroborate this conclusion is the rotation to Euclidean, which as it turns out changes nothing.

However we pursue here further to find the explicit eigenvalues of the spectrum.
After using  $\dot q=\partial \mathscr H/\partial p$ to write $p$ in terms of $q,\dot q$, repeated use of equations of motion and integration by parts leads to 
\begin{equation}\label{It2Op1}
\left.\delta \mathcal{S}^{(2)}\right|_{\Delta = 0} = \int d\eta d^3\vec{x}\sqrt{\gamma} \left\{\frac{\kappa^2 a^2}{8\mathcal K^2}\left(\dot q^2+\frac83\kappa\dot\varphi_0^2 q^2\right)-\frac{d}{d\eta}\left[\frac{\kappa^2a^2q^2}{4\mathcal K^2}\frac{(3\mathcal K -\kappa \dot{\varphi}_0^2)}{\dot{a}/a}\right]\right\}~.
\end{equation}
We note that the bulk term can be generalized to the Lorentz invariant form $d^4x \sqrt{-g}(\partial_\mu q\partial^\mu q +\dots)$. 
As the last step in Minkowski we substitute $q\to q/a$, and we find
\begin{equation}
\left.\delta \mathcal{S}^{(2)}\right|_{\Delta = 0} =  
\int d\eta d^3\vec{x}\sqrt{\gamma}\left\{
\frac{\kappa^2 }{8\mathcal K^2}\left(-q\ddot q-\mathcal K q^2+\frac52\kappa\dot\varphi_0^2 q^2\right)+
\frac{\kappa^2 }{8\mathcal K^2}\frac{d}{d\eta}\left[
q\dot q+q^2\left(\frac{2\kappa \dot\varphi_0^2 - 3(\dot{a}/a)^2- 6\mathcal K}{3\dot{a}/a}
\right)
\right]
\right\}~.
 \label{FINOB}
\end{equation}
Finally, going to the Euclidean, $\tau=i\eta$ and with $()^\prime=d/d\tau$
\begin{equation}
\left.\delta \mathcal{S}_{\rm E}^{(2)}\right|_{\Delta = 0} =  
\int d\tau d^3\vec{x}\sqrt{\gamma}
\frac{\kappa^2 }{8\mathcal K^2}\left[\left( -q q^{\prime\prime} + \mathcal Kq^2 - \frac52\kappa f_a^2\theta^{\prime\,2} q^2\right)+
\frac{d}{d\tau}\left(q q^\prime-\frac{2\kappa f_a^2\theta^{\prime\,2} a^2  + 3a^{\prime2}
-6a^2\mathcal K}{3a a^\prime}q^2\right)\right]~.\label{FINOBEc}
\end{equation}
At this point, an important comment is in order. In addition to the usual analytical continuation, in eq.~(\ref{FINOBEc}) we performed 
the formal substitution $\varphi_0\to \imath f_a\theta$. This is because the starting point of our analysis, 
that is the quadratic action in eq.~(\ref{eq:FullQuadratic}), was derived in~\cite{Garriga:1997wz} considering a generic real scalar field $\varphi_0$.
In our case, on the contrary, we are interested in the case of an axion field $\phi$ 
(or, more generically, in the case of a Goldstone boson which admits a three-form description). 
Going from Minkowski to Euclidean, in the axion case one gets an extra minus sign  whenever a term quadratic in $\theta$ appears 
(as discussed in section~\ref{sec:BulkAction} 
and appendix~\ref{app:Minkowski2Euclidean}), and the imaginary factor in $\varphi_0\to \imath f_a\theta$ precisely accounts for this issue.
For instance, this replacement is indeed crucial to match eq.~(\ref{eq:ConformalEinstein}) starting from the corresponding Minkowski version in eq.~(\ref{eq:MinkowskiEOM}). 

Substituting the wormhole solution for the background fields $a$ and $\kappa f_a^2\theta^{\prime\,2}$ as given in eq.~(\ref{eq:AxionWormhole}), 
and considering the case of a closed Universe $\mathcal{K} = 1$, 
we have the following quadratic action
\begin{equation}\label{eq:ExplictOp}
\left.\delta \mathcal{S}_{\rm E}^{(2)}\right|_{\Delta = 0} =  
\int d\tau d^3\vec{x}\sqrt{\gamma}
\frac{\kappa^2 }{8\mathcal K^2}\left\{  q\left[ -\frac{d^2}{d\tau^2}+1-  \frac{15}{\cosh^2(2\tau)}\right]q+\frac{d}{d\tau}\left[qq^\prime-\frac{3-\cosh^2(2\tau)}{\cosh(2\tau)\sinh(2\tau)}q^2\right] \right\}~,
\end{equation}
and, consequently, the following operator in the bulk 
\begin{mymathbox}[ams gather, title=Homogeneous scalar fluctuations around the wormhole background, colframe=titlepagecolor]
\mathcal{O} \equiv  
 -\frac{d^2}{d\tau^2}+1-  \frac{15}{\cosh^2(2\tau)}~.
\label{eq:Schrod}
\end{mymathbox}

This action can be written as the square of a generalized momentum, as the analysis in Minkowski indicated, explicitly
\begin{equation}
\left.\delta \mathcal{S}_{\rm E}^{(2)}\right|_{\Delta = 0} =  \int d\tau d^3\vec{x}\sqrt{\gamma} \frac{\kappa^2}{8\mathcal K^2}\left[q^\prime - \frac{3-\cosh^2(2\tau)}{\cosh(2\tau)\sinh(2\tau)}q\right]^2~. \label{eq:posform}
\end{equation}

Remarkably, the spectral problem for the quadratic action describing fluctuations around the wormhole background 
reduced to the eigenvalue problem for a Schr\"odinger-type operator. In particular, the differential operator in eq.~(\ref{eq:Schrod}) describes the one-dimensional motion 
of a particle subject to the P\"oschl-Teller potential (see appendix~\ref{App:PT}). 

\begin{figure}[!htb!]
\minipage{0.5\textwidth}
  \includegraphics[width=.9\linewidth]{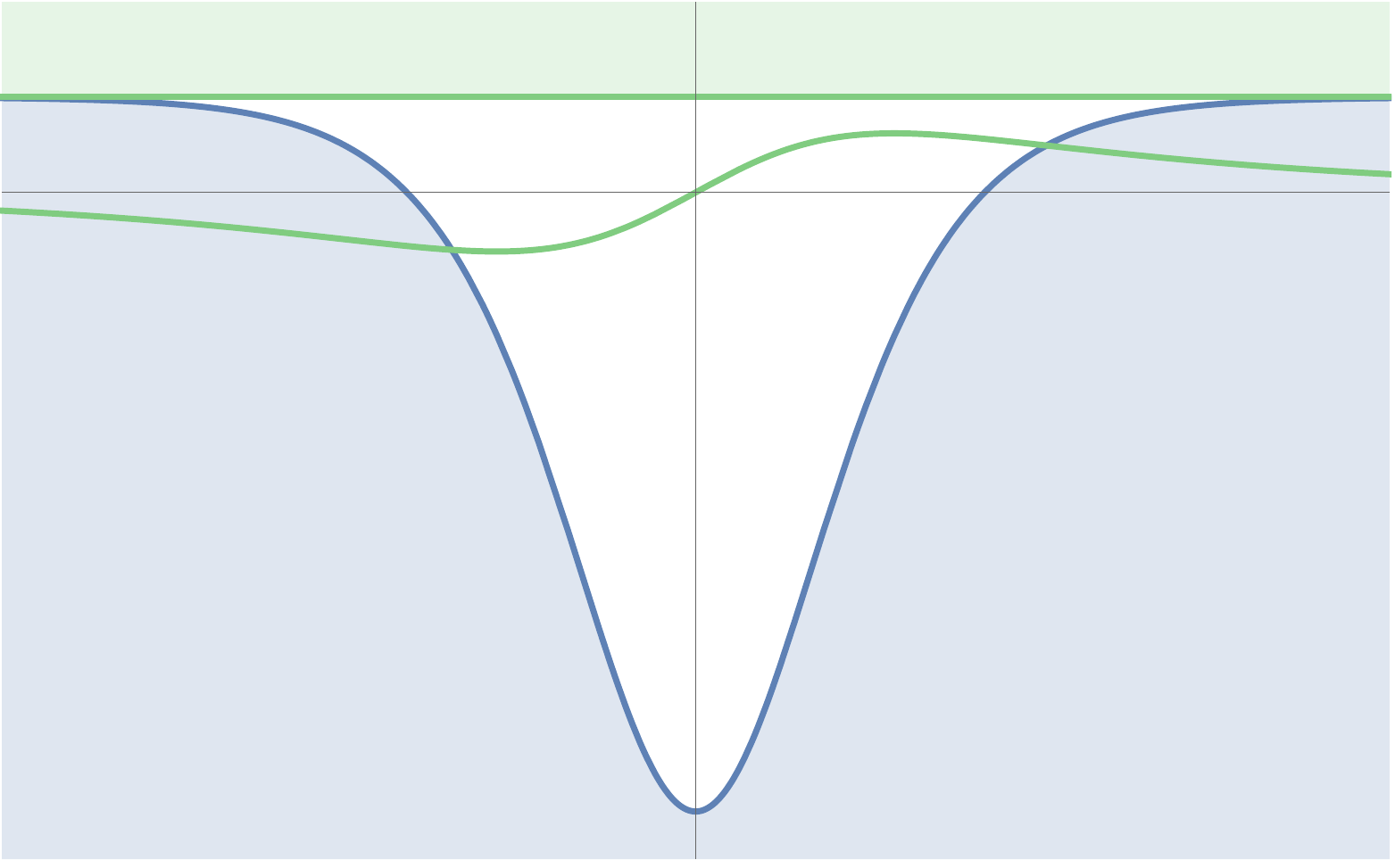}
\endminipage 
\hspace{.3cm}
\minipage{0.5\textwidth}
  \includegraphics[width=.9\linewidth]{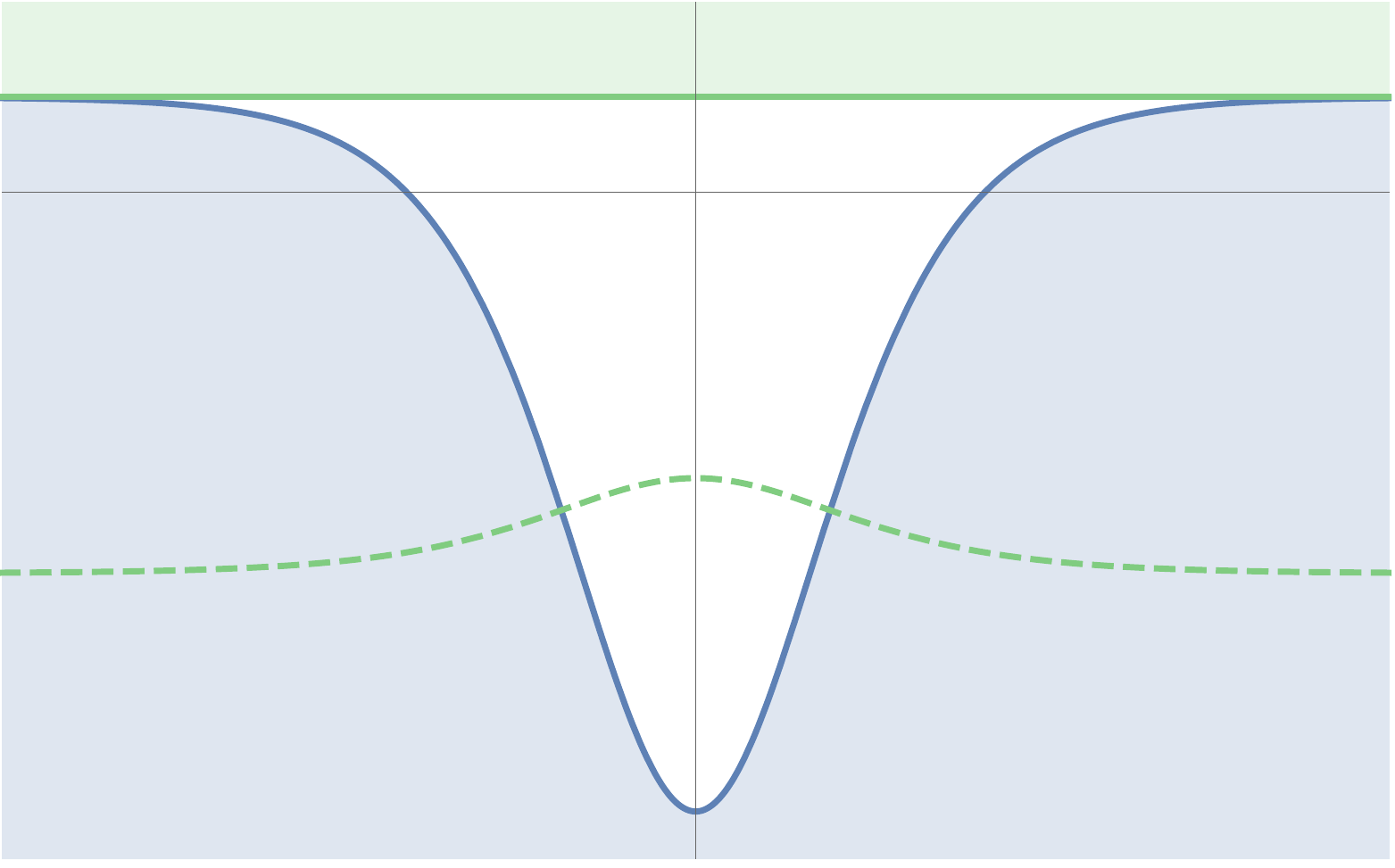}
\endminipage 
\vspace{-.2 cm}
\caption{\label{Fig:Schrod}\em 
Spectrum of the differential operator in eq.~(\ref{eq:Schrod}) that  coincides with  the P\"oschl-Teller potential well in Quantum Mechanics. The green band corresponds to the continuum spectrum.
In the left (right) panel we show the parity odd (parity even) eigenfunction
$q^{-}_{\lambda = 0}(\tau)$ ($q^{+}_{\lambda = -8}(\tau)$).
 We do not show the parity even eigenfunction $q^{+}_{\lambda = 0}(\tau)$ since it is not square-integrable. 
}
\end{figure}

The possible presence of negative eigenvalues is therefore related to the existence of   bound states.
Indeed, in general $\mathcal{O}$ has a discreet spectrum of bound states and for higher energies a continuum. The eigenvalues have definite `partiy' under
$\tau\to-\tau$ and this will be determining.
In figure ~\ref{Fig:Schrod} we show the odd (left panel) and even (right panel) discrete eigenvalues. 
Considering the eigenvalue equation  $\mathcal{O}q_\lambda(\tau) = \lambda q_\lambda(\tau)$, they correspond to 
\begin{align}
{\rm Odd\,eigenfunction:}~~&\left\{q^{-}_{\lambda = 0}(\tau) = \frac{\sinh(2\tau)}{\cosh^{3/2}(2\tau)}\right\}~,\label{eq:PTOdd}\\
{\rm Even\,eigenfunctions:}~~&\left\{q^{+}_{\lambda = -8}(\tau) = \frac{1}{\cosh^{3/2}(2\tau)}~,~~
q^{+}_{\lambda = 0}(\tau) =  \frac{3 - \cosh(4\tau)}{2\cosh^{3/2}(2\tau)}\right\}~.\label{eq:PTEven}
\end{align}
At this stage, one would na\"{\i}vely conclude that the differential operator in eq.~(\ref{eq:Schrod}) has one negative eigenvalue;  this is, the reader might have noticed, in 
contradiction with eq.~(\ref{eq:posform}).
The resolution of this conflict requires  closer inspection,
in particular, one needs to check the behaviour of the eigenfunctions at the boundaries in eq.~(\ref{FINOBEc}) 
and the transformation property under parity.
Let us start from the latter.
Parity under $\tau\to-\tau$ $(\sim\eta\to-\eta)$ can be assigned to the perturbations in eq.~(\ref{eq:Perturbations}), and carried on to the expression for $q$ in eq.~(\ref{qOurs}) we see that it has odd parity, a reminder of the fact that the canonical transformation has `swapped' momentum and coordinate.
In eqs.~(\ref{eq:PTOdd},\ref{eq:PTEven}), 
the only eigenfunction that is compatible with this requirement is $q^{-}_{\lambda = 0}(\tau)$, with eigenvalue $\lambda = 0$.
Based on this parity argument, 
we therefore discard the eigenfunctions $q^{+}_{\lambda = -8}(\tau)$ and $q^{+}_{\lambda = 0}(\tau)$.
Second, we have to check the behaviour of the eigenfunctions at the boundaries in eq.~(\ref{eq:ExplictOp}).
 This can be done in two ways: $i)$ direct computation in eq.~(\ref{eq:ExplictOp}) where one finds that the boundary term cancel at $\tau\pm\infty$ for $q^{-}_{\lambda=0}$ (it does cancel at $\tau=0$ as well if one instead considers a half-wormhole), $ii)$ alternatively substitution of $q^{-}_{\lambda=0}$ in eq.~(\ref{eq:posform}) yields $0$ which comprises both bulk and boundary contributions in eq~(\ref{eq:ExplictOp}). In this regard, let us note that the even solutions put in eq.~(\ref{eq:posform}) yield $\infty$ when integrating around the throat of the wormhole. In this sense only $q^{-}_{\lambda=0}$ is an eigenfunction of both eq.~(\ref{eq:ExplictOp}) and eq.~(\ref{eq:posform}) which adds to the evidence in favour of the odd eigenfunction. 

Notice that otherwise the presence of a zero eigenvalue can be inferred from the `time'-translation symmetry for the wormhole solution, as
 discussed in section~\ref{sec:BulkAction}.

All in all, we found that homogeneous scalar perturbations give a positive contribution 
to the  fluctuation determinant in eq.~(\ref{eq:FluctuationDet}). This result supports 
the interpretation of the wormhole as an instanton mediating tunneling transitions between degenerate vacua.
Equipped with this result, we can now move to compute the effective potential generated by gravity.

Before proceeding, let us comment about the discrepancy with the result presented in~\cite{Rubakov:1996cn} where a negative eigenvalue was found.
The metric studied in~\cite{Rubakov:1996cn} has the form $ds^2 = N^2(\rho)d\rho^2 + R^2(\rho)d\Omega^2$, with the 
wormhole solution corresponding to $N(\rho) = 1$ and $R(\rho) = R_{\rm wh}(\rho)$, where $R_{\rm wh}(\rho)$ satisfies the equation
$[R_{\rm wh}^{\prime}(\rho)]^2 = 1- 1/R_{\rm wh}^4(\rho)$ with $R_{\rm wh}(\rho = 0) = L$. 
 The authors of~\cite{Rubakov:1996cn}
focused the analysis only on homogeneous perturbations; they defined the perturbed metric element as
$ds^2 = [1 + n(\rho)]^2d\rho^2 + [R_{\rm wh}(\rho) + r(\rho)]^2$, and -- contrary to the explicit gauge-invariant formulation in this work -- 
they fixed the gauge with the choice $n(\rho) = 0$.
This procedure may lead to incorrect results since perturbations are truncated  
before the gauge fixing, thus preventing from the possibility to fix all possible gauge degrees of freedom. Furthermore, 
in~\cite{Rubakov:1996cn} the conformal factor problem 
was circumvented by using the Gibbons-Hawking-Perry rotation~\cite{Gibbons:1978ac}. In synthesis, the `wrong' negative kinetic term 
in the quadratic action for the perturbation $r(\rho)$ changes its sign as a consequence of the replacement $r\to \imath r$, with the new $r$ real.
As already mentioned above, this is an {\it ad hoc} prescription without a clear physical interpretation, 
and its na\"{\i}ve application may lead to misleading results~\cite{Mazur:1989by}. 

\section{Effective potential from Euclidean wormholes}\label{sec:EffectivePotential}

The non-linearly realized global $U(1)$  symmetry $\phi\to \phi + \alpha f_a$ is generated by the axion charge $\mathcal{Q}$.
The ground state -- corresponding to the bottom of the mexican hat potential -- is degenerate along the angular direction, 
and vacua are described by continuously connected field configurations with minimum energy.

\begin{figure}[!htb!]
\minipage{0.5\textwidth}
  \includegraphics[width=1.\linewidth]{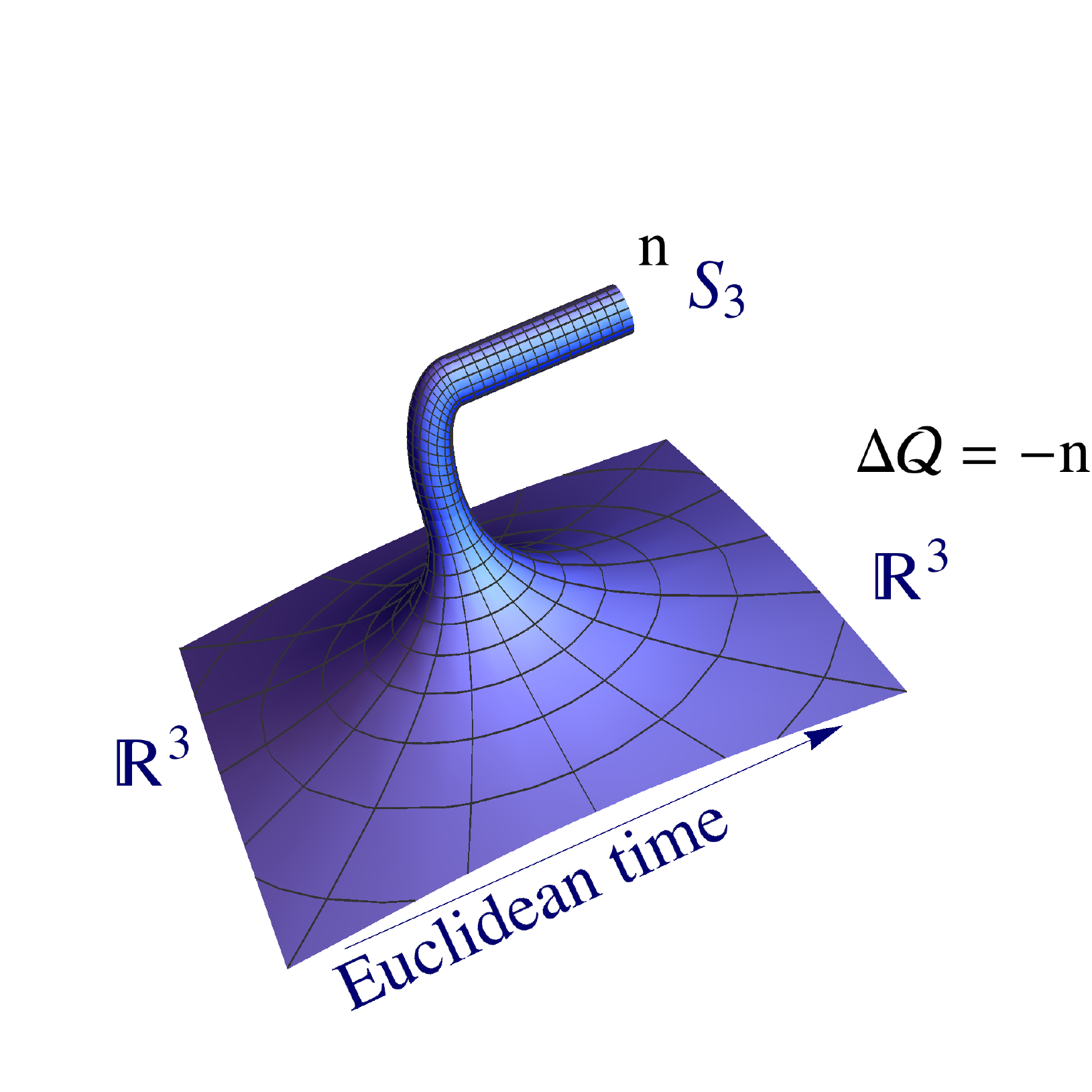}
\endminipage 
\hspace{-.05cm}
\minipage{0.5\textwidth}
  \includegraphics[width=1.\linewidth]{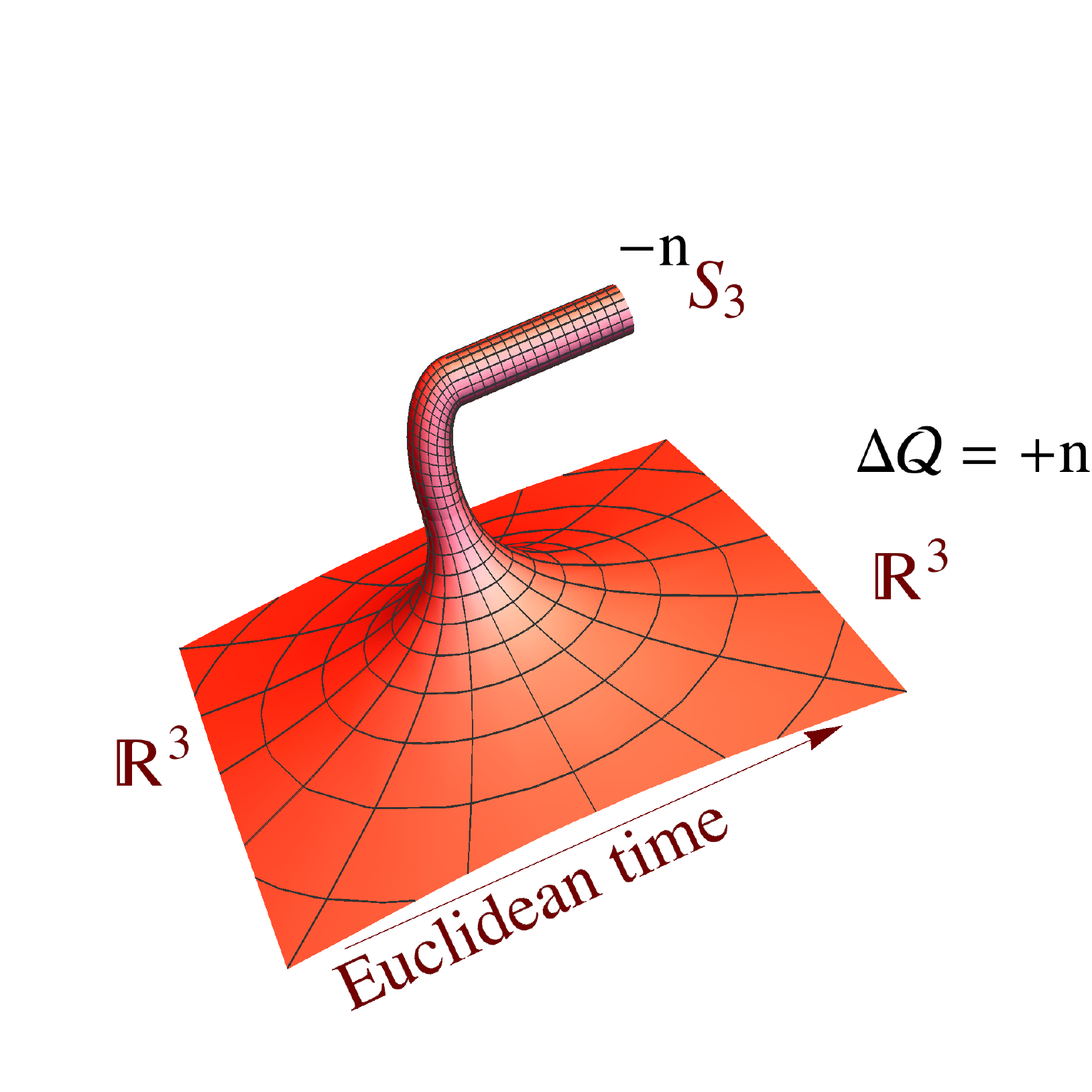}
\endminipage 
\vspace{-.2 cm}
\caption{\label{fig:Coleman}\em 
Half-wormholes as instantons 
in Euclidean space. In the left (right) panel we show an  instanton (anti-instanton) mediating the topology change 
$\mathbb{R}^3 \to \mathbb{R}^3 \oplus S_3$, and carrying away an axion charge $+n$ ($-n$). The axion charge is overall conserved
but  an observer on $\mathbb{R}^3$ experiences, as Euclidean time passes by, a change  $\Delta\mathcal{Q} = -n$ ($\Delta\mathcal{Q} = +n$).
The non-conservation of the axion charge $\mathcal{Q}$ on $\mathbb{R}^3$ implies an explicit breaking of the symmetry it generates.
}
\end{figure}

Half-wormholes induce quantum tunneling 
between classical vacuum states with different axion charge $\mathcal{Q}$.
The situation is schematically represented in fig.~\ref{fig:Coleman}.
As Euclidean time passes by, 
in the presence of an instanton (anti-instanton)
an observer on $\mathbb{R}^3$ experiences 
a change  $\Delta\mathcal{Q} = -n$ ($\Delta\mathcal{Q} = +n$) since 
there is a net flux of axion charge equal to $+n$ ($-n$) through the wormhole throat~\cite{Giddings:1987cg,Lee:1988ge,Rey:1989mg}.
Eventually, we will consider the case with $n = 1$, since the instanton action in eq.~(\ref{eq:GIAction}) is minimized.
Consequently, from the point of view of the observer on $\mathbb{R}^3$, the axion charge is not conserved, and 
the associated symmetry explicitly broken down to the discrete gauge 
 symmetry $\phi \to \phi + 2k\pi f_a$, with $k\in \mathbb{Z}$, 
 which remains intact in the presence of wormhole instantons, as discussed in section~\ref{sec:BulkAction}.

\begin{figure}[!htb!]
\centering
  \includegraphics[width=.5\linewidth]{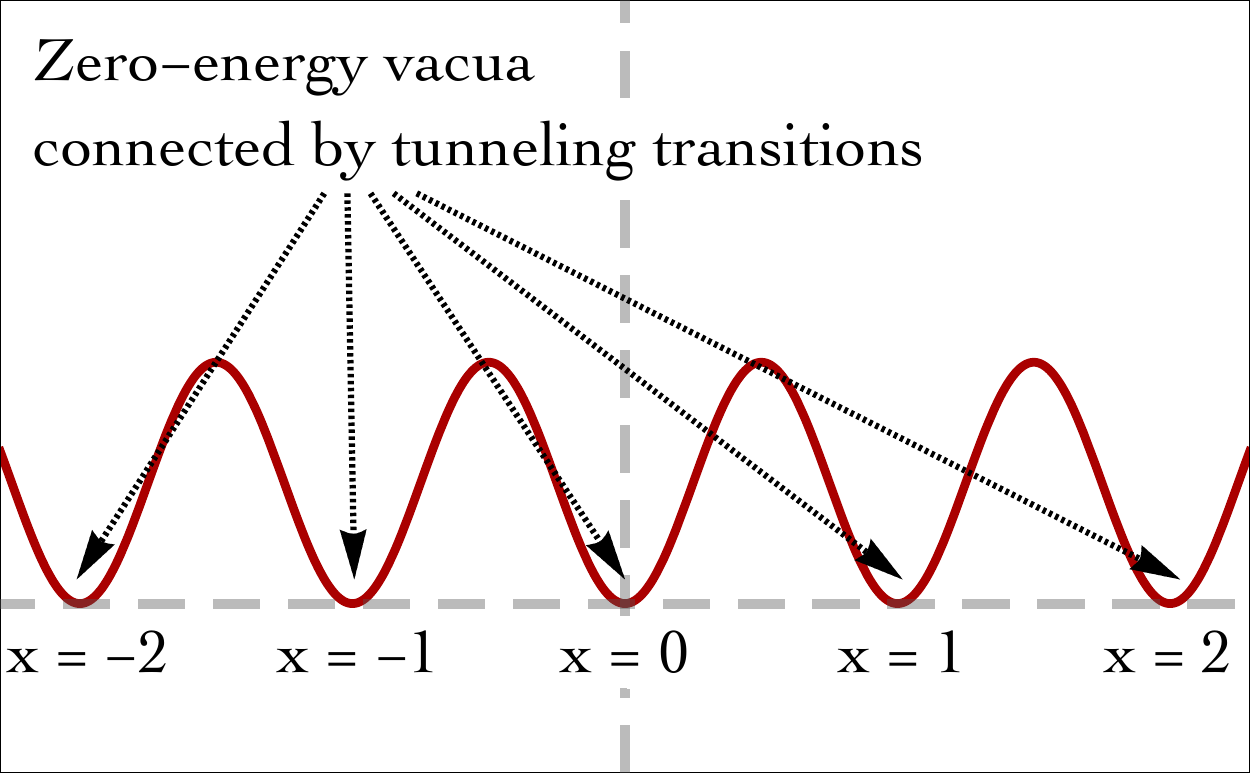}
\caption{\label{fig:Topology}\em 
Periodic potential in Quantum Mechanics. 
Non-perturbative instanton solutions tunnel between classically degenerate vacua.
}
\end{figure}

In the presence of wormhole instantons the situation is  similar to 
that of a particle moving in a one-dimensional periodic potential.
There are infinite classical minima at $x = j$, and each minimum corresponds to a degenerate 
ground state $|j\rangle$, as shown in fig.~\ref{fig:Topology}.
Instantons can begin at any initial position, $x=j$, and go to the next one, $x = j+1$.
As a result of these tunneling transitions, 
the true vacuum of the system is
a superposition of the degenerate ground states $|j\rangle$.
It is instructive to work out this analogy in more detail. In particular, we can compute the probability for the 
tunneling process $|j \rangle \to |k \rangle$ summing over all the possible instanton and anti-instanton configurations. 
We find~\cite{ColemanBook}
\begin{equation}\label{eq:PeriodicPotential}
\langle k|e^{-HT}|j \rangle = \sqrt{\frac{\omega}{\pi}}e^{-\omega T/2}\sum_{n = 0}^{\infty}\sum_{\bar{n} = 0}^{\infty}
\frac{1}{n!\bar{n}!}\left(
Ke^{-\mathcal{S}_0}T
\right)^{n+\bar{n}}\delta_{n-\bar{n},k-j}~,
\end{equation}
where $\omega \equiv V^{\prime\prime}(0)$, $K$ 
is the determinant factor describing quantum fluctuations around the classical instanton trajectory.
Notice that for a single instanton/anti-instanton path with action  $\mathcal{S}_0$ the contribution to the 
transition amplitude is~\cite{ColemanBook} 
\begin{equation}\label{eq:SingleInstanton}
\sqrt{\frac{\omega}{\pi}}e^{-\omega T/2}Ke^{-\mathcal{S}_0}~.
\end{equation}
Eq.~(\ref{eq:PeriodicPotential}) include multi-instanton/anti-instanton  
solutions using the dilute-gas approximation. 
According to this approximation, strings of widely separated instantons and anti-instantons centered at 
$\tau_1,\dots, \tau_n$, and satisfying the condition $-T/2 < \tau_1 < \dots < \tau_n < T/2$, 
are distributed arbitrarily along the time direction. 
Multi-instantons are not exact classical solutions, but they represent the leading term for
the tunneling amplitude between distant wells.
The contribution of a multi-instanton solution consisting in $n$ well-separated objects takes the same form as in eq.~(\ref{eq:SingleInstanton}) 
but with $K \to K^n$, $\mathcal{S}_0 \to n\mathcal{S}_0$. A similar result, with $n$ substituted by $\bar{n}$, is valid for an anti-instanton string.
In addition, the integration over the freely-distributed position gives the factor
\begin{equation}
\int_{-T/2}^{T/2}dt_n\int_{-T/2}^{t_n}dt_{n-1}\dots \int_{-T/2}^{t_2}dt_1 = \frac{T^n}{n!}~.
\end{equation}
The Kronecker delta in eq.~(\ref{eq:PeriodicPotential}) takes into account the fact that the total number of instantons minus the total number
of anti-instantons must equal the change in $x$ between the initial and final position eigenstates. Using the Fourier series 
representation
\begin{equation}
\delta_{a,b} = \int_0^{2\pi} \frac{d\bar{\zeta}}{2\pi}e^{\imath \bar{\zeta}(a - b)}~,
\end{equation}
we find 
\begin{equation}\label{eq:DiluteGas}
\langle k|e^{-HT}|j \rangle =  \sqrt{\frac{\omega}{\pi}}e^{-\omega T/2}\int_0^{2\pi} \frac{d\bar{\zeta}}{2\pi}
e^{\imath \bar{\zeta} (j-k)}\exp\left(2Ke^{-\mathcal{S}_0}T\cos\bar{\zeta} \right)~.
\end{equation}
From eq.~(\ref{eq:DiluteGas}), we read the energy eigenstates (Bloch waves, using the language of solid state systems) and eigenvalues
\begin{equation}\label{eq:PeriodicSpectrum}
|\zeta  \rangle = \left(
\frac{\omega}{\pi}
\right)^{1/4}\frac{1}{\sqrt{2\pi}}\sum_{n}e^{-\imath n\zeta}|n\rangle~,~~~~~E(\zeta) = \frac{\omega}{2} + 2Ke^{-\mathcal{S}_0}\cos\zeta~.
\end{equation}
The periodic potential contains a translation symmetry -- analogue in r\^ole to the gauge symmetry $\theta \to \theta + 2k\pi$ in eq.~(\ref{eq:DiscreteShift}), 
left unbroken by wormhole instantons -- that forces the eigenstates to be
shift-invariant. It is indeed possible to introduce an operator $\mathcal{T}$ that generates an elementary translation $\mathcal{T}| j\rangle = | j + 1 \rangle$.  
$\mathcal{T}$ commutes with the Hamiltonian, and both operators can be diagonalized simultaneously.
Each energy eigenstate in eq.~(\ref{eq:PeriodicSpectrum}) is characterized by an angle $\zeta$,  eigenvalue of $\mathcal{T}$. 
It is indeed immediate to check that 
the states $|\zeta  \rangle$ are eigenstates of $\mathcal{T}$, with $\mathcal{T}|\zeta  \rangle = e^{\imath \zeta}|\zeta  \rangle$.

A rigorous treatment in the context of wormhole physics was proposed in~\cite{Rey:1989mg,Giddings:1988cx,Coleman:1988cy} (see also~\cite{Preskill:1988na,Hawking:1990ue,Hawking:1988ae}).
\begin{figure}[!htb!]
\centering
  \includegraphics[width=.8\linewidth]{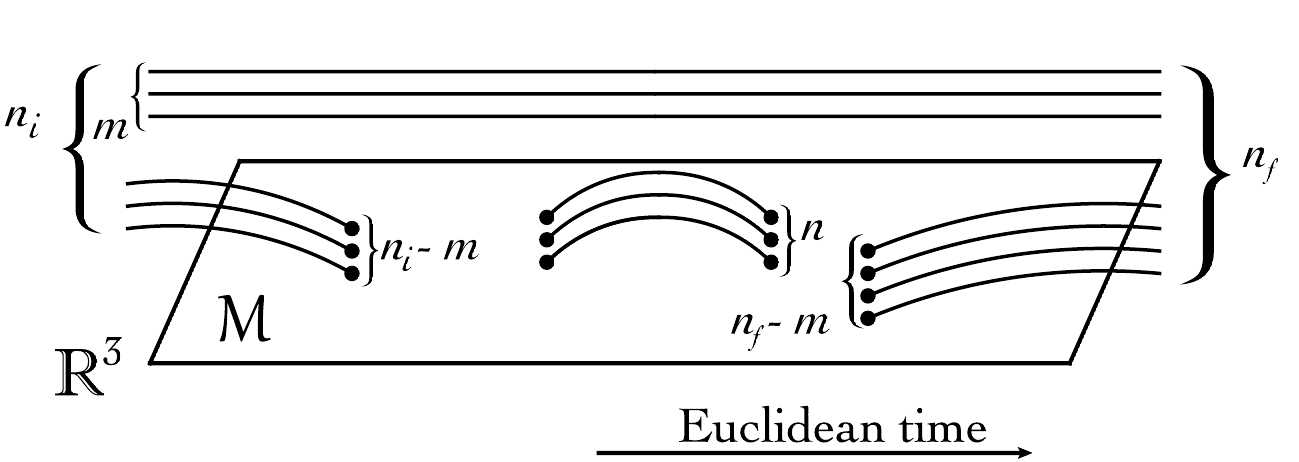}
\caption{\label{fig:WormholeTopology}\em 
The computation of the transition amplitude
 between a state $|n_i\rangle$ with $n_i$ instantons 
to a state $|n_f\rangle$ with $n_f$ instantons
 requires the sum over all possible four-geometries. We show here the most general one~\cite{Giddings:1988cx,Coleman:1988cy}
 which includes $m$ non-interacting wormholes, $n$ wormholes that are 
emitted and reabsorbed, $n_i - m$ half-wormholes in the initial and  $n_f - m$ half-wormholes in the final state.
From a local perspective, an effective potential on the space-time volume $\mathcal{M}$ is generated (see text for details).
}
\end{figure}
The goal of these papers was to compute the influence of wormholes on physics at low energy, or at distances large compared to the thickness of the 
wormhole throat, $L$.
The key idea 
is that at energy scales below $L^{-1}$ wormholes can be integrated out.\footnote{Notice that 
the validity of this approach relies on the fact that large wormholes, with size $L^{\prime}>L$, can be safely neglected. 
Na\"{\i}vely, by looking at the action in eq.~(\ref{eq:GIAction}), one would conclude that this is indeed the case: The action of a 
wormhole of size $L$ scales as $L^2M_{\rm Pl}^2$, and larger wormholes will be strongly suppressed by $e^{-\mathcal{S}}$. 
It is important to remark that this conclusion might not be so obvious. Wormholes received in the past a lot of attention 
in connection with a possible solution of the cosmological constant problem~\cite{Coleman:1988tj,Klebanov:1988eh}. 
The major objection to this proposal, 
known precisely as `large wormhole problem', refers to the fact that large wormholes are also characterized by 
large densities in the Euclidean space, and, despite the suppression provided by their action,  they may lead to strong non-local interactions
over arbitrarily large distances, a  prediction of fatal consequences~\cite{Fischler:1988ia}. A possible solution to this obstruction was proposed in~\cite{Coleman:1989ky,Coleman:1990tz}.  The point is that 
if the wormhole solutions  carry a non-zero
value of some conserved charge -- as in the case we are considering in this work -- large 
wormholes are destabilized by small ones as a consequence of the charge non-conserving interactions introduced by the latter (see eq.~(\ref{eq:MainPotential}) below),
 and, as a result, only small wormholes survive. 
This argument was refined in~\cite{Ridgway:1991gg} in the axion case.
}
In the resulting 
low-energy effective theory, the Lagrangian density takes the form 
\begin{equation}\label{eq:EffectiveWormholeLagrangian}
\mathcal{L} = \mathcal{L}_0\left[\Phi(x),\dots\right] + \sum_{i}\mathcal{O}_{i}\left[\Phi(x),\dots\right]\mathcal{A}_i~.
\end{equation}
The second term 
captures the effect of topological fluctuations due to wormhole physics. 
In full generality, the sum over $i$ represents the possible presence of instantons of different type (for instance, instantons with different charge $\mathcal{Q}$). 
$\mathcal{O}_{i}(x) \equiv \mathcal{O}_{i}\left[\Phi(x),\dots\right]$ are generic functions
of fields and their derivatives, and $\mathcal{A}_i$ are combinations of creation and annihilation operators describing emission 
and absorption of half-wormhole geometries of type $i$. 
Notice that $\mathcal{A}_i$ do not depend on space-time position since wormholes do not carry any momentum.
Furthermore, wormholes do not carry away any quantity coupled to gauge fields, and $\mathcal{O}_{i}$ must 
be a Lorentz scalar, singlet under charge and color.  However, as we learned in section~\ref{sec:EuclideanWormholeSolution}, wormholes carry off
axion charges, and as a consequence we expect the operator $\mathcal{O}_{i}(x)$ to transform non-trivially under the global $U(1)$ symmetry.
We shall return on this point later.

Before proceeding, let us give a more quantitative understanding.
The fact that integration over wormhole geometries gives rise to the structure in eq.~(\ref{eq:EffectiveWormholeLagrangian}) can be 
understood with an explicit computation~\cite{Giddings:1988cx,Coleman:1988cy}.
In practice, one can compute the probability for the transition between a state $|n_i\rangle$ with $n_i$ instantons 
to a state $|n_f\rangle$ with $n_f$ instantons, $\langle n_f|e^{-HT}|n_i\rangle$. 
For simplicity we do not distinguish here between instantons and anti-instantons, and we only consider instantons of unit charge.
The computation should include a sum over all possible four-geometries, and in fig.~\ref{fig:WormholeTopology} we show the most general 
of such configurations~\cite{Giddings:1988cx,Coleman:1988cy}. This configuration includes $m$ disconnected 
wormholes that do not interact with the Euclidean space-time, and $n$ wormholes that are 
emitted and reabsorbed (see also fig.~\ref{fig:Wormhole2}). 
The amplitude for this geometry is 
weighted by the factor $e^{-\mathcal{S}_{\rm inst}(2n + n_f + n_i - 2m)}$, where $\mathcal{S}_{\rm inst}$
is the instanton action (with unit charge) computed in eq.~(\ref{eq:Fullaction}). 
Furthermore, we indicate with $K\sqrt{g}d^4x$ the amplitude for inserting a single wormhole end in an infinitesimal volume of $\mathcal{M}$.
For simplicity, we treat $K$ as a constant while, in general, it contains combinations of fields defined on $\mathcal{M}$. 
The final result -- in analogy with eq.~(\ref{eq:PeriodicPotential}) -- is
\begin{equation}\label{eq:TransitionWormhole}
\langle n_f|e^{-HT}|n_i\rangle = \sum_{n=0}^{\infty}\sum_{m=0}^{{\rm min}(m_i,n_f)} \frac{\sqrt{n_i!\,n_f!}}{m!}
\frac{(Ke^{-S_{\rm inst}}VT)^{2n+n_i + n_f - 2m}}{2^n n!(n_i - m)!(n_f - m)!}~,
\end{equation}
where the volume  factor $VT = \int_{\mathcal{M}}d^4x\sqrt{g}$  
comes from an integration over the location of each instanton in the dilute gas approximation.
The crucial observation  in~\cite{Giddings:1988cx,Coleman:1988cy} is that the same result presented in eq.~(\ref{eq:TransitionWormhole}) can be directly 
obtained 
from the left-hand side of eq.~(\ref{eq:TransitionWormhole}) using the Hamiltonian $H = Ke^{-S_{\rm inst}}V(a + a^{\dag})$, 
after introducing creation and annihilation operators $a^{\dag}$ and $a$
subject to the commutation relation $[a,a^{\dag}]=1$, and defined by $|n\rangle = \sqrt{1/n!}(a^{\dag})^n|0\rangle$.
This result shows the validity of the assumption made in eq.~(\ref{eq:EffectiveWormholeLagrangian}).\footnote{The full wormhole configuration with both ends in $\mathbb R^3$ has no boundary term at the throat and therefore its action does not match $2\mathcal{S}_{\rm inst}$ as in eq.~(\ref{eq:Fullaction}), this can be accounted for in the summation formula but
the correction is sub-leading $e^{-2\mathcal{S}_{\rm inst}}$ vs. $e^{-\mathcal{S}_{\rm inst}}$.}

\begin{figure}[!htb!]
\centering
  \includegraphics[width=1\linewidth]{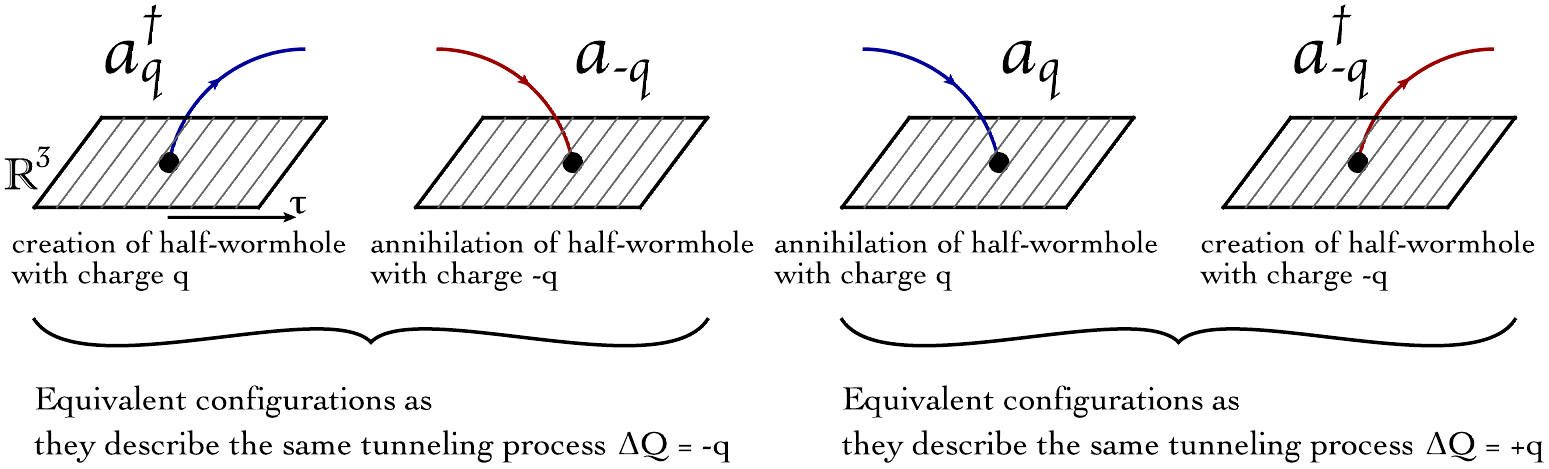}
\caption{\label{fig:WormholeTopology2}\em 
Creation and annihilation operators for instanton (blue) and anti-instantons (red) with commutation relations 
$[a_q, a_{q^{\prime}}] = [a_q^{\dag}, a_{q^{\prime}}^{\dag}]=0$ and $[a_q, a_{q^{\prime}}^{\dag}] = \delta_{qq^{\prime}}$.
}
\end{figure}

The effective wormhole action -- writing explicitly the presence of field operators $\mathcal{O}_{i}(x)$ in the spirit of eq.~(\ref{eq:EffectiveWormholeLagrangian}) --  
is therefore~\cite{Rey:1989mg} 
\begin{eqnarray}\label{eq:EffectiveWormholeAction}
\mathcal{S}_{\rm wh} &=& \int d^4x\sqrt{g}\sum_{q}K_q e^{-\mathcal{S}_{\rm inst}}\left[
\left(a^{\dag}_q + a_{-q}\right)\mathcal{O}_{-q}(x) + \left(a^{\dag}_{-q} + a_{q}\right)\mathcal{O}_{q}(x)
\right]\nonumber \\
&=& \int d^4x\sqrt{g}\sum_{q}K_q e^{-\mathcal{S}_{\rm inst}}
\mathcal{O}_{S}(x)
\left[
\left(a^{\dag}_q + a_{-q}\right)\exp\left(-\frac{\imath q\phi}{f_a}\right)+ \left(a^{\dag}_{-q} + a_{q}\right)\exp\left(\frac{\imath q\phi}{f_a}\right)
\right]~,
\end{eqnarray}
where $a^{\dag}_q + a_{-q}$ ($a^{\dag}_{-q} + a_{q}$) describes the creation of a half-wormhole with charge $q$ ($-q$) or, equivalently, the annihilation 
of a half-wormhole with charge $-q$ ($q$), as illustrated in fig.~\ref{fig:WormholeTopology2}. The operators $a_q$ and $a_q^{\dag}$ obey the usual 
commutation relations $[a_q, a_{q^{\prime}}] = [a_q^{\dag}, a_{q^{\prime}}^{\dag}]=0$ and $[a_q, a_{q^{\prime}}^{\dag}] = \delta_{qq^{\prime}}$.

As anticipated before, the operator $\mathcal{O}_{q}(x)$ transforms non-trivially under the global $U(1)$ symmetry, and it is possible to show that 
 $\mathcal{O}_{q}(x) = \exp(\imath q\phi/f_a)\mathcal{O}_{S}(x)$, where $\mathcal{O}_{S}(x)$ is a singlet.
 Under the symmetry transformation $\phi \to \phi + \alpha f_a$, it follows that $\mathcal{O}_{q}(x) \to  \exp(\imath q\alpha)\mathcal{O}_{q}(x)$.
 As shown in~\cite{Rey:1989mg},  this transformation property ensures conservation of the global $U(1)$ charge 
 when the sum over all possible topologies is considered (that is, for instance, $\mathbb{R}^3 \oplus S_3$ for the creation 
 of a single instanton, see fig.~\ref{fig:Coleman}).
 
 Finally, in order to derive a viable effective action, it is crucial to understand the r\^ole of the creation and annihilation 
 operators in eq.~(\ref{eq:EffectiveWormholeAction}).
 To this end, we can apply the lesson learned from Quantum Mechanics. 
 At the beginning of section~\ref{sec:Stability}, we discussed the double-well potential.
 In perturbation theory, the system has two degenerate vacua $|\pm k\rangle$, and, 
 after including instanton solutions, the Hamiltonian is diagonalized by  $|+k\rangle \pm |- k\rangle$, with the true vacuum 
 corresponding to the antisymmetric combination. 
 This example shows that in the presence of instanton with no negative modes 
 the quantum vacuum is constructed as a coherent superposition of classical vacua.
 The same conclusion remains valid for the periodic potential studied at the beginning of this section, with the true vacuum defined by the superposition
  in eq.~(\ref{eq:PeriodicSpectrum}).
  A similar situation arises in Quantum Field Theory.
  The QCD Lagrangian has a discrete set of degenerate classical minima, labelled by integers $n$ -- dubbed winding number -- and indicated with $|n\rangle$. 
  Indeed, considering the 
  Chern-Simons current $\mathcal{K}^{\rm CS}_{\mu}$ and the corresponding charge $\mathcal{K}^{\rm CS}$
  \begin{equation}
  \mathcal{K}^{\rm CS}_{\mu} = 2\epsilon_{\mu\nu\alpha\beta}\left[
  A_{\nu}^{a}(\partial_{\alpha}A_{\beta}^{a}) + \frac{g_s}{3}f_{abc}A_{\nu}^a A_{\alpha}^b A_{\beta}^c
  \right]~,~~~~~~\mathcal{K}^{\rm CS} = \frac{g_s^2}{32\pi^2}\int \mathcal{K}_0^{\rm CS}(\vec{x},t)d^3\vec{x}~,
  \end{equation}
  it is possible to show that integer values $\mathcal{K}^{\rm CS} = n$ correspond to non-equivalent pure gauge field configurations with zero energy.
  Yang-Mills instantons tunnel between two such classical vacua, resulting in a non-zero matrix element $\langle n+1|e^{-HT}|n\rangle$.
  It means that in the presence of instantons the $|n\rangle$ vacua do not diagonalize the Hamiltonian, in analogy with eq.~(\ref{eq:DiluteGas}) and 
  eq.~(\ref{eq:TransitionWormhole}). On the contrary, the true vacuum is the coherent superposition $|\theta\rangle = \sum_n e^{\imath n\theta}|n\rangle$.
  We expect the same situation in the presence of gravitational instantons. 
  To this end, we introduce the operators $C_q \equiv a^{\dag}_q + a_{-q}$ and $C^{\dag}_q \equiv a^{\dag}_{-q} + a_{q}$ with 
  commutation relations $[C_q, C_{q^{\prime}}] = [C^{\dag}_q, C^{\dag}_{q^{\prime}}] = [C_q, C^{\dag}_{q^{\prime}}] = 0$.
  It is possible to simultaneously diagonalize the operators $C_q$ and $C_q^{\dag}$, and we define the eigenvalue equations
   $C_q|\alpha \rangle = \alpha_q |\alpha \rangle$, $C^{\dag}_q|\alpha \rangle = \alpha_q^{*} |\alpha \rangle$.
  
  Tunneling transitions bring the system in the coherent state $(a^{\dag}_{-q} + a_q)|\alpha \rangle = \alpha_q e^{\imath \delta_q}|\alpha \rangle$, 
  defined in analogy with the harmonic oscillator
   as the eigenstate of the annihilation operator with complex eigenvalue $\tilde{\alpha}_q \equiv \alpha_q e^{\imath \delta_q}$.
   We can therefore replace creation and annihilation operators in eq.~(\ref{eq:EffectiveWormholeAction}) with their eigenvalues, and obtain
   \begin{mymathbox}[ams gather, title=Effective potential from Euclidean wormholes, colframe=titlepagecolor]\label{eq:MainPotential}
 \mathcal{S}_{\rm wh} = \int d^4x\sqrt{g}\sum_{q}K_q e^{-\mathcal{S}_{\rm inst}}\alpha_q \mathcal{O}_S\cos\left(
   \frac{q\phi}{f_a}  + \delta_q
   \right)~,\\
   \mathcal{O}_S = \mathds{1} +  aL^2 \mathcal{R} + bL^4 (\partial_{\mu}\phi)(\partial^{\mu}\phi) + \dots~,\label{eq:FinalEffective}
\end{mymathbox}
   where in the last line we explicitly wrote the most general combination of operators singlet under the $U(1)$ global symmetry.
   In eq.~(\ref{eq:FinalEffective}),  we used as mass suppression scale the cutoff $1/L$, and $a$, $b$ and $\alpha_q$ are expected to be order one numbers.
   In eq.~(\ref{eq:MainPotential}) the crucial point is that  
   breaking effects are {\it always} proportional to the factor $e^{-\mathcal{S}_{\rm inst}}$, a trademark representing
    their non-perturbative origin. 
    We remark that this point seems to be quite often overlooked in the literature, especially as far as phenomenological applications are concerned.
   Very often breaking effects generated by gravity are treated in a na\"{\i}ve way, 
   along the line of the discussion outlined in the introduction, see eq.~(\ref{eq:Kamion}).
   On the contrary, the presence of the suppression factor $e^{-\mathcal{S}_{\rm inst}}$ plays a crucial r\^ole. 
       In the second part of this paper we shall discuss in detail the most relevant phenomenological applications.
       
\section{Phenomenological implications}\label{sec:Pheno}

In this section we discuss several phenomenological situations 
characterized by the presence of light axions with a decay constant such that non-perturbative gravity effects become relevant.

\subsection{The QCD axion}\label{eq:QCDAxion}

We start discussing the implication of eq.~(\ref{eq:MainPotential}) for the QCD axion.
We consider only wormholes with unit charge, since they give the dominant contribution.
The effective axion potential, including both QCD and non-perturbative gravitational contributions, is
schematically given by 
\begin{equation}\label{eq:QCDPotential}
V(a) = \Lambda_{\rm QCD}^4\cos\left(\frac{\phi}{f_a}\right) + (1/L)^4e^{-\mathcal{S}_{\rm inst}}\cos\left(
\frac{\phi}{f_a} + \delta_1
\right)~,
\end{equation}
where we estimated the prefactor of the gravitational contribution to be of order $(1/L)^4$. 
This is nothing but an order-of-magnitude estimate based on dimensional arguments but it does not affect the final result 
since the exponential term $e^{-\mathcal{S}_{\rm inst}}$ dominates. 
In this regard, it is worth emphasizing the size of this suppression. Considering the benchmark 
value $f_a = 10^{10}$ GeV for a QCD axion in the classical window, we have $\mathcal{S}_{\rm inst} \simeq 1.7\times 10^8$.
The explicit breaking generated by gravity is therefore negligible for all purposes. 

Let us explore the implications of eq.~(\ref{eq:QCDPotential}) in more detail.
First, notice  that we defined the axion field so that the low energy QCD contribution to the axion
potential is minimized at $\phi = 0$, and we redefined accordingly the phase $\delta_1$ in eq.~(\ref{eq:MainPotential}).
 The phase shift 
$\delta_1$ arises  from a mismatch between the gravitational and the low energy QCD term.
In the absence of CP violation in the gravitational sector, the phase $\delta_1$ is just 
proportional to ${\rm Arg\,Det}[Y_u Y_d]$, with $Y_{u,d}$  the Yukawa matrices for up- and down-type quarks, and it arises from the chiral
rotation needed to move
the phase of the fermion mass matrix into the  $\theta$-term.\footnote{
The possibility to have CP violation from gravitational effects is a subtle question, and, to the best of our knowledge, the situation is the following.
Gravity is CP-conserving at the perturbative level, and the only source of CP violation may arise in connection with non-perturbative physics.
In the Standard Model plus gravity,  CP-violating effects 
are related to the three terms
\begin{equation}\label{eq:EH}
\mathcal{L}_{\cancel{{\rm CP}}} = \underbrace{\frac{\theta_{\rm QCD}}{32\pi^2}G_{\mu\nu}\tilde{G}^{\mu\nu}}_{\rm Yang-Mills\,instantons} + 
\underbrace{\frac{\theta_{\rm EM}}{32\pi^2}F_{\mu\nu}\tilde F^{\mu\nu}
+ \frac{\theta_{\rm grav}}{32\pi^2}\mathcal{R}_{\mu\nu\rho\sigma}\tilde{\mathcal{R}}^{\mu\nu\rho\sigma}}_{\rm Eguchi-Hanson\,gravitational\,instantons}~,
\end{equation}
where $\tilde{\mathcal{R}}_{\mu\nu\rho\sigma} = \epsilon_{\mu\nu\alpha\beta}\mathcal{R}^{\alpha\beta}_{~~~\rho\sigma}/2$.
Notice that,  in addition to the standard Ricci term of the Einstein-Hilbert action, 
one needs to add the topological contribution proportional to the CP-odd combination $\mathcal{R}_{\mu\nu\rho\sigma}\tilde{\mathcal{R}}^{\mu\nu\rho\sigma}$. 
As  well known, these three terms are total derivatives, and they do not contribute to the equation of motion.
However, in QCD $\theta_{\rm QCD}$ becomes  a fundamental physical parameter at the non-perturbative level. 
The reason is ultimately related to the fact that the third homotopy group of $SU(3)_{\rm C}$ is non-trivial, since $\pi_3[SU(N)] = \mathbb{Z}$, and 
non-equivalent pure gauge filed configurations with zero energy fall into topologically distinct classes.
QED in Minkowski space does not possess this property since $\pi_3[U(1)] = 0$.
However, in the presence of gravity this is not true anymore.
Eq.~(\ref{eq:EH}) admits non-perturbative gravitational instanton solutions, dubbed Eguchi-Hanson instantons~\cite{Eguchi:1978xp,Eguchi:1978gw}, 
that provide a non-trivial background metric in which $\theta_{\rm EM}$ becomes physical.
The Eguchi-Hanson gravitational instantons are not related to wormhole physics, and in general  their contribution is 
suppressed by the large action $\mathcal{S}_{\rm EH} = P^2\pi/\alpha_{\rm EM}$,
 where $P$ is the electric charge of the instanton~\cite{Holman:1992ah}. In this paper we do not investigate such non-perturbative solutions.
 Recently, in~\cite{Dvali:2005an,Dvali:2013cpa,Dvali:2017mpy,Dvali:2016uhn,Dvali:2016eay} CP-violating effects related to $\mathcal{R}\tilde{\mathcal{R}}$ were explored 
using the language of differential forms.
In the three-form formulation the strong CP problem is equivalent to the dynamical generation of a mass gap for the Chern-Simons three-form of QCD, $\mathcal{C}$.
This is exactly what the axion solution does, since it provides a pseudo-scalar degrees of freedom that is eaten up by $\mathcal{C}$ which in turn becomes massive.
In this picture, the only way  to re-introduce the strong CP problem in the presence of the axion 
is to re-establish a massless pole in the propagator of  $\mathcal{C}$.
This can be accomplished by introducing an additional massless three-form, $\mathcal{C}_{\rm G}$, since in this case the axion will be able to
give a mass only to a linear combination of $\mathcal{C}$ and $\mathcal{C}_{\rm G}$. 
Gravity naturally provides a candidate for  $\mathcal{C}_{\rm G}$, that can be identified with the gravitational Chern-Simons three-form 
whose field strength equals $\mathcal{R}\tilde{\mathcal{R}}$~\cite{Dvali:2005an}.
}
In the following, we interpret $\delta_1$ as an arbitrary order one phase.
\begin{figure}[!htb!]
\minipage{0.5\textwidth}
  \includegraphics[width=1.\linewidth]{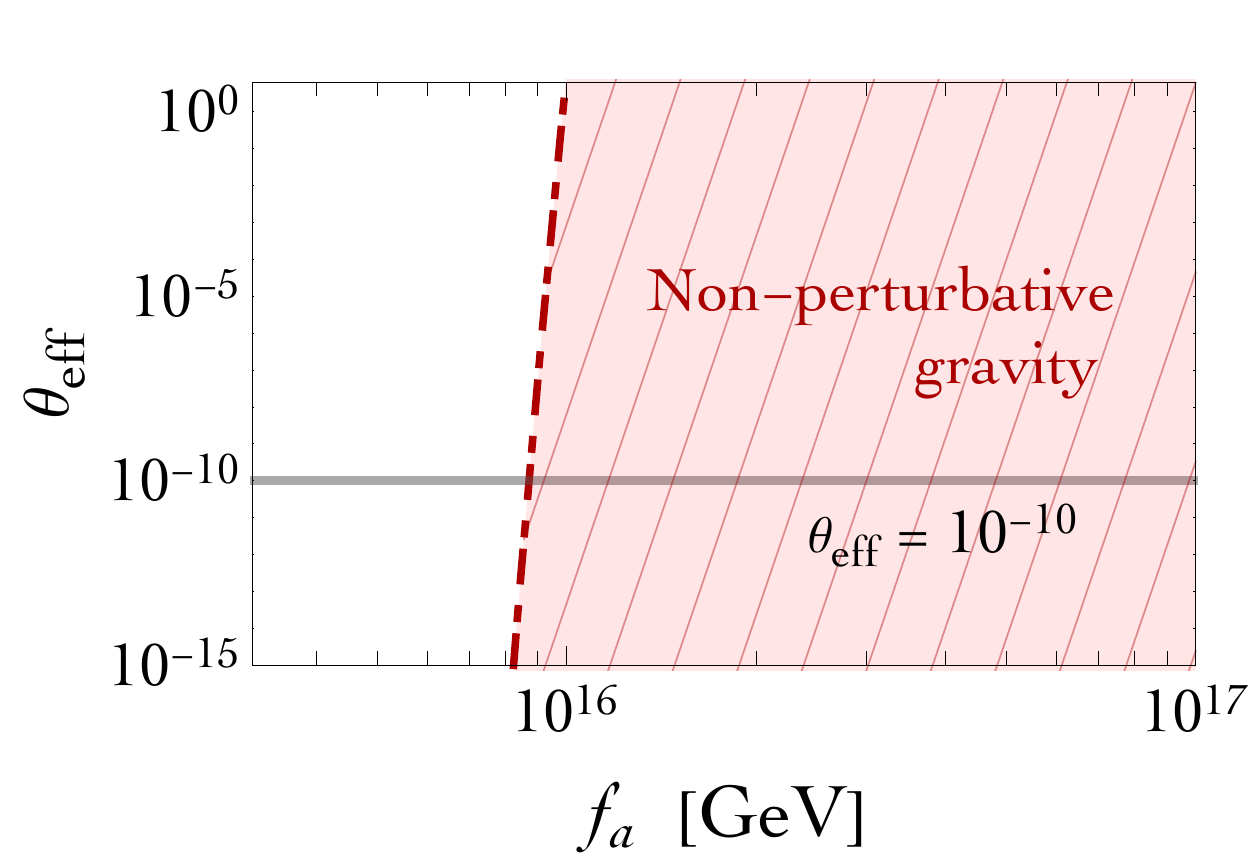}
\endminipage 
\hspace{.3cm}
\minipage{0.5\textwidth}
  \includegraphics[width=1.\linewidth]{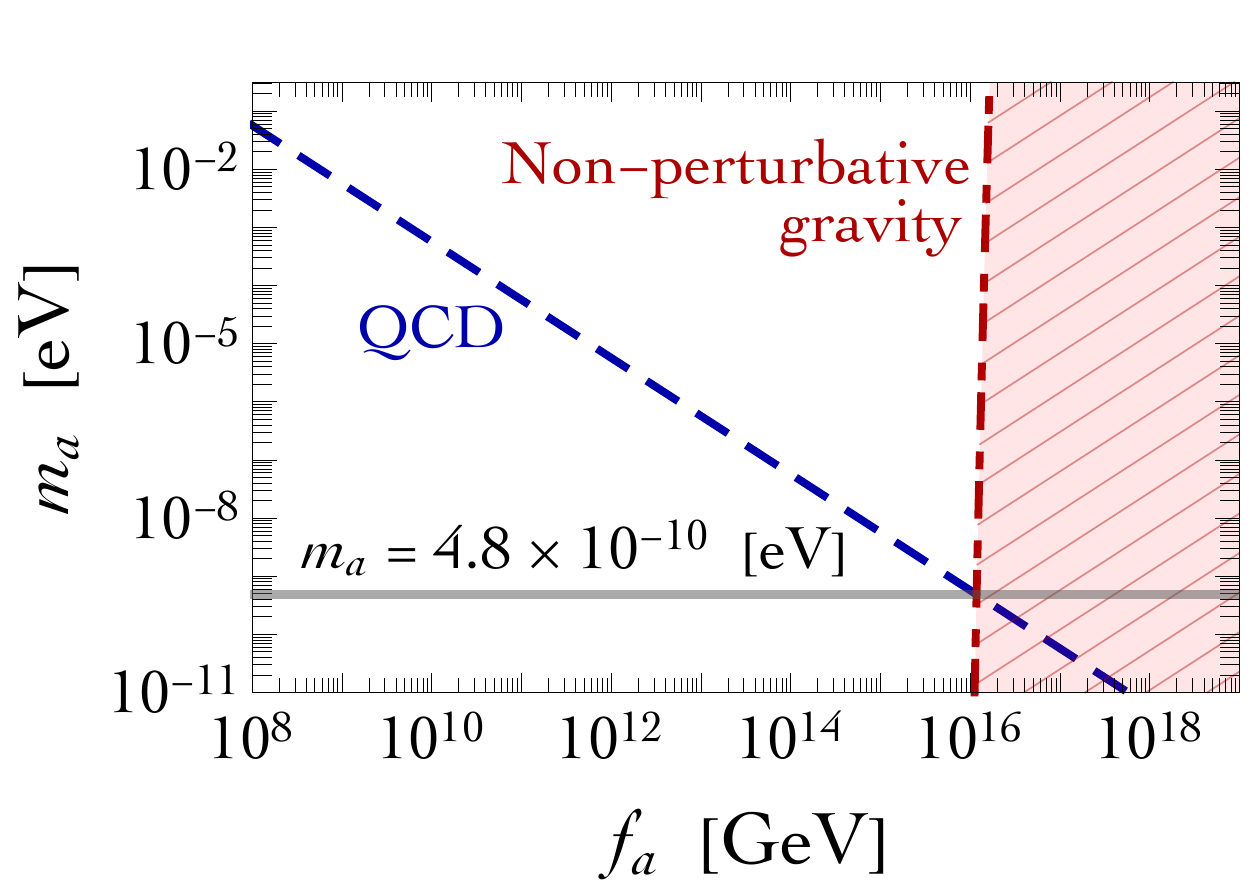}
\endminipage 
\vspace{-.2 cm}
\caption{\label{fig:QCD}\em 
Non-perturbative gravity corrections (red dot-dashed line) on $\theta_{\rm eff}$ (left panel) and the axion mass $m_a$ (right panel)
as a function of the axion decay constant $f_a$. 
In the regions shaded in red non-perturbative gravity dominates. In the right panel, 
the dashed blue line represents the contribution to the axion mass generated by QCD, computed according to~\cite{diCortona:2015ldu}.
}
\end{figure}

The effect of the gravitational correction is twofold:
It shifts the axion mass, and produces a non-zero $\theta_{\rm eff}$.
We find
\begin{eqnarray}
m_a^2 &\simeq &  \frac{\Lambda_{\rm QCD}^4}{f_a^2} + \frac{(1/L)^4}{f_a^2}e^{-\mathcal{S}_{\rm inst}}~,\label{eq:GravityMass}\\
\theta_{\rm eff} &\simeq &  \frac{(1/L)^4}{\Lambda_{\rm QCD}^4}\,\sin\delta_1\,e^{-\mathcal{S}_{\rm inst}}~.\label{eq:GravityThetaEff}
\end{eqnarray}
 where we remind the reader that the action scales with $M_{\rm Pl}/f_a$.  The gravitational correction to the axion mass features a very interesting property if compared with the contribution generated by QCD.
As $f_a$ increases, the QCD axion mass $\Lambda_{\rm QCD}^4/f_a^2$ decreases but the gravitational term becomes more and more important, and 
eventually it overcomes the former. This means that we expect a lower bound on the axion mass.
This is shown in the right panel of fig.~\ref{fig:QCD} where we shaded in red the region in which non-perturbative 
gravitational corrections dominate. We find the lower bound on the axion mass (or, equivalently, the upper bound on the axion decay constant)
\begin{mymathbox}[ams gather, title=Lower bound on the QCD axion mass, colframe=titlepagecolor]
 m_a \gtrsim 4.8\times 10^{-10}\,{\rm eV}~,~~~~ f_a \lesssim 10^{16}\,{\rm GeV}~.\label{eq:QCDbound}
\end{mymathbox}
In the left panel of fig.~\ref{fig:QCD} we show the impact of the non-zero $\theta_{\rm eff}$ in eq.~(\ref{eq:GravityThetaEff}).
For reference, we consider the experimental bound $\theta_{\rm eff} < 10^{-10}$ (horizontal gray line).
We find that if $f_a \gtrsim 10^{16}\,{\rm GeV}$ then the contribution induced by gravity 
becomes too large. Notice that, by accident, the bounds on $f_a$ extracted from $m_a$ and $\theta_{\rm eff}$ are of the same order, and 
assuming $\delta_1 = 0$ does not change quantitatively our conclusion.

\begin{figure}[!htb!]
\centering
  \includegraphics[width=.55\linewidth]{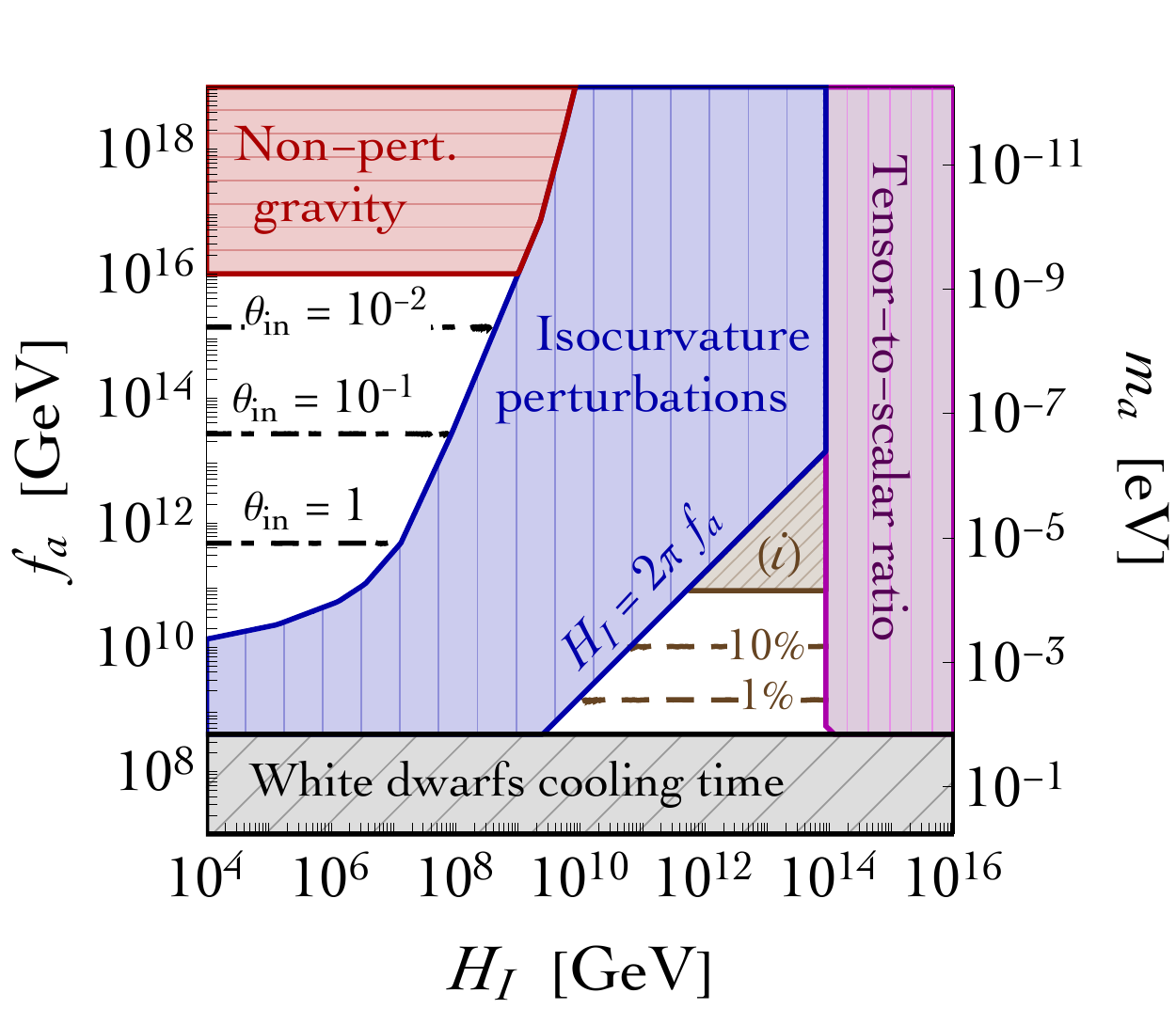}
\caption{\label{fig:AxionMasterPlot}\em 
QCD axion parameter space (see text for details). 
The region shaded in red with horizontal meshes is excluded by non-perturbative gravitational effects, eq.~(\ref{eq:QCDbound}).
}
\end{figure}
In fig.~\ref{fig:AxionMasterPlot} we show the impact of the bound in eq.~(\ref{eq:QCDbound})
on the QCD axion parameter space. As customary, the condition $f_a \lessgtr H_I/2\pi$, with $H_I$ the Hubble expansion rate at the end of inflation
distinguishes between the cases in which the PQ symmetry is broken before ($>$) or after ($<$) the end of inflation.
The first possibility defines the so-called anthropic axion window, in which  the possibility to reproduce the observed 
 dark matter relic density $\Omega_a h^2 \simeq 0.1$
relies on a fine-tuned choice of the initial misalignment angle $\theta_{in}$ (black dot-dashed lines in 
fig.~\ref{fig:AxionMasterPlot} for different $\theta_{in}$). Notice that a substantial part of the anthropic axion window is excluded  
by the presence of  axion isocurvature perturbations in the early Universe (blue region with vertical meshes) that we 
compute following~\cite{Visinelli:2009zm,Visinelli:2014twa} (see also~\cite{Visinelli:2009kt,Visinelli:2017imh} for recent progress for non-standard cosmologies and generalization to axion-like particles). 
The region $f_a < H_I/2\pi$ defines the classical axion window. In the region ({\it i}) shaded in brown with diagonal meshes 
$\Omega_a h^2 > 0.1$, and the Universe is over-closed. The dashed brown 
lines reproduce $1\%$ and $10\%$ of the observed dark matter abundance. In this region a sizable contribution from decay of topological defects
is expected~\cite{Kawasaki:2014sqa}. For completeness, we also show the region excluded by white dwarf 
cooling time~\cite{Raffelt:1985nj} and upper limit on the tensor-to-scalar ratio~\cite{Ade:2015lrj}.
The bound in eq.~(\ref{eq:QCDbound}) applies in the anthropic axion window, and 
disfavors a region of the parameter space (shaded in red with horizontal meshes) previously allowed. 
We remark that the bound in eq.~(\ref{eq:QCDbound}) was derived without specific assumption
but only considering  minimal coupling of the axion field with Einstein gravity.

We now move to briefly discuss few scenarios of phenomenological relevance 
in which non-perturbative gravitational corrections to the mass of the QCD axion may play an important r\^ole.

\subsubsection{QCD axions and black hole superradiance}

Spin and mass measurements of stellar-size black holes exclude the QCD axion mass window~\cite{Arvanitaki:2014wva}
\begin{equation}\label{eq:SuperradianceBound}
6\times 10^{-13} \lesssim m_a\,[{\rm eV}] \lesssim  2\times 10^{-11}~,
\end{equation}
corresponding to $3\times 10^{17}\lesssim f_a\,[{\rm GeV}] \lesssim 10^{19}$.
This is because 
superradiance effects become efficient~\cite{Brito:2015oca,Arvanitaki:2010sy}
when the Compton wavelength of the axion is comparable with the horizon size of the black hole. 
The axionic field forms a quasi-stationary configuration around the black hole at the expense of its rotational energy, 
giving birth to a quasi-bound system that shares remarkable similarities -- such as energy orbitals and level transitions --  with the hydrogen atom. 
From $\lambda_{\rm Compton} = h/m_ac \sim R$, we have
\begin{equation}
m_a \sim 6\times 10^{-12}\left(\frac{30\,{\rm km}}{R}\right)\,{\rm eV}~,
\end{equation}
where $R$ is the typical radius of a stellar-mass black hole, thus justifying the mass range excluded in eq.~(\ref{eq:SuperradianceBound}).
At the two sides of this interval, and within 
the mass range $m_a =[10^{-14}, 10^{-10}]$ eV, 
stellar black hole superradiance in the presence of the QCD axion may produce in the next few years 
spectacular signatures -- both direct and indirect -- in gravitational wave detectors such as Advanced LIGO~\cite{Arvanitaki:2016qwi,Giudice:2016zpa}.
Indirect signatures refer to the observation of gaps
 in the spin-mass distribution of  final state black holes produced  by binary black hole mergers.
 Direct signatures refer to monochromatic gravitational wave signals 
 produced 
 during the dissipation of the scalar condensate after the superradiant condition is saturated.
 
 As far as  direct signatures are concerned,  a careful assess of the detection prospects in Advanced LIGO and LISA was recently proposed 
 in~\cite{Brito:2017wnc,Brito:2017zvb}.
 The outcome of the analysis is that, considering optimistic astrophysical models for black hole populations, the 
 gravitational wave signal produced by superradiant clouds of scalar bosons  
with mass in the range 
 $m_a = [2\times 10^{-13}, 10^{-12}]$ eV is observable -- i.e. it is characterized by a signal-to-noise ratio larger than the experimental threshold -- by Advanced LIGO.
 Notice that this region seems to be ruled out if one considers at face value the bound in eq.~(\ref{eq:SuperradianceBound}).
 However, it is worth emphasizing that the bound in eq.~(\ref{eq:SuperradianceBound}) is most likely only indicative since it is based on 
 black hole spin measurements 
 that are extracted indirectly from X-ray observations of accretion disks in X-ray binaries. We only have very few of such measurements at our disposal, 
 and it is difficult to extract a bound with significant statistical confidence. 
 The discussion of this matter however escapes the scope of the present work.
    
As clear from the right panel of fig.~\ref{fig:QCD} and from the parameter space in fig.~\ref{fig:AxionMasterPlot}, 
the QCD axion in the mass range $m_a =[10^{-14}, 10^{-10}]$ eV violates the bound in eq.~(\ref{eq:QCDbound}), and 
fits into a region where non-perturbative gravitational effects dominate over QCD. 
We therefore conclude that -- working under the very same hypothesis, that is an axion minimally coupled to Einstein gravity -- 
the phenomenologically interesting mass range  $m_a =[10^{-14}, 10^{-10}]$ eV  motivated by black hole superradiance 
is theoretically forbidden for the QCD axion. 
This is not, of course, a lapidary conclusion.
To be more optimistic, observing  the QCD axion in connection with 
black hole mergers  at  the Advanced LIGO could imply an evidence for
 modifications of Einstein gravity. 
 Alternatively, a scalar boson different from the QCD axion could still leave its imprint in the texture of gravitational wave signatures. 
 \begin{figure}[!htb!]
\centering
  \includegraphics[width=.6\linewidth]{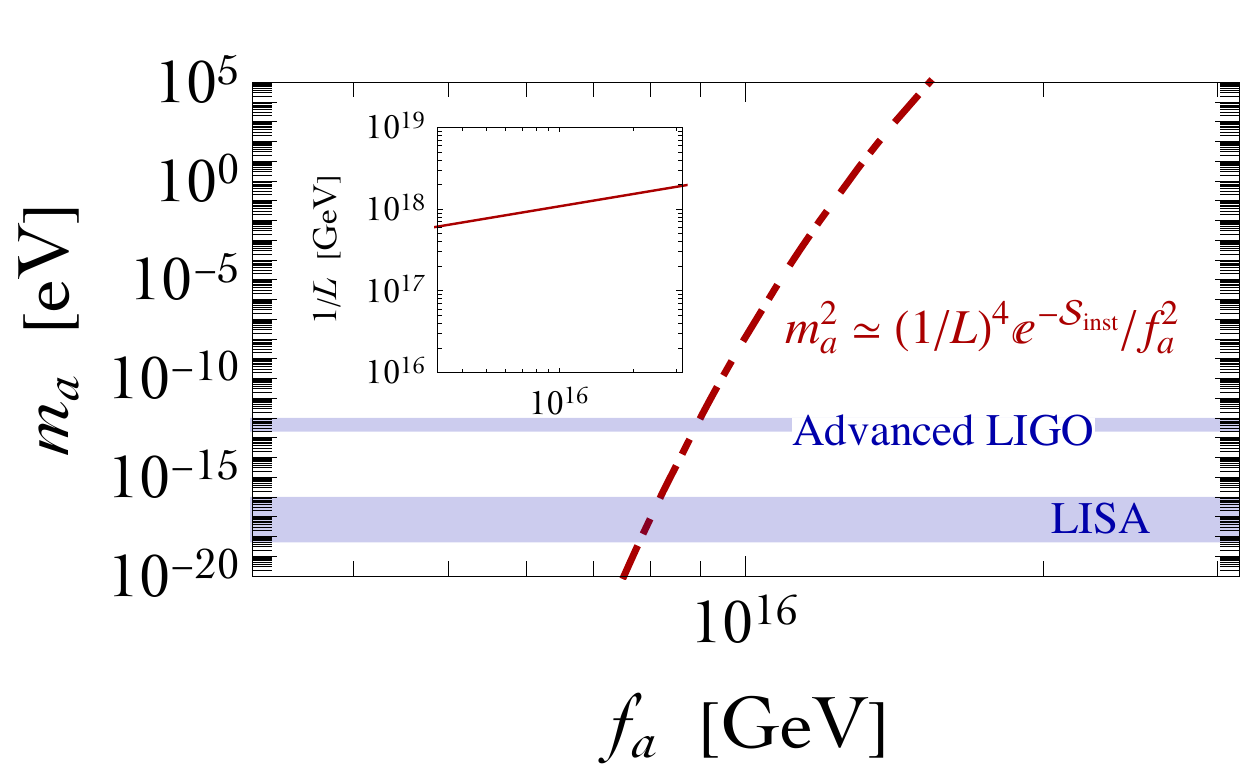}
\caption{\label{fig:ScalarBosonMass}\em 
Gravitational mass generated by non-perturbative Euclidean wormholes as a function of the Goldstone boson decay constant $f_a$. 
In the inset plot, we show, in the same interval of $f_a$, the inverse of the wormhole size $L$ (see eq.~(\ref{eq:Wormhole})).
The horizontal blue lines represent the mass range favored by the analysis in~\cite{Brito:2017wnc,Brito:2017zvb} considering both 
Advanced LIGO ($m_a = [2\times 10^{-13}, 10^{-12}]$ eV) and LISA ($m_a = [5\times 10^{-19}, 5\times 10^{-16}]$ eV) gravitational wave detectors.
}
\end{figure}
In this last case,  the mass term generated by gravity can be used to justify the lightness of the scalar boson. 
By turning off the QCD contribution in eq.~(\ref{eq:GravityMass}), we show the gravitational mass 
$m_a^2 \simeq  (1/L)^4e^{-\mathcal{S}_{\rm inst}}/f_a^2$ in fig.~\ref{fig:ScalarBosonMass}.
Clearly, if $f_a \simeq 10^{16}$ GeV it is possible to span, thanks to the exponential factor $e^{-\mathcal{S}_{\rm inst}}$ a 
large range of allowed values, including the regions favored by the analysis in~\cite{Brito:2017wnc,Brito:2017zvb}.
  
\subsubsection{Axion stars as black hole seeds}

The idea is to consider axion 
stars as black hole seeds, which are supermassive in the case of ultralight axions~\cite{Helfer:2016ljl}.
By studying numerically the collapse of axion stars in the context of general relativity, 
the authors of~\cite{Helfer:2016ljl} identified the critical value of the axion decay constant $f_{\rm TP}\sim 0.06\,M_{\rm Pl}$
above which black hole formation occurs. The mass of the typical black hole formed from axion star collapse is
$M_{\rm BH}\sim 3.4\left(f_a/0.12\,M_{\rm Pl}\right)^{1.2}M_{\odot}$. 

As before, the condition $f_a \gtrsim f_{\rm TP}\sim 0.06\,M_{\rm Pl}$
is not compatible with the bound in eq.~(\ref{eq:QCDbound}).
We stress that this bound was derived 
considering a QCD axion minimally coupled to Einstein gravity, therefore  
without any additional assumptions compared to the ones in~\cite{Helfer:2016ljl}.
Said differently, it would be interesting to include in the analysis of~\cite{Helfer:2016ljl} 
non-perturbative 
 gravitational corrections to the axion potential since they play a relevant r\^ole in the region of parameter space 
 in which black hole formation may occur.

\subsection{Ultralight scalars as cosmological dark matter}

Dark matter must behave sufficiently classically to be confined on galaxy scales.
If we suppose dark matter to be a boson with mass $m_a$ and velocity $v$, 
we can require to behave classically down to the typical size of Milky Way 
satellite galaxies, and obtain the condition
\begin{equation}\label{eq:DeB}
\lambda_{\rm De\,Broglie} = \frac{1}{m_a v}\lesssim 1~{\rm kpc}
~~~~\Longrightarrow~~~~
m_a \gtrsim 10^{-22}\,\left(\frac{120\,{\rm km/s}}{v}\right)\,{\rm eV}~.
\end{equation}
For a typical halo with size $R\sim {\rm kpc}$, and mass $M \sim 10^9\,M_{\odot}$, 
we expect 
a virial velocity $v \sim \sqrt{G_N M/R} \sim 70$ km/s. From eq.~(\ref{eq:DeB}), we extract 
a lower limit for the mass of bosonic dark matter of about $10^{-22}$ eV. The dark matter saturating this value is known as
{\it fuzzy dark matter} (FDM)~\cite{Hu:2000ke}.
This kind of ultralight dark matter could form a Bose-Einstein condensate on galactic scales, 
providing a  possible solution to the tensions 
that arise when the standard cold dark matter paradigm is probed into the deep 
non-linear regime at redshift $z \sim 0$~\cite{Moore:1999gc,BoylanKolchin:2011de,BoylanKolchin:2011dk}.
Apart from this motivation, an ultralight dark matter condensate features peculiar observational astrophysical 
properties, and it catalyzed increasing attention in the dark matter 
community (see, e.g., ~\cite{Chavanis:2011uv,Boehmer:2007um,Harko:2011xw,Sikivie:2009qn,Schive:2014dra,Harko:2011zt,Hui:2016ltb,Diez-Tejedor:2017ivd}).

From a particle physics perspective, the extreme lightness of the FDM
 is well suited 
by an axion-like particle. Furthermore, reproducing the observed value of relic abundance via the misalignment mechanism 
gives a clue about the value of its decay constant $f_a$.

For completeness, let us quickly sketch the computation. We consider here an axion-like field $a(t, \vec{x}) = f_a\theta(t, \vec{x})$ 
with potential $V(a) = \mu^4\left(1 - \cos\theta\right)$. In the Friedmann-Robertson-Walker cosmology $ds^2 = -dt^2 +R^2(t)d\vec{x}^2$  
the evolution of the axion field is given by $\ddot{a}(t, \vec{x}) + 3H\dot{a}(t, \vec{x}) - \triangle a(t, \vec{x})/R^2 + 
dV(a)/da = 0$, where $H \equiv \dot{R}/R$, 
and the dot  indicates derivative w.r.t. time. Neglecting higher order in the potential, we have 
$\ddot{a}(t, \vec{x}) + 3H\dot{a}(t, \vec{x}) - \triangle a(t, \vec{x})/R^2 + m_a^2a(t, \vec{x}) = 0$, with $m_a^2 \equiv \mu^4/f_a^2$. 
As customary, we can define the time $t_1$ at which the condition $m_a = 3H(t_1)$ is satisfied.
Notice that for simplicity we are considering an axion mass that is, to a first approximation, temperature-independent.  
The time $t_1$ separates two regimes. 

For $t< t_1$, we can neglect the mass term, $H \gg m_a$. Introducing the Fourier decomposition
$a(t,\vec{k}) = \int d^3\vec{x}e^{i\vec{k}\cdot\vec{x}}a(t,\vec{x})$ we obtain $\ddot{a}(t,\vec{k}) + 3H\dot{a}(t,\vec{k}) -
 (k^2/R^2)a(t,\vec{k}) = 0$. 
 The Fourier modes $a(t,\vec{k})$ separate into modes outside (that is for $k/R \ll H$) and inside (that is for $k/R \gg H$) the horizon.
 Inside the horizon, we can not neglect the $k^2/R^2$ term. Solving the equation of motion for these modes, 
  it is possible to show that  
 the corresponding solutions oscillate with frequency $k/R$, and the amplitude decreases with time as $1/R$.
The modes that are confined inside the horizons until $t\simeq t_1$ can therefore be neglected. 
 As far as the modes outside the horizon are concerned, 
 we can 
 neglect the $k^2/R^2$ term, and the equation of motion is solved by 
 $a(t,\vec{k}) = c_1(\vec{k}) + c_2(\vec{k})t^{-1/2}$. The modes that are confined outside the horizon until $t\simeq t_1$ are therefore  
 frozen at some constant initial value,
 and they are collectively  called zero modes $a_0$. 
 
 Let us now move to discuss the second regime,  $t\gtrsim t_1$, focusing on the zero modes.
 When the axion mass becomes non-negligible, the evolution of the zero modes is described by the equation $\ddot{a}_0 + 3H\dot{a}_0 + m_a^2 a_0 = 0$.
The zero modes, previously frozen, at $t\gtrsim t_1$ start 
oscillate with frequency set by $m_a$. From the definition of axion energy density $2\rho_{a_0} = \dot{a}_0^2 + m_a^2 a_0^2$
we find, using the equation of motion, $\dot{\rho}_{a_0} = -3H\dot{a}_0^2$. Since $a_0$ oscillates with period $m_a$, we can average over an oscillation, 
and estimate $\dot{a}_0 \simeq m_a a_0$. Consequently, $\rho_{a_0} \simeq m_a^2 a_0^2$ and $\dot{\rho}_{a_0} \simeq -3H\rho_{a_0}$.
After integration, the latter equation gives the scaling $\rho_{a_0}(t) \sim m_a/R^3(t)$. This is the key argument defining the physics of the  
misalignment mechanism: During the cosmological evolution of the axion field, for 
 $t\gtrsim t_1$ the number of axions in a comoving volume is conserved. We can therefore write the energy density 
 stored in the axion zero modes today as
 \begin{equation}\label{eq:Misalignment}
 \rho_{a_0}(t_0) = \rho_{a_0}(t_1)\left[
 \frac{R(t_1)}{R(t_0)}
 \right]^3~,~~~~\rho_{a_0}(t_1) \simeq m_a^2 f_a^2\theta_{in}^2~,
 \end{equation}
 where the dependence from the so-called initial misalignment angle $\theta_{in}$ is manifest. 
 
 The rest of the computation makes use of some basic thermodynamic concepts to
  rephrase  eq.~(\ref{eq:Misalignment}) in terms of observable quantities.
 From entropy conservation,  $g_*^s(T)R^3 T^3 = {\rm const}$, we have 
 \begin{equation}\label{eq:thermodynamic}
 \left[
 \frac{R(t_1)}{R(t_0)}
 \right]^3 = \frac{s(T_0)}{\frac{2\pi^2}{45}g_*(T_1)T_1^3}~,~~~~~{\rm with}~~~s(T_0) = \frac{4}{3}\left[
 \frac{g_*^s(T_0)}{g_*(T_0)}
 \right]\left(
 \frac{\rho_R(T_0)}{T_0}
 \right)~,
 \end{equation}
 where $h \equiv H_0/(100\,{\rm km/s/Mpc})$, with $H_0 = 67.8\,{\rm km/s/Mpc}$~\cite{Ade:2015xua}, is the reduced Hubble constant, 
 $T_{1}$ is the temperature at time $t_1$, $T_0 = 2.726$ $^{\circ}$K is the present temperature of the Universe,  
 $\rho_{\rm crit} = 3H_0^2/8\pi G_N$ is the critical energy density, $g_*(T)$ ($g_*^s(T)$) is the effective number of degrees of freedom (effective 
 number of degrees of freedom in entropy) at temperature $T$, and $\rho_{R}(T_0)$ is the present 
 energy density in radiation.
 In eq.~(\ref{eq:thermodynamic}) $s(T_0)$ is the entropy density at present time, and the relation between $s(T_0)$ and $\rho_{R}(T_0)$  follows from 
 \begin{equation}
 s(T) = \frac{2\pi^2}{45}g_*^s(T)T^3~,~~~~\rho_R(T) = \frac{\pi^2}{30}g_*(T)T^4~.
 \end{equation}
 
 The energy density in radiation can be estimated from
  the two Planck measurements~\cite{Ade:2015xua} of the current matter density $\Omega_{m}$ and the redshift of radiation-matter equality $z_{eq}$.
  From $\rho_m(z_{eq}) = \rho_{R}(z_{eq})$, which is equivalent to $\Omega_{m}(z_{eq} + 1)^3 
  = \Omega_{R}(z_{eq} + 1)^4$, and using $\Omega_{m} = 0.3$, $z_{eq} = 3365$, it follows that $\Omega_{R}h^2 \simeq 4.3\times 10^{-5}$.
  
  The value of $T_1$ follows from the condition $m_a = 3H(t_1)$ taking into account that we can write the Hubble parameter as a function of the temperature as
  $H(T) = \sqrt{\pi^2 g_*(T)/90 M_{\rm Pl}^2}T^2$. Parametrically, we have the relation $M_{\rm Pl}m_a \simeq T_1^2$.
  For a FDM candidate with $m_a \simeq 10^{-22}$ eV, it follows that $T_1 \simeq 10^3$ eV.
  At such small value of the temperature, we can approximate $g_*(T_1) \approx 1$. Furthermore, we also have $g_*(T_0) \simeq g_*^s(T_0)$.
  
  All in all, defining $\Omega_a \equiv \rho_{a_0}(t_0)/\rho_{\rm crit}$, and assuming $\theta_{\rm in}\simeq 1$, we find the parametric order-of-magnitude scaling
  \begin{equation}\label{eq:FDM}
  \Omega_a h^2 \approx 0.1\,\left(
  \frac{f_a}{10^{17}\,{\rm GeV}}
  \right)^2\left(
\frac{m_a}{10^{-22}\,{\rm eV}}  
  \right)^{1/2}~.
  \end{equation}
  
  Eq.~(\ref{eq:FDM}) points towards a very specific interplay between $f_a$ and $m_a$, and it is interesting to further investigate its origin.
  In~\cite{Hui:2016ltb} this relation was justified referring to string theory models in which one expects
  an explicit breaking of the axionic shift symmetry due to
   the existence of 
  worldsheet or membrane instantons~\cite{Svrcek:2006yi}.
  Even without invoking string theory constructions, 
  we point out that  the mass term generated by gravity in eq.~(\ref{eq:MainPotential})
  provides a nice explanation for the relic abundance in eq.~(\ref{eq:FDM}). 
  Indeed, using $m_a^2 \simeq (1/L)^4 e^{-\mathcal{S}_{\rm inst}}/f_a^2$, 
    the observed abundance can be reproduced with $f_a \simeq 8\times 10^{15}$ GeV and $m_a \simeq 2.5\times 10^{-18}$ eV.
    Notice that  this value falls within the mass range that will be explored by the LISA gravitational wave interferometer, see fig.~\ref{fig:ScalarBosonMass}.
    This is an interesting observation, 
    since it shows  that a dark matter candidate with nothing more than gravitational interactions
may explain in a natural way the lightness of its mass. 

We finally note that, as fig.~\ref{fig:ScalarBosonMass} shows,  a value of the axion decay constant  above $f_a \simeq 10^{16}$ GeV
would generate an unacceptably large contribution in eq.~(\ref{eq:FDM}).

\section{Comments on the r\^ole of possible UV completions}\label{sec:UV}

In this section we discuss possible UV completions of the axion Goldstone mode, and their consequences on wormhole physics.

\subsection{Dynamical radial mode}\label{sec:DynamicalRadialMode}

We start our investigation considering the simplest UV completion of the axion theory.
This is the case of a $U(1)$ global symmetry that is spontaneously broken by the VEV $\langle |\Phi|\rangle = f_a/\sqrt{2}$ of a
complex scalar field $\Phi(x) = [f(x)/\sqrt{2}]e^{\imath \phi(x)/f_a}$. The mexican hat potential responsible for the spontaneous symmetry breaking 
is $V(f) = \lambda(f^2 - f_a^2)^2/4$. The radial field $f(x)$ gets a mass $M_f^2 = \lambda f_a^2$.
The physics discussed in the previous sections corresponds to the case in which the radial mode does not participate in the dynamics, and its value 
remains frozen at $f = f_a$. We expect this situation to be strictly true if $M_{f} > 1/L$.  
For the critical value $f_a \simeq 10^{16}$ GeV 
that we identified in section~\ref{eq:QCDAxion}, 
we have $1/L \simeq 10^{18}$ GeV. 
It is therefore hard to imagine that the radial field $f$ plays no dynamical r\^ole, since we expect  its mass
 -- for reasonable small coupling $\lambda$ -- to lie below the scale set by the wormhole throat.
In order to validate our conclusions, it is  necessary to generalize the wormhole solutions, and 
compute their action,  to the case in which $f$ is a dynamical field~\cite{Kallosh:1995hi,Abbott:1989jw}.
In order to make contact with the existing literature, we look for spherically symmetric solutions, and
 we use the Euclidean metric ansatz $ds^2 = dr^2 + R^2(r)d\Omega^2_{3,1}$.
The Ricci scalar is $\mathcal{R} = -6[-1+(R^{\prime})^2 + RR^{\prime\prime}]/R^2$, and the Einstein
 field equations, together with the equation of motion for the radial field $f$, are~\cite{Lee:1988ge,Abbott:1989jw}
 \begin{eqnarray}
 (R^{\prime})^2 &=& 1 - \frac{8\pi R^2}{3M_{\rm Pl}^2}\left[
 -\frac{(f^{\prime})^2}{2} + V(f)  + \frac{n^2}{8\pi^4f^2 R^6}
 \right]~,\label{eq:Radial1} \\
 -1 + (R^{\prime})^2 + 2RR^{\prime\prime} &=& -  \frac{8\pi R^2}{M_{\rm Pl}^2}
 \left[
 \frac{(f^{\prime})^2}{2} + V(f) - \frac{n^2}{8\pi^4f^2 R^6}
 \right]~,\label{eq:Radial2}\\
 f^{\prime\prime} + \frac{3R^{\prime}}{R}f^{\prime} &=& \frac{dV}{df} - \frac{n^2}{4\pi^4f^3R^6}~,\label{eq:Radial3}
 \end{eqnarray}
 where we already substituted $f^2(\theta^{\prime})^2 = n^2/4\pi^4f^2R^6$, with $n$ the quantized axion charge along the wormhole throat.
 Instead of eqs.~(\ref{eq:Radial1},\ref{eq:Radial2}), in our numerical investigation we found more useful to use the combination 
 $R^{\prime\prime} = (4\pi R/3M_{\rm Pl}^2)[-2(f^{\prime})^2 - 2V(f) +n^2/2\pi^4 f^2 R^6]$.
 We introduce the dimensionless variables~\cite{Kallosh:1995hi,Abbott:1989jw}
 \begin{equation}\label{eq:Dimensionless}
 \rho \equiv rM_{\rm Pl}\sqrt{\frac{3\lambda}{8\pi}}~,~~~~A\equiv RM_{\rm Pl}\sqrt{\frac{3\lambda}{8\pi}}
 ~,~~~~F\equiv\frac{f}{M_{\rm Pl}}\sqrt{\frac{8\pi}{3}}~,
 \end{equation}
 and the differential equations become  
 \begin{equation}\label{eq:Shooting}
 A^{\prime\prime} = \frac{A}{2}\left[
 -2(F^{\prime})^2 - \frac{1}{2}(F^2 - F_a^2)^2 + \frac{4Q}{F^2A^6}
 \right]~,~~~~~~~F^{\prime\prime} + \frac{3A^{\prime}}{A}F^{\prime} = F(F^2 - F_a^2)^2 - \frac{2Q}{F^3A^6}~,
 \end{equation}
 with $Q\equiv n^2\lambda^2/8\pi^4$. 
 Finally, in terms of the dimensionless variables defined in eq.~(\ref{eq:Dimensionless}), 
 the action -- including the Gibbons-Hawking-York boundary term -- reads
 \begin{equation}\label{eq:ActionRadial}
 \mathcal{S} = \frac{2\pi^2}{\lambda}\int_0^{\infty}d\rho\left[
 A^3(F^{\prime})^2 + 2AA^{\prime}\left(1-A^{\prime}\right)
 \right]~.
 \end{equation}
 
 We solve eq.~(\ref{eq:Shooting}) by means of a shooting 
 method that we validate against the results of~\cite{Kallosh:1995hi}.
\begin{figure}[!htb!]
\minipage{0.5\textwidth}
  \includegraphics[width=1.\linewidth]{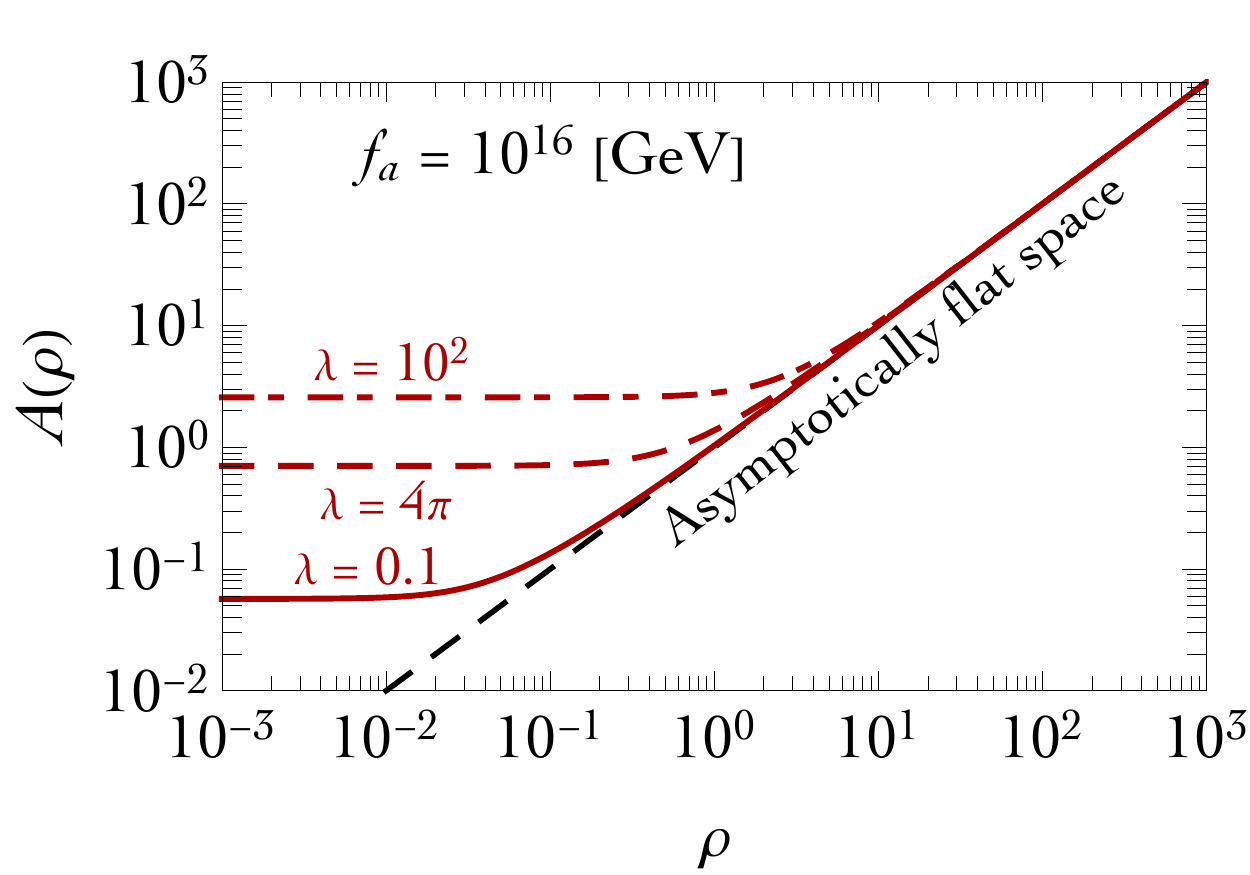}
\endminipage 
\hspace{.3cm}
\minipage{0.5\textwidth}
  \includegraphics[width=1.\linewidth]{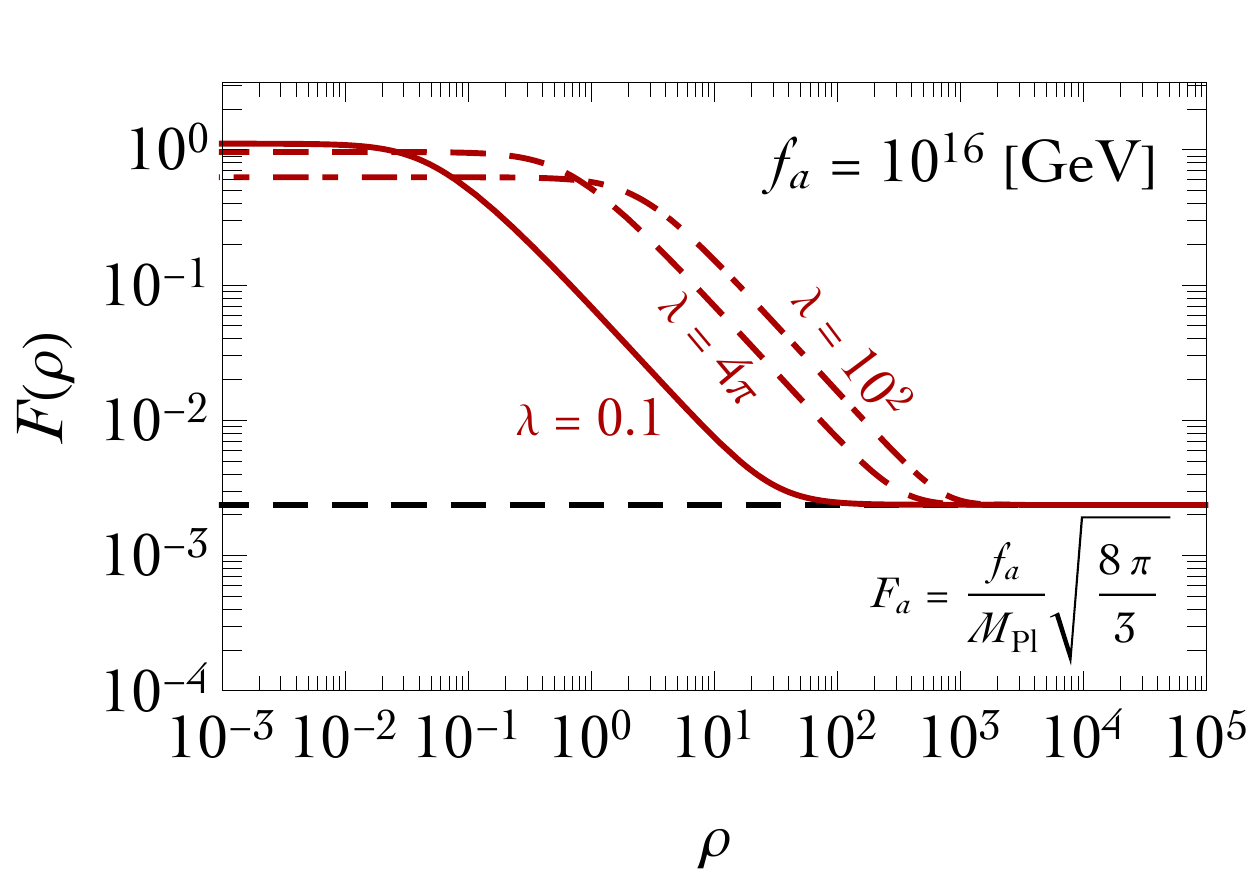}
\endminipage 
\vspace{-.2 cm}
\caption{\label{fig:RadialMode}\em 
Left panel. Wormhole geometry in terms of the rescaled metric factor $A(\rho)$ as a function of the dimensionless 
coordinate $\rho$ for different values of $\lambda$. The dashed black line refers to the flat Euclidean limit $A(\rho) = \rho$. 
Right panel. Rescaled radial field $F(\rho)$ as a function of the dimensionless 
coordinate $\rho$ for different values of $\lambda$. The dashed black line refers to the VEV $F_a$.
}
\end{figure}
The procedure goes as follows.
First, for a given initial guess for $F(\rho = 0)$ we compute $A(0)$ from the equation $A^4(0) - A^6(0)[F(0)^2 - F_a^2]^2/4 - Q/F^2(0) = 0$.
Second, we solve the system in eq.~(\ref{eq:Shooting}) with the four boundary data 
$F(0)$, $A(0)$, $A^{\prime}(0) = 0$, $F^{\prime}(0) = 0$. Finally, we check the asymptotic condition $F(\infty) = F_a$, and
we tune the initial guess $F(0)$ until we find it satisfied.
We show our result in fig.~\ref{fig:RadialMode}, where we focused on the critical value $f_a = 10^{16}$ GeV 
and wormhole with unit charge $n=1$.
In the left (right) panel we show the dimensionless metric function $A(\rho)$ (the dimensionless radial field $F(\rho)$) 
as a function of the rescaled distance $\rho$ for 
three different values of $\lambda$. As far as the geometry is concerned, we see that, asymptotically, 
the solution recovers the flat space $A(\rho) = \rho$ while for small $\rho$
 deviations describing the wormhole geometry emerge.
Going through the wormhole throat, 
the radial field $F(\rho)$ substantially deviates 
w.r.t. the its asymptotic value $F_a$. As expected, including the radial mode with mass $M_f < 1/L$ as dynamical degree of freedom 
alters the wormhole solution 
found for non-propagating $f$. It is therefore important to quantify such deviation by computing the value of the action in 
eq.~(\ref{eq:ActionRadial}).
\begin{figure}[!htb!]
\minipage{0.5\textwidth}
  \includegraphics[width=.95\linewidth]{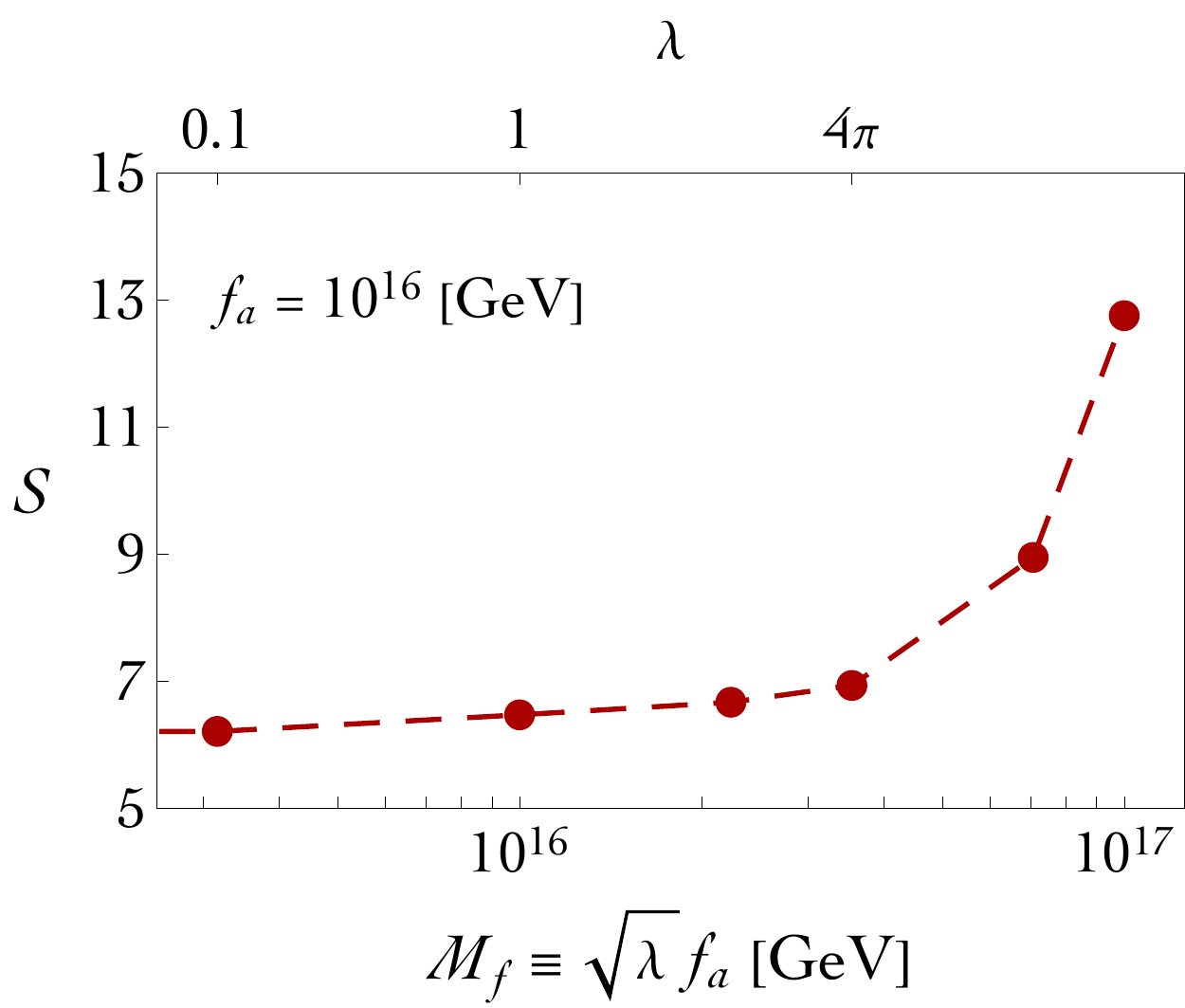}
\endminipage 
\minipage{0.5\textwidth}
  \includegraphics[width=1.\linewidth]{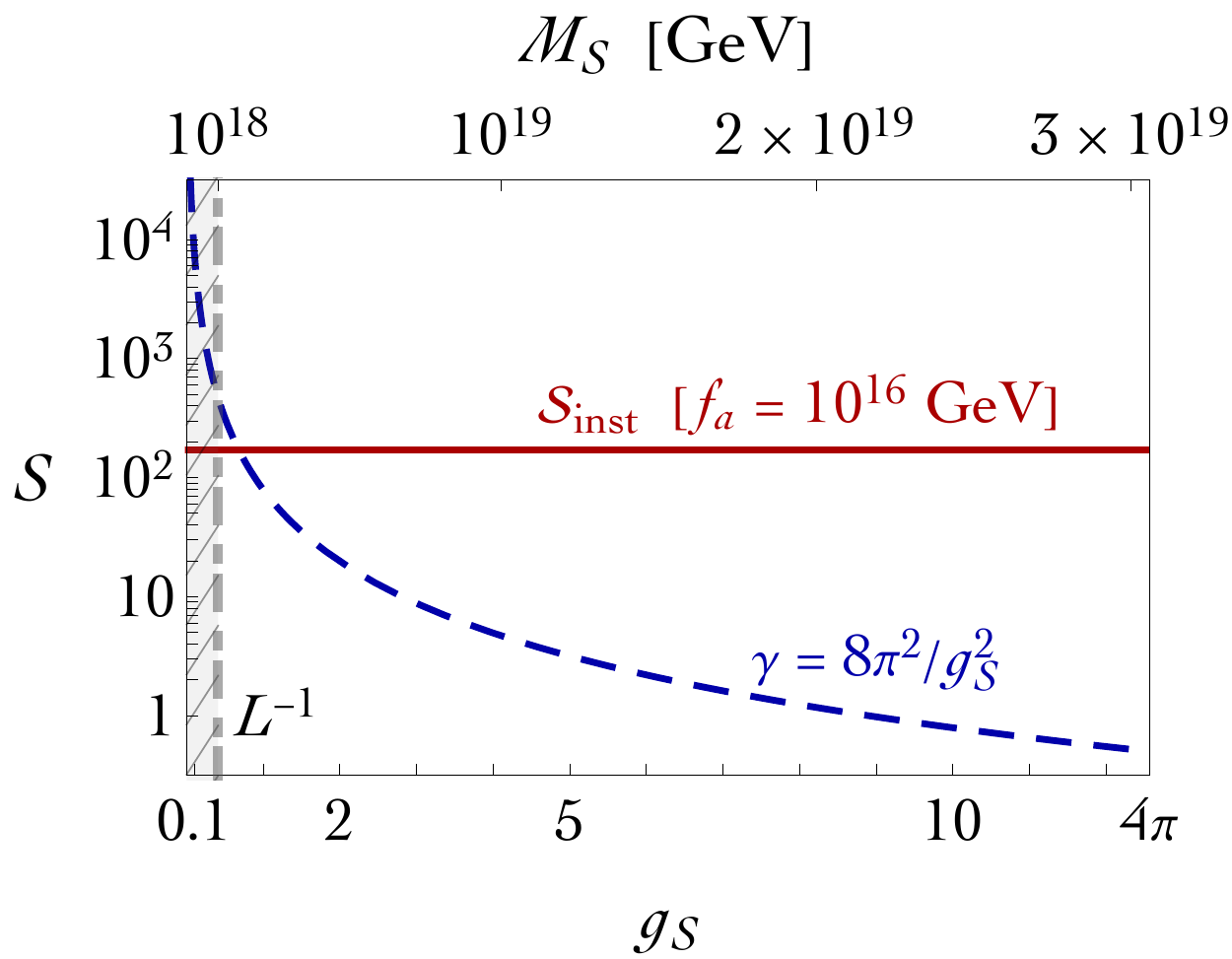}
\endminipage 
\vspace{-.2 cm}
\caption{\label{fig:FullAction}\em 
Left panel. On-shell action in eq.~(\ref{eq:ActionRadial}) as a function of the mass of the radial mode $M_f$.
Right panel. Instanton action as a function of the string coupling $\mathrm{g}_S$. 
On the top x-axis we show the value of the string mass scale $M_S$ computed in eq.~(\ref{eq:StringMassHeterotic}).
In solid red, we show the on-shell 
action $\mathcal{S}_{\rm inst}$ for the instanton at the critical value $f_a = 10^{16}$ GeV. The blue dashed line
represents the topological contribution $\gamma = 8\pi^2/\mathrm{g}_S^2$.
The region shaded in gray corresponds to $M_S < L^{-1}$.
}
\end{figure}
We show our result in the left panel of fig.~\ref{fig:FullAction}, in which we plot the on-shell action $\mathcal{S}$ 
as a function of the mass of the radial mode $M_f$. 
From eq.~(\ref{eq:Fullaction}), we have that the instanton action 
with the radial mode frozen at $f_a = 10^{16}$ GeV is $\mathcal{S}_{\rm inst}\simeq 170$.
Including the dynamics of the radial mode results in a net decrease of the action, in agreement with the result of~\cite{Kallosh:1995hi}.
Let us  give a qualitative understanding of this effect. As derived in eq.~(\ref{eq:GIAction}), the wormhole action is proportional to $M_{\rm Pl}^2 L^2$, and it decreases 
if the size of its throat gets smaller. 
The inclusion of a light dynamical radial mode has precisely this effect, as can be seen from the left panel in fig.~\ref{fig:FullAction}: Going 
towards smaller values of $\lambda$  the size of the wormhole shrinks, and, consequently, the wormhole action decreases. 
For illustrative purposes, in fig.~\ref{fig:FullAction} we extend the plot to large values of $\lambda$ 
in order to show that, as $M_f$ moves towards the threshold $1/L$ above which it can 
be safely integrated out, the action increases.  

The inclusion of the radial mode in the $U(1)$ model 
shows that the presence of dynamical degrees of freedom belonging to the UV completion of the axion Lagrangian 
does not weaken the strength of the non-perturbative 
global symmetry breaking induced by gravity.
On the contrary, the total action in fig.~\ref{fig:FullAction} is 
significantly smaller if compared to the one in which the radial mode does not fluctuate around its VEV.

We therefore conclude that the bound derived in section~\ref{eq:QCDAxion} presumably represents a conservative estimate,
since extra dynamics related to the UV completion of the axion Lagrangian may even strengthen the global symmetry breaking.

\subsection{Higher-curvature operators and string theory}\label{sec:HigherCurv}

In this section we explore the impact of higher-curvature operators on the wormhole solution. 
To be more concrete, we include the Gauss-Bonnet term
\begin{equation}\label{eq:GB}
\Delta\mathcal{S}_{\rm E} = -\frac{\gamma}{32\pi^2}\int d^4x \sqrt{g}\left(
\mathcal{R}_{\mu\nu\lambda\delta}\mathcal{R}^{\mu\nu\lambda\delta}  - 4\mathcal{R}_{\mu\nu}\mathcal{R}^{\mu\nu} + \mathcal{R}^2
\right)~,
\end{equation}
that represents the $4^{\rm th}$  order in the derivative expansion of the gravitational action.
We limit our analysis to
 the Gauss-Bonnet term since it is 
the unique ghost-free quadratic curvature invariant~\cite{Zwiebach:1985uq}. 
Furthermore, as we shall discuss later, the presence of the effective higher-curvature correction in eq.~(\ref{eq:GB}) is a peculiar prediction of many 
string theories -- bosonic~\cite{Zwiebach:1985uq}, heterotic~\cite{Gross:1986mw}, and type-I string~\cite{Tseytlin:1995bi} (while it vanishes in type-II superstring theory~\cite{Gross:1986iv}).

 In $D=4$ dimensions, the Gauss-Bonnet term reduces to a total derivative.
As a consequence, it does not enter in the equation of motion but it gives a non-vanishing topological contribution to the wormhole action. 
Before proceeding, we notice that the validity of the derivative expansion in the gravitational action imposes the cutoff 
$\Lambda \sim M_{\rm Pl}/\sqrt{\gamma/4\pi}$. It is therefore natural to require the condition $\Lambda > 1/L$, in 
order to ensure the validity of the derivative expansion all the way down the wormhole throat.
We find that this condition corresponds to 
\begin{equation}
\gamma < \frac{4M_{\rm Pl}}{\sqrt{3\pi} f_a}~.
\end{equation}
To fix ideas, if $f_a = 10^{16}$ GeV we have $\gamma \lesssim 1600$.
We now compute the value of the Gauss-Bonnet term on the wormhole solution.
Using the metric in eq.~(\ref{eq:WormholeMetric}), a direct computation gives
\begin{equation}
\mathcal{R}_{\mu\nu\lambda\delta}\mathcal{R}^{\mu\nu\lambda\delta}  - 4\mathcal{R}_{\mu\nu}\mathcal{R}^{\mu\nu} + \mathcal{R}^2 =
\frac{24\alpha^{\prime}(-1 + \alpha^2)}{r^3\alpha^5}~,
\end{equation}
and the action in eq.~(\ref{eq:GB}) gives 
\begin{equation}\label{eq:Bonnet}
\Delta\mathcal{S}_{\rm E} = \frac{3\gamma}{2}\int_L^{\infty}\frac{\alpha^{\prime}(1-\alpha^2)}{\alpha^4}
= \frac{3\gamma}{2}\left|
\frac{1}{\alpha} - \frac{1}{3\alpha^3}
 \right|_{r = L}^{r = \infty} = \gamma~.
\end{equation}
Let us focus on the case of the QCD axion, discussed in section~\ref{eq:QCDAxion}. 
At the critical value $f_a = 10^{16}$ GeV, the instanton action is $\mathcal{S}_{\rm inst} \simeq  170$.
As a rule of thumb, we can say that if $\gamma \gtrsim 170$ the higher-curvature topological correction 
due to the Gauss-Bonnet term in eq.~(\ref{eq:Bonnet}) significantly alters the instanton action, increasing its value, and making its
 impact on axion physics harmless.
 Although qualitatively correct, 
 the real point here is to understand the physical implications 
 of this numerical condition. 
On a pure gravitational ground, $\gamma$ is nothing but a dimensionless number, 
and it seems difficult to justify the relation  $\gamma \gtrsim 170$. 
Reinterpreting the effective action in eq.~(\ref{eq:GB}) having in mind 
the UV physics responsible for its generation would be much more useful.
The lack of a quantum theory of gravity, however, makes any argument highly speculative.
For lack of a better choice, we can ask string theory to help us with this task. 
In string theory the Gauss-Bonnet term often appears with the identification~\cite{Green:1987sp,Green:1987mn}
\begin{equation}\label{eq:GBString}
\gamma = \frac{\pi\alpha^{\prime}M_{\rm Pl}^2}{4}~,
\end{equation}
where $\alpha^{\prime} = l_S^2$ is related to the string scale $l_S$. 
We can gain more insight considering the case of  heterotic string theory
in which the string gauge coupling $\mathrm{g}_S$, the string mass scale $M_S$, and the parameter $\alpha^{\prime}$ are related by~\cite{Gross:1984dd}
\begin{equation}\label{eq:StringMassHeterotic}
M_S = \frac{2}{\alpha^{\prime}} = \frac{2}{l_S^2} = \frac{\mathrm{g}_S M_{\rm Pl}}{\sqrt{8\pi}}~,
\end{equation}
in close analogy with the relation discussed in eq.~(\ref{eq:StringMassScale}). From eq.~(\ref{eq:GBString}), we find 
$\gamma = 8\pi^2/\mathrm{g}_S^2$. Notice that, presented in this form,  the topological contribution to the action precisely matches the typical 
instanton action in the 
Yang-Mills case. We are now in the position to compare $\gamma = 8\pi^2/\mathrm{g}_S^2$ and $\mathcal{S}_{\rm inst}$.
We show our result in the right panel of fig.~\ref{fig:FullAction}, 
in which we compare the half-wormhole action $\mathcal{S}_{\rm inst}$ computed at the critical value $f_a = 10^{16}$ GeV 
that we identified in section~\ref{eq:QCDAxion}, 
with the action $\gamma = 8\pi^2/\mathrm{g}_S^2$ as a function of the string coupling.
On the top y-axis we put the values of the string mass scale $M_S$, computed according to eq.~(\ref{eq:StringMassHeterotic}).
The region shaded in gray corresponds to the condition $M_S < 1/L$.

Clearly,  in the presence of a weak string coupling our computation can be completely invalidated. The string mass scale
becomes lower than $L^{-1}$ -- thus breaking the effective description -- and the topological action generated by string theory dominates over the gravitational instanton contribution.

However, we also stress that in the presence of a moderately strong string coupling the UV completion of GR cannot fix the problem since the gravitational instanton
contribution dominates over the topological string term. 

The lack of a quantum theory of gravity makes any speculation quite far-fetched, and the only 
intent of the plot in the right panel of fig.~\ref{fig:FullAction} is
 to provide a fair example in which the gravitational action based on Einstein gravity captures 
 the relevant non-perturbative gravitational corrections.

\subsection{Higher-dimensional operators}\label{sec:HDO}

As discussed in the introduction, 
many authors proposed and tailored suitable extensions of the PQ symmetry in order to protect it 
from the presence of power-suppressed 
higher dimensional operators like those in eq.~(\ref{eq:Kamion}). 
Clearly, these 
attempts were motivated by a na\"{\i}ve understanding of the breaking effects generated by gravity.
As revisited in this paper, gravity breaks global symmetries at the non-perturbative level, 
and the effective operators originated from this 
physics are always suppressed by the exponential of the wormhole action, as exemplified in eqs.~(\ref{eq:MainPotential},\ref{eq:FinalEffective}).

However, it 
is conceivable that the $U(1)$ PQ symmetry arises as an 
accidental global symmetry  from a more
fundamental theory. 
In this realization, because of its accidental nature, higher-dimensional operators 
may source 
an explicit breaking.
In order to preserve the solution of the strong CP problem without introducing any degree of fine-tuning, 
 it is necessary to protect the accidental symmetry from breaking effects induced by dangerous irrelevant operators. 
 This can be achieved in a natural way by supporting the accidental PQ symmetry with a gauge symmetry.
 In this way, as we shall see, 
 one can prevent the presence of symmetry breaking operators up to (in principle arbitrarily) high dimensions, 
 thus obtaining a global PQ symmetry of very good quality, though accidental.
 
In this context, it is interesting to see what happens to non-perturbative breaking effects. 
In order to answer this question, let us discuss the specific construction put forward in~\cite{Fukuda:2017ylt}.
 The model features the presence of two sectors with two anomalous PQ symmetries, $U(1)_{\rm PQ}$ and $U(1)_{\rm PQ}^{\prime}$, realized 
 through the complex fields $\sqrt{2}\phi = f_a e^{\imath\tilde{a}/f_a}$ and $\sqrt{2}\phi^{\prime} = f_b e^{\imath\tilde{b}/f_b}$.
 Notice that for simplicity we do not consider explicitly the radial components.
 The domain of the two phase fields are $\tilde{a}/f_a = [0,2\pi)$ and $\tilde{b}/f_b = [0,2\pi)$.
 The crucial observations are the following.
 \begin{itemize}
\item[$\circ$] It is possible to define a linear combination of $U(1)_{\rm PQ}$ and $U(1)_{\rm PQ}^{\prime}$ that 
is free from QCD anomaly. Consequently, it can be promoted to a gauge symmetry, $U(1)_{\rm PQ}^{\rm gauge}$.
\item[$\circ$] In the limit in which the only interactions between the two sectors are those dictated by the aforementioned gauge symmetry, 
the theory possesses an accidental $U(1)$ symmetry, and delivers a massless Goldstone field.  
\end{itemize}
 The gauge symmetry $U(1)_{\rm PQ}^{\rm gauge}$ can be characterized as follows. 
 The phases of  $\phi$ and $\phi^{\prime}$, with gauge charges $q$ and $q^{\prime}$, transform under $U(1)_{\rm PQ}^{\rm gauge}$ 
 according to the
  shifts 
  \begin{equation}\label{eq:GaugeTransformation}
  \tilde{a}/f_a \to \tilde{a}/f_a + q\alpha~,~~~~~~\tilde{b}/f_b \to \tilde{b}/f_b + q^{\prime}\alpha~,
 \end{equation}
   and from the kinetic Lagrangian we get
   \begin{eqnarray}
   \mathcal{L} &=& \left|D_{\mu}\phi\right|^2 + \left|D_{\mu}\phi^{\prime}\right|^2  = 
   \left|\partial_{\mu}\phi - \imath g qA_{\mu}\phi \right|^2 + \left|\partial_{\mu}\phi^{\prime} - \imath g q^{\prime}A_{\mu}\phi^{\prime} \right|^2  \nonumber \\
   &=& \frac{1}{2}(\partial_{\mu}\tilde{a})^2 + \frac{1}{2}(\partial_{\mu}\tilde{b})^2 + \frac{m_A^2}{2}A_{\mu}A^{\mu} 
   - gA_{\mu}\left[
   qf_a(\partial^{\mu}\tilde{a})  + q^{\prime}f_b(\partial^{\mu}\tilde{b}) 
   \right]~,
   \end{eqnarray}
   where the mass of the gauge field is $m_A^2 \equiv g^2(q^2 f_a^2 + q^{\prime\,2}f_b^2)$.
   The rotation
   \begin{equation}
\left(
\begin{array}{c}
  a  \\   
  b
\end{array}
\right) = \frac{1}{\sqrt{q^2 f_a^2 + q^{\prime\,2}f_b^2}}
\left(
\begin{array}{cc}
q^{\prime}f_b  &  -qf_a   \\
qf_a  &   q^{\prime}f_b 
\end{array}
\right)\left(
\begin{array}{c}
  \tilde{a}  \\   
  \tilde{b}
\end{array}
\right)~,
   \end{equation}
   brings the Lagrangian into the neat form
   \begin{equation}\label{eq:Accidental}
   \mathcal{L} = \frac{1}{2}(\partial_{\mu}a)^2  + \frac{m_A^2}{2}\left(A_{\mu} - \frac{1}{m_A}\partial_{\mu}b\right)^2~.
   \end{equation}
 The field $b$ is the would-be Goldstone boson eaten by the massive vector field $A_{\mu}$.  
 The theory in eq.~(\ref{eq:Accidental}) enjoys an accidental (non-linearly realized)
 $U(1)$ global symmetry under which the massless Goldstone field $a$ remains shift-invariant. 
 The Goldstone field $a$ plays the r\^ole of the axion, and its continuos global shift symmetry is analogue to 
 the one  discussed in eq.~(\ref{eq:ContinuosShift}).
 In addition, and again in close analogy with the discussion outlined in the introduction, 
 a subgroup of this continuos shift symmetry is gauged, meaning that there is a phase redundancy in the definition of $a$.
   In order to visualize this property, we consider, 
   following~\cite{Fukuda:2017ylt},  the specific values $q=2$ and $q^{\prime} = 3$.
   Notice that these two integers are relatively prime, meaning that their greatest common divisor is $1$.
This particular choice does not change the conclusion  of the present discussion but it simplifies the formulas.
   
In the right panel of fig.~\ref{fig:Accidenti} we show in solid red the gauge orbits 
described in the field space $(\tilde{a}/f_a, \tilde{b}/f_b)$ by the gauge transformations in eq.~(\ref{eq:GaugeTransformation}).
For instance, starting from the point $\tilde{a}/f_a = 0$,  $\tilde{b}/f_b = 0$ in field space (`pure gauge' configuration),
by changing the value of the gauge parameter $\alpha$ one moves along the corresponding red line in the direction of the red arrow.
At the point $\tilde{a}/f_a = 4\pi/3$, $\tilde{b}/f_b = 2\pi$, because of the $2\pi$-periodicity in $\tilde{b}/f_b$,
the gauge orbits reappears at  $\tilde{a}/f_a = 4\pi/3$, $\tilde{b}/f_b = 0$. Similarly, 
because of the $2\pi$-periodicity in $\tilde{a}/f_a$, the gauge orbit connects  the points in field space
$\tilde{a}/f_a = 2\pi$, $\tilde{b}/f_b = \pi$ and $\tilde{a}/f_a = 0$, $\tilde{b}/f_b = \pi$.

The axion field $a = (q^2 f_a^2 + q^{\prime\,2}f_b^2)^{-1/2}(q^{\prime}f_b\tilde{a} - qf_a\tilde{b})$ 
is gauge invariant, and describes the direction orthogonal to the gauge orbits. 
The crucial points is that configurations in field space connected by gauge orbits are physically equivalent, since related by gauge transformations.
As a consequence, by crossing the gauge orbits 
the axion field experiences a periodicity, and  it is possible to show that the domain of $a$ is given by~\cite{Fukuda:2017ylt,Kim:2004rp}  
\begin{equation}
a = \left[
0,\frac{2\pi f_a f_b}{\sqrt{q^2 f_a^2 + q^{\prime\,2}f_b^2}}
\right)~.
\end{equation}
In parallel with eq.~(\ref{eq:DiscreteShift}), we therefore conclude that the discrete symmetry 
\begin{equation}
a \to a + 2k\pi F_a~,~~~~~ k\in \mathbb{Z}~,~~~~~{\rm with}~~F_a \equiv \frac{f_a f_b}{\sqrt{q^2 f_a^2 + q^{\prime\,2}f_b^2}}~,
\end{equation}
represents  a gauge symmetry of the axion field $a$.
Finally, since the $U(1)_{\rm PQ}^{\rm gauge}$ gauge symmetry must be anomaly free, the anomalous QCD coupling 
should involve the combination
\begin{equation}
\mathcal{L}_{\rm QCD} = \frac{g_s^2}{32\pi^2}\left(
\frac{q^{\prime}\tilde{a}}{f_a}  -  \frac{q\tilde{b}}{f_b}
\right)G_{\mu\nu}\tilde{G}^{\mu\nu} = \frac{g_s^2}{32\pi^2}\frac{a}{F_a}G_{\mu\nu}\tilde{G}^{\mu\nu}~.
\end{equation}

The advantage of this construction is the following. The accidental $U(1)$ symmetry remains 
unbroken in the limit in which one only considers 
gauge interactions between the two sectors. More generally, additional interactions between the two sectors have no reason to respect 
the accidental $U(1)$ symmetry but they must be invariant under the gauge symmetry $U(1)_{\rm PQ}^{\rm gauge}$.
This means that operators of the form 
\begin{equation}\label{eq:EffectiveBreaking}
\mathcal{L}_{\cancel{U(1)}} = \frac{\mathcal{O}_1\mathcal{O}_2}{\Lambda^{d_{\mathcal{O}_1} + d_{\mathcal{O}_2} - 4}} + h.c.~,
\end{equation}
may source a breaking of the accidental $U(1)$ symmetry. 
The suppression scale $\Lambda$ may or may not coincide with the Planck scale.
$\mathcal{O}_1$ and $\mathcal{O}_2$ are generic operators with 
mass-dimension $d_{\mathcal{O}_1}$ and $d_{\mathcal{O}_2}$ 
made out of fields in the two sectors, and they have equal and opposite $U(1)_{\rm PQ}^{\rm gauge}$ charge in order to respect 
the gauge symmetry. 
The lowest dimensional $U(1)$ symmetry breaking operator depends on the charge assignment under  $U(1)_{\rm PQ}^{\rm gauge}$, 
and, in the context of explicit models~\cite{Fukuda:2017ylt}, it is possible to show that judicious choices
suppress $U(1)$ breaking effects down to an acceptable level.

Having set the framework, we can now go back to the original motivation of this section, and explore the r\^ole and impact of non-perturbative wormhole solutions.

As noticed before, eq.~(\ref{eq:Accidental}) contains an accidental (non-linearly realized)
 $U(1)$ global symmetry under which the massless Goldstone field $a$ remains shift-invariant.
 The subgroup $a \to a + 2k\pi F_a$, $k\in \mathbb{Z}$ is a gauge symmetry.
 The conditions for the existence of Euclidean wormhole solutions discussed in section~\ref{sec:EuclideanWormholeSolution} 
 are therefore fulfilled, and non-perturbative gravitational effects lead to an explicit breaking of  the axionic shift symmetry
generating a potential as in eq.~(\ref{eq:MainPotential}). The wormhole action in eq.~(\ref{eq:Fullaction}) is controlled by the decay constant 
$F_a$ that sets the periodicity of $a$. 
In conclusion, the outcome of this qualitative 
discussion is that the protection due to the gauge symmetry $U(1)_{\rm PQ}^{\rm gauge}$ may suppress irrelevant operators like those in eq.~(\ref{eq:EffectiveBreaking}) but 
cannot remove non-perturbative breaking terms. 
The idea behind this construction is to use gauge symmetry to protect the axion by means of 
a local discrete symmetry; however,  since wormhole solutions already respect this symmetry, their presence is unaffected.
We expect this argument to be fairly general.

\subsection{Relaxation of the electroweak scale and the clockwork mechanism }\label{sec:Relaxion}

The relaxation of the electroweak scale addresses
 the issue of naturalness from a 
cosmological perspective~\cite{Graham:2015cka}.
The idea that lies at the heart of the model is that the mass parameter in the Higgs potential
$V(|H|) = \mu^2H^{\dag}H + \lambda(H^{\dag}H)^2$
changes during inflation as a function of  the classical value of a slow-rolling axion-like scalar field, the relaxion $\phi$. 
To achieve a large separation of scales between the electroweak scale $v$ and the cutoff  $M$ of the theory
one needs a compact field space of size $2\pi F$, with $F > M/g$. The condition $M \gg v$ implies $g \ll 1$, 
and, for reasonably large values of $M$, one comes up against the problem of having trans-Planckian field
excursions, typically $F > M/g > M_{\rm Pl}$~\cite{Gupta:2015uea}.

An elegant solution to this problem, dubbed clockwork mechanism, was proposed in~\cite{Choi:2015fiu,Kaplan:2015fuy}.
In a nutshell, the clockwork is a theory where a $U(1)^{N+1}$ global symmetry is explicitly broken 
in such a way to preserve a single $U(1)_{\phi}$ symmetry under which the original $U(1)_{i = 0,\dots, N}$ fields $\phi_i$ 
rotate synchronously, $\delta\phi_i = q\delta\phi_{i+1}$, where $q$ is the clockwork factor.
 This construction delivers a single Goldstone boson with large field range $f_{\rm eff} \approx q^N f \gg f$, 
 where $f$ is the symmetry breaking scale of the $U(1)^{N+1}$ symmetry.

As already noticed in~\cite{Gupta:2015uea}, 
the effective operators generated by gravity in eq.~(\ref{eq:MainPotential}) with $F>M_{\rm Pl}$ would spoil the simplest realization 
of the relaxation mechanism
by introducing large corrections to the slow-roll potential of $\phi$.
It is therefore interesting to address the following question:
Is the clockwork mechanism protected against non-perturbative gravitational corrections?
To answer this question we can start from a simple realization of the mechanism with only two axions~\cite{Choi:2015fiu}.
The Lagrangian density of the model is 
\begin{equation}\label{eq:Clockwork1}
\mathcal{L}_{\rm CW}^{(2)} = \frac{1}{2}(\partial_{\mu}\phi_1)(\partial^{\mu}\phi_1)+
 \frac{1}{2}(\partial_{\mu}\phi_2)(\partial^{\mu}\phi_2) + \underbrace{\Lambda^4\cos\left(
 \frac{\phi_1}{f_1} + q\frac{\phi_2}{f_2}
 \right)}_{\equiv V(\phi_1,\phi_2)}~,
\end{equation}
where the two shift symmetries $\phi_{i=1,2}/f \to \phi_{i=1,2}/f_{i=1,2} + c_{i=1,2}$ are explicitly broken by the potential $V(\phi_1,\phi_2)$ down to 
the residual $U(1)_{\phi}$ symmetry under which the two axion fields transform according to 
\begin{equation}\label{eq:ResidualU1}
U(1)_{\phi} =
\left\{
\begin{array}{c}
  \phi_1/f_1 \to \phi_1/f_1  + qc~, \\
 \phi_2/f_2 \to \phi_2/f_2  - c~,
\end{array}
\right.~
\end{equation}
with $c$ arbitrary continuos parameter. 
In full generality, we are considering different (but comparable $f_1\sim f_2$) decay constants $f_{1,2}$.
The potential $V(\phi_1,\phi_2)$ is endowed with the flat direction  $\phi \equiv (qf_1\phi_1 - f_2\phi_2)/f_{\rm eff}$, with 
$f_{\rm eff}\equiv \sqrt{q^2f_1^2 + f_2^2}$, as one can see by diagonalizing the mass matrix  
\begin{equation}\label{eq:DiagonalCW}
M^2 =\Lambda^4
\left(
\begin{array}{cc}
 1/f_1^2  & q/f_1f_2  \\   
 q/f_1f_2 &  q^2/f_2^2  
\end{array}
\right)~~~~\Longrightarrow~~~~
\Lambda^4\left(
\begin{array}{cc}
 0 & 0  \\
 0 &  1/f_1^2 + q^2/f_2^2
\end{array}
\right)~.
\end{equation}
The canonically normalized massive direction is $\phi_H \equiv (f_2\phi_1 +qf_1\phi_2)/f_{\rm eff}$.

The crucial point is to identify the periodicity of the flat direction $\phi$.
The length of a periodic flat direction is determined by the minimal shift $\Delta\phi$
under which the field configuration $\phi$ comes back to its original value.
   \begin{figure}[!htb!]
\minipage{0.5\textwidth}
  \includegraphics[width=.9\linewidth]{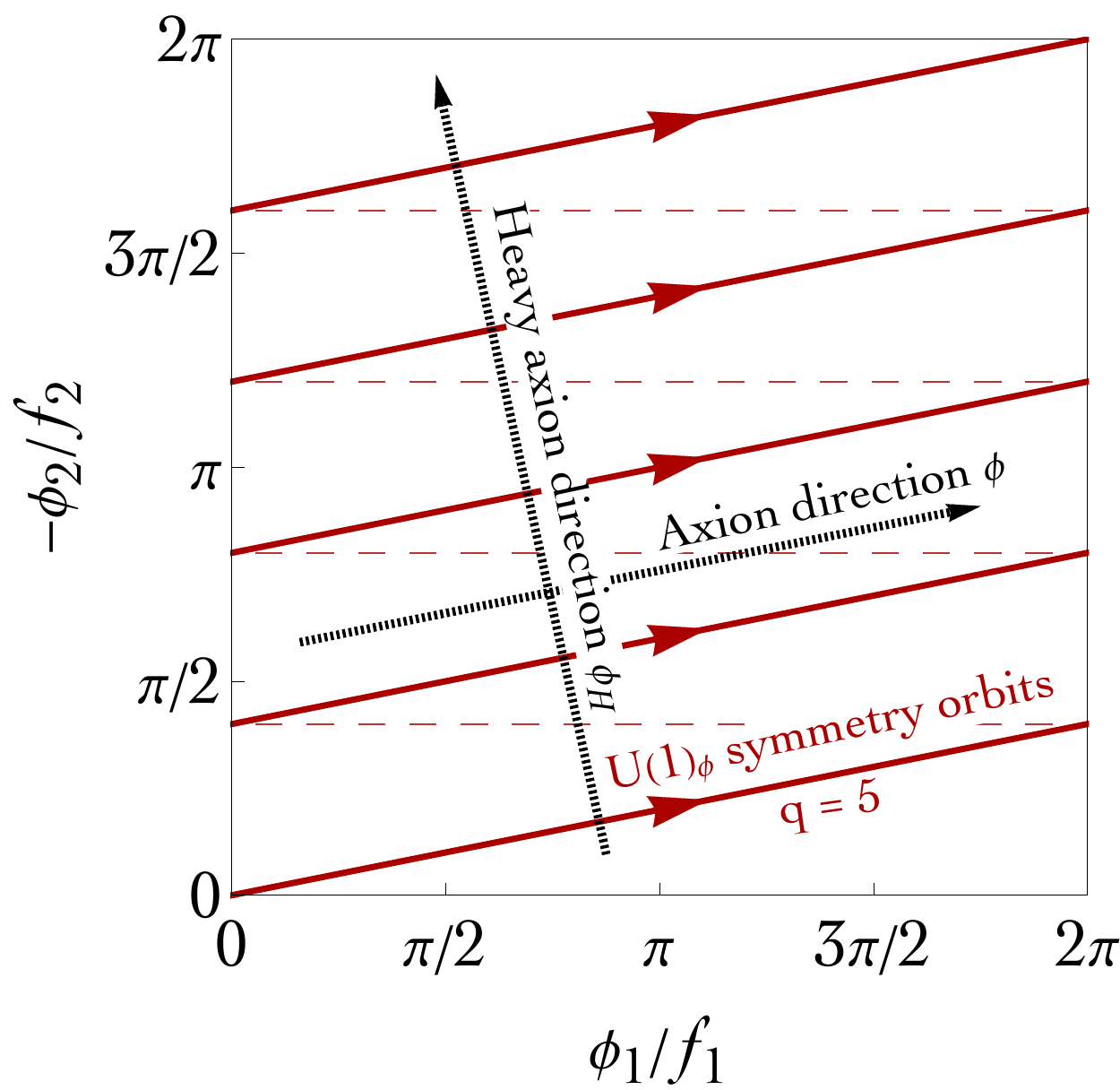}
\endminipage 
\hspace{.3cm}
\minipage{0.5\textwidth}
  \includegraphics[width=.9\linewidth]{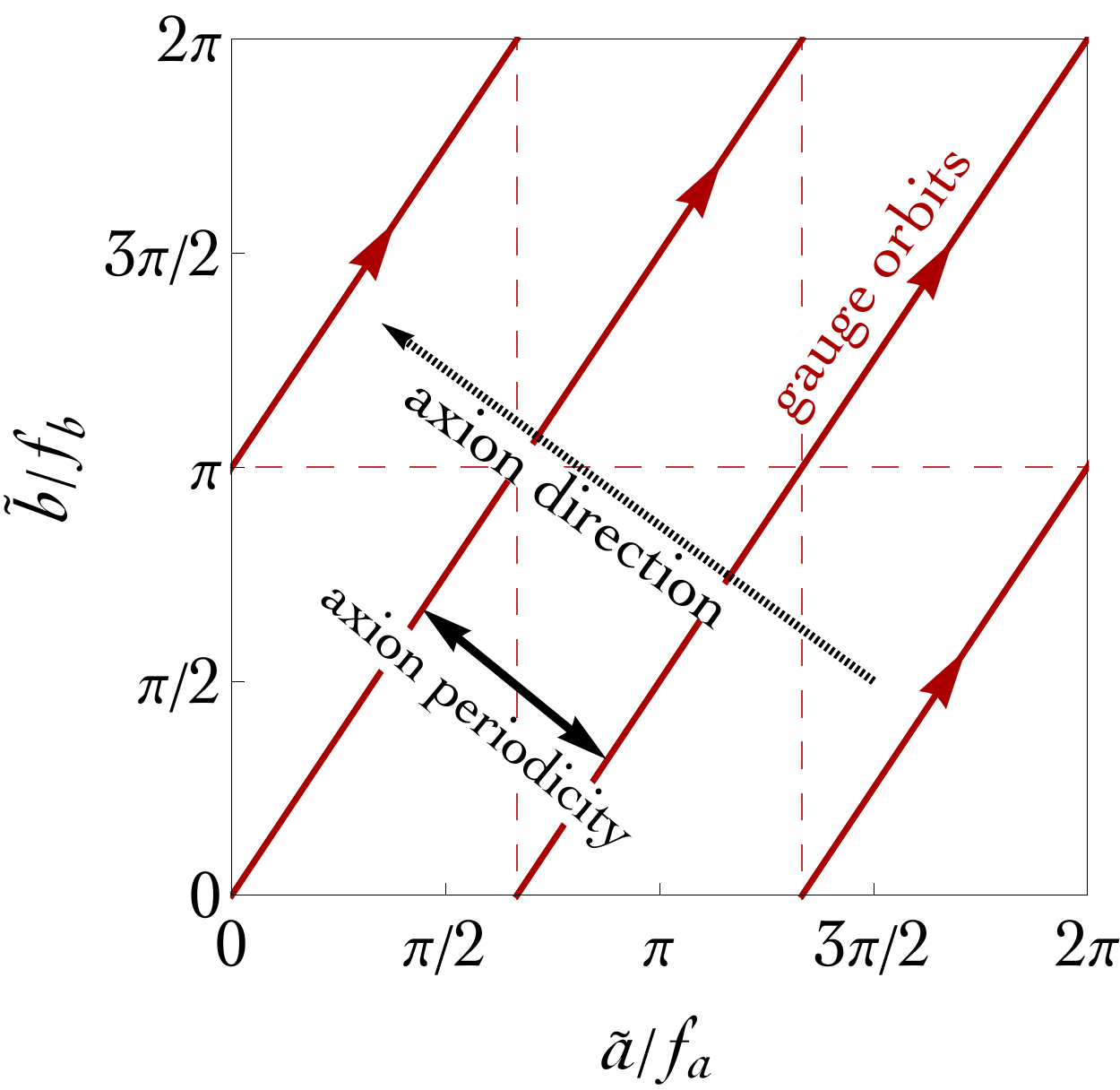}
\endminipage 
\vspace{-.2 cm}
\caption{\label{fig:Accidenti}\em 
Left panel. The red lines follow
the $U(1)_{\phi}$ orbits described by the symmetry transformation in eq.~(\ref{eq:ResidualU1}) with $q= 5$.
For illustrative purposes, we also show the direction in field space corresponding to the axion $\phi$ and the massive mode $\phi_H$.
Right panel. The red lines follow the gauge orbits described by the symmetry transformation in eq.~(\ref{eq:GaugeTransformation}).
The axion direction is orthogonal to the gauge orbits (black arrow, see section~\ref{sec:HDO} for details).
}
\end{figure}
We can visualize the periodicity of the flat direction $\phi$ by means of a simple plot, as shown in the left panel of fig.~\ref{fig:Accidenti}.
Let us start from the point in field space $(\phi_1 = 0, \phi_2 = 0)$, with $\phi = 0$, and follow the $U(1)_{\phi}$ orbits (solid red lines)
obtained from the symmetry transformation in eq.~(\ref{eq:ResidualU1}) by continuously changing the parameter $c$.
For illustration, we take $q = 5$. 
Because of the periodicities $\phi_{i=1,2} \equiv \phi_{i=1,2} + 2\pi f_{i=1,2}$, the $U(1)_{\phi}$ orbits are wrapped 
in the elementary domain $\phi_1/f_1 = [0,2\pi]$, $-\phi_2/f_2 = [0,2\pi]$.
Notice that $\phi_H$ is invariant under the symmetry transformation in eq.~(\ref{eq:ResidualU1}), and it describes the 
direction orthogonal to the $U(1)_{\phi}$ orbits. The axion direction, on the contrary, is aligned with 
the $U(1)_{\phi}$ orbits (black arrows in fig.~\ref{fig:Accidenti}). The axion field comes back to the origin 
after a distance in field space
\begin{equation}
\Delta\phi = 2\pi \sqrt{q^2f_1^2 + f_2^2}\equiv 2\pi f_{\rm eff}~,
\end{equation}
which is enhanced by the winding number $q$ if compared with the original periodicities $2\pi f_1\sim 2\pi f_2$.

In terms of mass eigenstates, the Lagrangian density takes the simple form
\begin{equation}
\mathcal{L}_{\rm CW}^{(2)} = \frac{1}{2}(\partial_{\mu}\phi)(\partial^{\mu}\phi) +
\frac{1}{2}(\partial_{\mu}\phi_H)(\partial^{\mu}\phi_H) -\frac{m_H^2}{2}\phi_H^2 + \dots~,
\end{equation}
with $m_H^2$ defined by the non-zero eigenvalue in eq.~(\ref{eq:DiagonalCW}), and where $\dots$ represents higher order in the potential for $\phi_H$.
  The massless scalar field $\phi$ corresponding to the unbroken $U(1)_{\phi}$ symmetry admits, 
in full analogy with the construction we put forward in section~\ref{sec:EuclideanWormholeSolution}, wormhole solutions
whose quantization condition is set by the periodicity  $2\pi f_{\rm eff}$. 
 We remark that $\phi$ represents the only direction in field space that can accommodate wormhole solutions as presented here. To understand this point  one might consider wormhole solutions
for the fields  $\phi_{1,2}$ with decay constants $f_{1,2}\ll f_{eff}$ defined by the 
Lagrangian in eq.~(\ref{eq:Clockwork1}) and see if the potential $V(\phi_1,\phi_2)$ can be treated as a small perturbation. 
However, it is possible to see\footnote{The wormhole solutions far from the throat ($\tau\to\pm \infty$) tend asymptotically to a constant value of the axion field, as displayed in fig.~\ref{fig:InstantonSolution}, and the axion field varies by $\sqrt{6}\pi/2f^2\kappa$ as it goes through the throat. This implies that (even if the one starts at the bottom of the potential at $\tau=-\infty$) the asymptotic value of the field will not sit at the bottom of the potential. This surplus of potential energy,
 as one integrates large scales away from the throat, will give the wormhole a diverging action.} that gravitational instantons  computed for the free field $\phi_{1,2}$ do not minimize the 
potential $V(\phi_1,\phi_2)$. As a consequence of the latter, instead of being a small correction, the potential gives a large contribution to the action when integrating over large scales away from the throat, and  this 
makes  these solutions, on balance, 
irrelevant.

This conclusion can be generalized to the case of $N+1$ axions.
The Lagrangian density is 
\begin{equation}\label{eq:Clockwork2}
\mathcal{L}_{\rm CW}^{(N+1)} =  \frac{1}{2}\sum_{j=0}^{N}(\partial_{\mu}\phi_j)(\partial^{\mu}\phi_j)
+\underbrace{\sum_{j=0}^{N-1}\Lambda_j^4 \cos\left(
\frac{\phi_i}{f_i} + q\frac{\phi_{i+1}}{f_{i+1}}
\right)}_{\equiv V(\phi_i)}~.
\end{equation}
The periodicity of the flat direction is given by~\cite{Choi:2015fiu,Kaplan:2015fuy}
\begin{equation}
f_{\rm eff} = \sqrt{\sum_{i=0}^{N}\left(\prod_{j=i}^{N-1}q^2\right)f_i^2} 
\overset{f_{i=0,\dots,N} \equiv f}{=}
f\sqrt{\frac{-1 + (q^2)^{N+1}}{-1+ q^2}}
\approx
 q^N f~,
\end{equation}
where in the equality we assumed for simplicity equal decay constant $f_{i=0,\dots,N}\equiv f$.
As before, it is possible to construct wormhole solutions in connection with 
the Goldstone boson $\phi$ associated with the unbroken non-linearly realized  $U(1)_{\phi}$ symmetry. 
The periodicity of $\phi$, $\Delta\phi = 2\pi f_{\rm eff} \approx 2\pi q^N f$,  is completely  fixed by the clockwork construction, and it
 sets the quantization condition discussed in section~\ref{sec:BulkAction}.


 From this discussion, it is clear that the instanton action is controlled by the ratio $M_{\rm Pl}/f_{\rm eff}$, and that 
non-perturbative breaking effects prevent from the possibility to have trans-Planckian field excursions:
 Gravity breaks explicitly the $U(1)$ direction left unbroken by the clockwork construction.

 However, in order to provide a rock-solid conclusion, it is important to keep 
in mind -- in close analogy with what discussed in section~\ref{sec:DynamicalRadialMode} --  the possible r\^ole played by UV dynamics.
For instance, the Lagrangian density in eq.~(\ref{eq:Clockwork2}) could be generated by 
the dynamics describing $N + 1$ complex scalar fields $\Phi_i = [(\rho_i + f_i)/\sqrt{2}]e^{\imath \phi_i/f_i}$ subject to the potential 
\begin{equation}
\mathcal{V}(\Phi_i) = \sum_{i=0}^{N}\left(
-m_i^2|\Phi_i|^2 + \frac{\lambda_i}{4}|\Phi_i|^4
\right) + \sum_{i=0}^{N-1}\left(
\epsilon_i \Phi_i^{\dag}\Phi_{i+1}^3  + h.c.
\right)~,
\end{equation}
with $m_i \sim f_i$, $\lambda_i \sim 1$, after integrating out the radial modes. 
In this setup, one gets $\Lambda_i \equiv (\epsilon_i f_i f_{i+1}^3/2)^{1/4}$.
The situation in similar in spirit to the setup explored in section~\ref{sec:DynamicalRadialMode} with the additional complication
given by the presence of more fields and explicit breaking terms controlled by the order parameters $\epsilon_i$.
 The results of section~\ref{sec:DynamicalRadialMode} suggest that the presence of extra fields does modify the 
geometry of the wormhole solution. However, 
this could result  -- as explicitly proved in section~\ref{sec:DynamicalRadialMode} 
considering the presence of a dynamical  radial mode -- in a decrease of the wormhole action, with a consequent tightening of the
 amount of symmetry breaking.

Within the present study, no firm conclusion can be established.  Nevertheless, we argue that the possible presence of Euclidean wormhole solutions represents the 
 most concrete threat against the possibility to engineer trans-Planckian field excursions via the clockwork mechanism.
We think this is an interesting open question, and we leave it for future investigation.

\section{Wormhole solutions for a generic Goldstone coset}\label{sec:GenericCoset}

Let us start with a lightning review of the treatment of Goldstone bosons from the coset  $\mathfrak G/\mathfrak H$  in 
the formalism of CCWZ~\cite{Coleman:1969sm,Callan:1969sn}. The broken part of the 
Lie algebra will be denoted by the generators $X_A$ whereas the un-broken symmetry generators are $T_a$ with the following structure
\begin{equation}
\left[ T_a\,,T_b\right]=  \imath f_{ab}^{\,\,\,\,\,c}\,T_c~,~~~~~~~~~~~~~~~ 
\left[ T_a\,,X_B\right]= \imath f_{aB}^{\,\,\,\,\,C}\,X_C~, \label{LieComm}
\end{equation}
where the last equation follows for compact groups and we take the generators to be hermitian. The Goldstones
parametrize the broken part of the group; one can write $\eta=\exp(iX\theta)$ with $\eta\in\mathfrak G/\mathfrak H$ in terms
of $n={\rm dim} (\mathfrak G)-{\rm dim}(\mathfrak H)$ $\theta$ coordinates although different parametrizations are useful and often used. 
The transformation properties of $\eta$ are
\begin{equation}
\eta\to \mathbf G\, \eta\, \mathfrak h^{-1}(\eta)~,\label{GBtrnsf}
\end{equation}
with $\mathbf G\in \mathfrak G$ and $\mathfrak h\in\mathfrak H$. This expression  
follows from the result that any element of the group can be factorized into a broken times an unbroken element 
and the composition law $\mathbf G \eta= \eta'\mathfrak h(\eta)$.
The basic building block is the Lie-dragged derivative which is a one-form $\omega$ living in the broken Lie algebra, or tangent 
space of the broken group manifold. Explicitly
\begin{equation}
\eta^{-1} d \eta= \imath\,\omega^A X_A+\imath\,V^aT_a~,~~~~~~~~~~~ d=dx^\mu\partial_\mu~,
\end{equation}
where $\omega^A=\omega^A_\mu dx^\mu$. 
In terms of the $\theta$ coordinates one has the vierbein and metric  (we use the convention of summed repeated indices unless otherwise stated)
\begin{equation}
\omega^A_\mu=e_{i}^A(\theta) \partial_\mu\theta^i~,~~~~~~~~~~~~
G_{ij}(\theta)=e_{i}^Ae_{j}^B\delta_{AB}~.
\end{equation}
In order to extract the symmetry transformations in terms of the $\theta$ coordinates we make eq.~(\ref{GBtrnsf}) infinitesimal
\begin{equation}
\delta_\epsilon\eta=(i\epsilon^a T_a+i\epsilon^A X_A)\eta-i\eta(\epsilon \cdot F(\eta))^aT_a~,~~~~~~~~~~~~
\eta^{-1}\delta_\epsilon\eta =\eta^{-1}(i\epsilon T+i\epsilon X)\eta-i(\epsilon\cdot F(\eta))^aT_a~,
\end{equation}
and we project out the non-linear piece $F(\eta)$ by restricting to broken generators only. We find
\begin{equation}
\eta^{-1} X_A \eta\big|_X = X_B e_i^B\xi_A^i 
,~~~~~~~~~~~~~~~~~~~~~~~~~~
\eta^{-1} T_a \eta\big|_X = X_B e_i^B\xi_a^i 
,
\end{equation}
where by $\big|_X$ we denote projection onto the broken generators. The above result yields the Killing vectors in $\theta$-space ($\delta_\epsilon\theta^i=\epsilon_A\xi_A^i$); there are
$n$ Killing vectors $\xi_A$ associated to the coset and dim$(\mathfrak H)$ Killing vectors $\xi_a$ for the unbroken symmetries. 
The transformation of the one-form $\omega^A$ under the unbroken part of the group is specially simple since
the symmetry is linearly realized
\begin{equation}
\delta_{\epsilon}\omega^A\to \epsilon^a f_{aB}^{\,\,\,\,\,A}\omega^B~,
\end{equation}
where we have used the second relation in eq.~(\ref{LieComm}). In  coordinate space the Killing vectors by construction satisfy
\begin{equation}\label{KillMetDef} 
\frac{\partial \xi^k}{\partial \theta^i}G_{kj}+G_{ik}\frac{\partial \xi^k}{\partial \theta^j}+\left(\xi^k\frac{\partial}{\partial \theta^k}\right)G_{ij} = 0~,
\end{equation}
The Noether currents for the broken generators read
\begin{equation}\label{eq:NoetherGeneralCoset}
J^\mu_A=\sqrt{g}\,f^2\xi_A ^i e_i^B e^B_j \partial^\mu\theta^j~,
\end{equation}
where we introduced the Goldstones decay constant $f$ and equivalent expressions apply for the unbroken currents.
We note that, if a coordinate  system can be found for which $\xi^i_Ae^B_i=\delta^A_{\,\,B}$, the Noether currents read $\sqrt{g}\omega^\mu_A$ and their conservation can be written as the equation of motion 
for a one-form although the Bianchi identity is not satisfied in general.

We now turn to the action. By using  the variational principle, one should be able to find non-trivial solutions of the equation of motion, 
and subsequently obtain the corresponding  non-perturbative
action by substituting back the solution in the original action -- in parallel with what was done in section~\ref{sec:BulkAction}.
It is often remarked how the axion gets an extra sign in its kinetic term when turning the action from Minkowski to Euclidean; here we observe that regardless of the sign, starting from the variational principle in Euclidean space yields a vanishing wormhole action, as we shall show below. 
This is remedied either using a three-form description -- as done in section~\ref{sec:BulkAction} -- or introducing Lagrange multipliers~\cite{Lee:1988ge,Grinstein:1988ja}. 
Since a general coset does not seemingly accept a three-form description we opt for the second option and write an action
\begin{align}
\mathcal{S}_{\rm E} =\int d^4x \left[\sqrt{g}\left(-\frac{\mathcal{R}}{2\kappa}+\frac{f^2}{2}\partial_\mu\theta \,G \, \partial^\mu \theta\right)+\lambda^A(x)\partial_\mu (\sqrt{g}\,f^2 \xi_A \, G \, \partial^\mu \theta)\right]
\end{align}
where $\kappa=8\pi G_N$, we have suppressed scalar indices which otherwise are summed univocally to build invariants
and the Lagrange multiplier's ($\lambda^A(x)$) coefficient is the divergence of the Noether currents ($J^\mu=\sqrt{g}f^2 \xi G\partial^\mu\theta $) which vanishes on-shell. Here we have used $\theta$ coordinates yet the following derivation does not require 
explicit expressions for the metric $G$ or the Killing vectors; in terms of the one form $\omega$ the
kinetic term of the Goldstones reads simply $\omega_A^\mu \omega_\mu^A/2$.

The variational principle yields the equations of motion $\left.\delta \mathcal{S}_{\rm E}\right|_{\delta g_{\mu\nu}} = 0$, 
$\left.\delta \mathcal{S}_{\rm E}\right|_{\delta\theta} = 0$, $\left.\delta \mathcal{S}_{\rm E}\right|_{\delta\lambda^A} = 0$ with 
\begin{eqnarray} \label{EinstEoM}
&&\left.\delta \mathcal{S}_{\rm E}\right|_{\delta g_{\mu\nu}} = \\ && \int d^4x\frac{\sqrt{g}}{2\kappa}\left[
\mathcal{R}^{\mu\nu}-\frac{g^{\mu\nu}}{2}\mathcal{R}+\kappa f^2\left(\frac{g^{\mu\nu}}{2}\partial_\rho\theta G\partial^\rho\theta-\partial^\mu\theta G\partial^\nu \theta\right)-\kappa f^2\left(g^{\mu\nu}\partial_\rho\lambda \xi G\partial^{\rho}\theta-2\partial^{\mu}\lambda \xi G \partial^\nu\theta\right)\right]\delta g_{\mu\nu}~,\nonumber \\ \label{GBEoM}
&&\left.\delta \mathcal{S}_{\rm E}\right|_{\delta\theta} = \\&& \int d^4x \,f^2\left[
-\delta\theta  \partial_\mu  \left(\sqrt{g}\, G \partial^\mu\theta\right)+\frac{\sqrt{g}}{2}\partial_\mu \theta \left(\delta \theta\frac{\partial}{\partial \theta}\right)G \partial^\mu\theta +\delta\theta \partial_\mu \left(\sqrt{g} G\xi\partial^\mu\lambda\right) - \partial_\mu \lambda \left(\delta \theta\frac{\partial}{\partial \theta}\right)\xi  G\partial^\mu\theta \right]~,\nonumber \\ \label{LagMulEoM}
&&\left.\delta \mathcal{S}_{\rm E}\right|_{\delta\lambda^A}  = \int d^4x\,f^2\delta\lambda^A\partial_\mu\left( \sqrt{g}\, \xi_A G \partial^\mu \theta\right)~,
\end{eqnarray}
Taking the trace in Einstein equations, if we omit $\lambda$ terms, yields $-\mathcal{R}+\kappa f^2 \partial_\mu \theta G\partial^\mu\theta=0 $ and a vanishing action (regardless of the sign that one can change with $\kappa f^2$). 
The r\^ole of the extra $\lambda$ term, as we shall see in the following,  is exactly to turn this result into a properly defined variational principle for wormhole solutions with non-vanishing action.
In this regard, our result is the generalization of the construction put forward in~\cite{Lee:1988ge} to the case of a generic coset.

Most of the computation now involves manipulations of the equations of motion in order to put them in a clean and readable form, and we  proceed as follows. 
In the equation for the Goldstones we perform the substitution $\delta\theta^i=\partial_\nu\lambda^B\xi^i _B$, and, after repeated use of the definition of Killing vectors with respect to the metric in eq.~(\ref{KillMetDef}), we find
\begin{equation}\label{GBEoMd1}
\partial_\nu\lambda^B\partial_\mu\left[\sqrt{g}\xi_BG\left( \xi\partial^\mu\lambda-\partial^\mu\theta \right)\right]
+\partial_\mu \lambda^A \partial_\nu\lambda^B \left(\xi_A\frac{\partial \xi_B}{\partial\theta}-\xi_B\frac{\partial \xi_A}{\partial\theta}\right)G\partial^\mu\theta
=0~,
\end{equation}
where the last term  is antisymmetric in the summed indices $A,B$ and proportional to the structure constants. This means that, for spherically symmetric solutions which depend only on the radius $r$ as the ones we are interested in, such a term
cancels, since $\partial_r\lambda^A\partial_r\lambda^B$ is symmetric in $A,B$.
From now on, we focus on this type of solutions, $\theta(r)^i$, $\lambda^A(r)$, and we adopt 
the spherically symmetric metric in eq.~(\ref{eq:Ansatz}).
As remarked above eq.~(\ref{GBEoMd1}) reduces in this case to
\begin{equation}
\partial_r\lambda_B\partial_r\left[\sqrt{g}\xi_BG\left( \xi_A\partial^r\lambda_A-\partial^r\theta \right)\right]=0~,
\end{equation}
solved by $\xi_A\partial_\mu \lambda^A=\partial_r \theta$. 
On passing, we note that this substitution in Einstein equations yields $\mathcal{R}+\kappa f^2\partial_r\theta G\partial^r\theta=0$, and a non-zero on-shell action. 
This relation for $\lambda$ back into the Goldstone equation of motion eq.~(\ref{GBEoM}) yields, substituting $\delta\theta=\partial_r \theta$
\begin{equation}\label{eq:Int1}
\frac12 \sqrt g\partial_r\theta\left(\partial_r G\right) \partial^r \theta-\sqrt g\partial_r\theta(\partial_r G\xi)\partial^r\lambda=0~.
 \end{equation}
We now use the eq.~(\ref{LagMulEoM}) for the variation of the Lagrange multiplier to substitute $\partial_r\xi$. Eq.~(\ref{eq:Int1}) above becomes 
\begin{equation}
-\frac12 \sqrt g \partial_r\theta
\left(\partial_r G\right) \partial^r \theta+\partial_r\theta (\partial_r \sqrt{g} G\partial^r\theta)=0~,
\end{equation}
and, after introducing a total derivative
\begin{equation}
\frac12\partial_r(\partial_r\theta \sqrt g G \partial^r\theta)+\frac12\partial_r\theta \sqrt g G \partial^r\theta\frac{g_{rr}}{\sqrt g}\partial_r \left(\frac{\sqrt g}{g_{rr}}\right)=0~,
\end{equation}
which can be integrated to find
\begin{align}\label{eq:GoldstoneKin}
\sqrt g \partial_r\theta  G \partial^r\theta=C\frac{g_{rr}}{\sqrt g}=C\frac{\alpha}{\beta^3}~,
\end{align}
where for consistency the constant $C\geq0$. Finally, we substitute the Goldstone boson kinetic term in eq.~(\ref{eq:GoldstoneKin}) back 
in the Einsteins equations
\begin{eqnarray}
-\frac{3\mathcal K \alpha^2}{\beta^2}+\frac{3(\partial_r\beta)^2}{\beta^2}+\frac{\kappa f^2}{2}C\frac{\alpha^2}{\beta^6} &=& 0~, \\
\hat g_{ij}\left[-\mathcal K -\frac{2\beta\partial_r\beta\partial_r\alpha}{\alpha^3}+\left(\frac{\partial_r\beta}{\alpha}\right)^2+2\frac{\beta\partial_r^2\beta}{\alpha^2}\right]-\kappa f^2\hat g_{ij}\frac{\beta^2}{2\alpha^2}C\frac{\alpha^2}{\beta^6} &=& 0~,
\end{eqnarray}
which, remarkably, have the same structure and solution for the metric as found in  the axion case, see eqs.~(\ref{eq:Einstein1}-\ref{eq:Einstein2}). The positive-definite constant $C$ can be given in terms of the $n$ Noether currents which are constants of motion $\sqrt{g}\xi_A G\partial^r\theta=C_A$; this in certain cases takes a simple form as the sum of the squares but  a general expression for all cases is not known to us.

Let us now move to discuss the most important aspect of this computation, that is the potential that these 
wormhole solutions  generate as a consequence of the explicit symmetry breaking that they induce.
 As emphasized in the discussion we put forward in the first part of this letter, 
 the crucial point in this respect is to understand the correct quantization condition. 
 In the simple case of the $U(1)$ symmetry, the quantization condition is related to the  periodicity of the axion field. 
In the generalization to the coset space $\mathfrak G/\mathfrak H$,  the fact that the Goldstone manifold is compact discretizes the possible charges and, as a consequence,  
the wormhole action. 
However, the discussion of the geometry a generic coset  can be forbiddingly complicated.
To shed light on the issue, we shall consider in the rest of this section an explicit example.
In order to make an educated choice, as ever, symmetry is helpful,  in particular the unbroken symmetry. 
 In this respect the maximal unbroken symmetry allowed is $O(n)$ for $n$ Goldstones, which is the largest linearly realised symmetry that the associated
 tangent space, $\omega_\mu^A$, admits. 
A breaking pattern that yields this unbroken symmetry is $O(n+1)/O(n)$, which we will adopt as an explicit case to exemplify our discussion. One has $n$ broken generators
$X_A$ and $n(n-1)/2$ unbroken ones $T_a$ that read -- we will use antisymmetric generators and therefore there is no need for factors of $\imath$ in the following:
\begin{equation}
(X_A)_{\alpha\beta}=\delta_{\beta}^{n+1}\delta_{\alpha}^{A}-\delta_{\alpha}^{n+1}\delta_{\beta}^{A}~,~~~~~~~~~~~~~~
(T_a)_{\alpha\, n+1}=(T_{a})_{n+1\,\beta}=0~,~{\rm with}~~T_a^T=-T_a~.
\end{equation}
The Goldstone fields can be taken as $\theta^i$ with $i=1,..,n$ but,  motivated by the symmetries of the unbroken group,  we shall use in the
following  spherical coordinates $\theta=\rho \, u(\varphi)$, with $u\cdot u=1$. We therefore have one radius $\rho$ and $n-1$ angles
$\varphi^i$. The Goldstone matrix reads
\begin{equation}
\eta=e^{X\cdot \theta}=e^{\rho X\cdot u}=\left(\begin{array}{cc}
1-(1-\cos\rho) u u^T& \sin\rho \, u\\
-\sin\rho \, u^T&\cos\rho
\end{array}
\right)~,
\end{equation}
where we note that $\rho+2\pi$ gives the same group element as $\rho$. This yields the one-form
\begin{equation}
\eta^{-1}d\eta\big|_X=\omega^AX_A=u^AX_Ad\rho+\sin \rho  \frac{\partial u^A}{\partial \varphi^i} X_A d\varphi^i~,
\end{equation}
which, in turn, gives a kinetic term for the Goldstones
\begin{equation}
\frac{ f^2}2\omega_\mu^A\omega_A^\mu=\frac{f^2}2\left(\partial_\mu\rho\partial^\mu \rho+\sin^2\rho\, \partial_\mu\varphi^i \frac{\partial u^A}{\partial\varphi^i}\delta_{AB}\frac{\partial u^B}{\partial\varphi^j}\partial^\mu\varphi^j\right)~,
\end{equation}
where $(\partial u/\partial \varphi)^2$ is the metric in a $S_{n-1}$ sphere. Using eq.~(\ref{eq:NoetherGeneralCoset}), 
the Noether currents can be found to be
\begin{eqnarray}
J^\mu_A &=& f^2\sqrt g (\xi_A\cdot e)^B \omega_B^\mu=f^2\sqrt g\left( u_A \partial^\mu\rho+\sin\rho\cos\rho\frac{\partial u_A}{\partial \varphi^i}\partial^\mu\varphi^i\right)~,\\
J^\mu_a &=& f^2\sqrt g(\xi_A\cdot e)^B \omega_B^\mu= f^2\sqrt g\sin^2\rho\, u^T T_a\frac{\partial u}{\partial \varphi} \partial^\mu\varphi~,
\end{eqnarray}
and in this case we can find the explicit connection between kinetic term and Noether currents by taking the sum over the unbroken currents
\begin{equation}
\sqrt{g}f^2\omega_\mu^A \omega^\mu_A=\frac{1}{\sqrt{g}f^2}\left( J_{\mu}^A J^\mu_A+J_\mu^a J^\mu_a\right)~.
\end{equation}
We now proceed in parallel to section~\ref{sec:BulkAction}.
We denote an instanton solution with $\bar\rho$\,, $\bar\varphi$, and 
we consider a variation of the fields $\delta \rho$, $\delta \varphi^i$ around it.   
The first variation of the action reads, after using the equations of motion for the background: 
\begin{equation}
\delta \mathcal{S}= \int dr d\Omega_{3,1} \frac{d}{dr}\left[\sqrt gf^2 \left( \delta\rho \partial^r \bar\rho+\delta\varphi^i\sin^2\bar\rho \frac{\partial u}{\partial\varphi^i}\frac{\partial u}{\partial\varphi^j} \partial^r\bar\varphi^j\right)\right]~.
\end{equation}
If we take the variations to be independent from the solid angle $\Omega_3$, we can rewrite $\delta\mathcal S$ in terms of charges
$Q_{A(a)}=\int d\Omega_{3,1} J_{A(a)}^r$. We find
\begin{equation}
\delta\mathcal  S= \left[ \delta\rho \left(\bar u^A Q_A\right)+\delta\varphi^i \left(\bar u\cdot T_a\cdot \frac{\partial \bar u}{\partial \varphi^i}Q_a\right)\right]~,
\end{equation}
where there is a sum over broken $Q_A$ and unbroken $Q_a$ charges. The explicit form of this expression is not relevant, what is important to note is that if gravity is to respect the unbroken symmetry, which certainly would be the case if it were gauged, the solutions would have $Q_a=0$ and therefore $\varphi$ coordinates would {\it not} get a potential. 
On the other hand the radius $\rho$ gets a contribution proportional to broken charges $Q_A$. 
Furthermore,
the periodicity of this field, in full analogy with the axion case, discretizes the possible charges to be $\bar u^A Q_A=k$,
with $k\in \mathbb Z$. 
We can expect therefore a potential generated for the radial coordinate as follows
\begin{equation}\label{eq:PotentialGoldstoneGeneral}
V(\rho) = -K e^{-\mathcal S_{\rm inst}} \cos\rho~,
\end{equation}
which nevertheless produces a mass term for {\it all} Goldstone bosons. 
 To visualize this fact, notice that the 
 minimum of eq.~(\ref{eq:PotentialGoldstoneGeneral}) sits at $\rho=0$ were the spherical coordinate system is singular. It is therefore better to 
 go back to $\theta$ coordinates. We find
\begin{equation}
-V(\theta^i) = K e^{-\mathcal S_{\rm inst}} \cos\rho\simeq\frac{K }{2}e^{-\mathcal S_{\rm inst}} \rho^2= \frac{K }{2}e^{-\mathcal S_{\rm inst}}\sum_i(\theta^i)^2
\end{equation}
This means gravity produces a mass for all Goldstone bosons. 
Although obtained in the specific case of the $O(n+1)/O(n)$ symmetry breaking pattern, it is tempting to generalize this result to a general coset.
We postpone the verification of this conjecture to future investigation.

\section{Conclusions and outlook}\label{sec:Conclusions}

The explicit breaking of global symmetries induced by gravitational effects is a 
mesmerising phenomenon. Furthermore, far from academic, this issue could have direct implication in
phenomenology.

In this paper we considered 
the specific case of the global shift symmetry of a Goldstone boson, and 
we explored 
the non-perturbative breaking due to Euclidean gravitational instantons in the context 
of Einstein gravity (aka wormholes).

 This field has been studied in a decades-long endeavour, and in this regard the main novelties of our analysis are the following.

\begin{itemize}

\item[$\circ$] {\underline{Theoretical analysis}}. On the theory side, 
we addressed the problem of stability of wormhole solutions by 
computing the spectrum of the 
quadratic action, and we found a positive spectrum. 
This result is of fundamental importance since only in the absence of negative eigenvalues 
the gravitational instantons mediate tunneling transitions between degenerate vacua, 
in parallel with other known situations both in Quantum Mechanics and Quantum Field Theory.
In turn, this property allows to consistently compute the effective potential generated by gravitational instantons.
The latter, as a consequence  of its non-perturbative nature, is characterized by the suppression factor 
$e^{-\mathcal{S}_{\rm inst}}$, where $\mathcal{S}_{\rm inst}$ is the wormhole action 
 which scales with the Goldstone decay constant and  Planck mass as $\mathcal{S}_{\rm inst}\propto M_{\rm Pl}/f_a$.

\item[$\circ$] {\underline{Phenomenological analysis}}.
On the phenomenological side, we focused our analysis on the compelling case of the QCD axion.
By computing non-perturbative gravitational corrections to the QCD axion potential, 
we found the following lower bound on the mass of the QCD axion
\begin{equation}
 m_a \gtrsim 4.8\times 10^{-10}\,{\rm eV}~,~~~~ f_a \lesssim 10^{16}\,{\rm GeV}~.
\end{equation}
As an application, we discussed important consequences 
related to black hole superradiance, and we  showed that the mass range 
$m_a =[10^{-14}, 10^{-10}]$ eV
motivated by experimental searches and phenomenological considerations
is theoretically disfavored.

In addition to the QCD axion, 
we discussed the case of ultralight scalars as cosmological dark matter.
We showed that non-perturbative gravitational effects due to Einstein gravity 
generate a mass  term 
in agreement with the estimate of the relic abundance based on the misalignment mechanism.
In numbers, we found a dark matter candidate with mass 
$m_a \simeq 2.5\times 10^{-18}$ eV corresponding to the decay constant 
$f_a \simeq 8\times 10^{15}$ GeV.

Finally, we discussed to r\^ole of non-perturbative gravitational corrections 
in connection with the relaxation of the electroweak scale and the clockwork mechanism.

\end{itemize}

We concluded with a derivation of the
 of wormhole solution for a generic Goldstone coset, and we inspected the breaking pattern $O(n+1)/O(n)$ to
 find that all Goldstone bosons in the coset acquire a mass via gravitational effects.

\section*{Acknowledgments}

We thank Diego Blas, Georgi Dvali, Benjamin Grinstein, Matthew McCullough, Riccardo Rattazzi, Luca Vecchi and Giovanni Villadoro for discussions.
We also benefited from important feedback by Michele Frigerio and David Marsh.
Finally, we thank Christopher Nolan for inspiration~\cite{Nolan}.

\appendix

\section{Dual Formulation of broken symmetry phases}\label{app:A}

\subsection{Axions and forms}

Let us start considering the theory of a massless Goldstone field $\theta$ described by the Lagrangian density 
\begin{equation}\label{eq:AxionBasic}
\mathcal{L}_{\theta} = \frac{f_a^2}{2}(\partial_{\mu}\theta)(\partial^{\mu}\theta)~.
\end{equation}
For simplicity, we restrict here to flat space.
It is straightforward to check that $\mathcal{L}_{\theta}$ can be obtained from
\begin{equation}
\mathcal{L}_{J} = i(\partial_{\mu}\theta)J^{\mu} + \frac{1}{2f_a^2}J_{\mu}J^{\mu}~,
\end{equation}
after eliminating the non-dynamical current $J^{\mu}$ by means of its algebraic equation of motion $J^{\mu} = -if_a^2(\partial^{\mu}\theta)$. 
Equivalently, after integration by parts, we have 
\begin{equation}\label{eq:DualPicture}
\mathcal{L}_{J} = - i(\partial_{\mu}J^{\mu})\theta + \frac{1}{2f_a^2}J_{\mu}J^{\mu}~,
\end{equation}
where now the Goldstone boson field is non-dynamical, implying the constraint $\partial_{\mu}J^{\mu} = 0$.
The Lagrangian density in eq.~(\ref{eq:AxionBasic}) is therefore nothing but $\mathcal{L}_{J} = J_{\mu}J^{\mu}/2f_a^2$, subject to the current conservation constraint 
$\partial_{\mu}J^{\mu} = 0$. 
The latter can be formally solved by introducing the antisymmetric tensor field $B$, with $J^{\mu} \equiv -\frac{i}{2}\epsilon^{\mu\nu\rho\sigma}(\partial_{\nu}B_{\rho\sigma})$.
From these simple manipulations follows that the Goldstone theory admits a description in terms of an  antisymmetric tensor field. 
Using the condition $J^{\mu} = -if_a^2(\partial^{\mu}\theta)$, we have
\begin{equation}\label{eq:Axion3Form}
f_a^2(\partial^{\mu}\theta) = \frac{1}{2}\epsilon^{\mu\nu\rho\sigma}(\partial_{\nu}B_{\rho\sigma})~~~~~\Longrightarrow~~~~~
f_a^2\epsilon_{\mu\alpha\beta\lambda}(\partial^{\mu}\theta) = \frac{1}{2}\epsilon_{\mu\alpha\beta\lambda}\epsilon^{\mu\nu\rho\sigma}(\partial_{\nu}B_{\rho\sigma}) \equiv H_{\alpha\beta\lambda}~,
\end{equation}
where $H$ is the dual strength of the antisymmetric tensor field $B$, 
$H_{\mu\nu\rho} = \partial_{\mu}B_{\nu\rho} + \partial_{\nu}B_{\rho\mu} + \partial_{\rho}B_{\mu\nu}$.
Notice that this definition implies the Bianchi identity $\epsilon^{\mu\nu\rho\sigma}(\partial_{\rho}H_{\sigma\mu\nu})=0$.

Eq.~(\ref{eq:Axion3Form}), $f_a^2\epsilon_{\mu\alpha\beta\lambda}(\partial^{\mu}\theta) = H_{\alpha\beta\lambda}$, 
incarnates the dual description of the field $\theta$ in terms of a three-form. 

Notice that at this level of the analysis we  did not specify the parity transformation of the Goldstone field. 
On the contrary, the construction in eqs.~(\ref{eq:AxionBasic}-\ref{eq:Axion3Form}) remains valid both for scalar and pseudo-scalar bosons.
From the dual relation $2f_a^2(\partial^{\mu}\theta) = \epsilon^{\mu\nu\rho\sigma}(\partial_{\nu}B_{\rho\sigma})$,
it  follows that if $\theta$ is a pseudo-scalar (scalar) field, then the two-form $B_{\rho\sigma}$ is forced to transform as a tensor (pseudo-tensor).
Without additional  input one can not tell one option from the other, both perfectly consistent (for instance, in string-inspired situation the two-form $B_{\rho\sigma}$ can be identified with the Kalb-Ramond field that, after compactification in 4-dimensions,
 is an even parity tensor; having this specific dual 
 picture in mind, the field $\theta$ transforms  as a pseudo-scalar). 
 

 In the dual picture, the free Lagrangian density in eq.~(\ref{eq:DualPicture}) takes the form 
\begin{equation}
\mathcal{L}_{J} = \frac{1}{4f_a^2}\left(
\partial_{\nu}B_{\rho\sigma}
\right)\left(\partial^{\nu}B^{\rho\sigma}\right) - \frac{1}{2f_a^2}\left(
\partial^{\sigma}B_{\sigma\rho}
\right)\left(\partial_{\nu}B^{\nu\rho}\right)~.
\end{equation}
The antisymmetric tensor $B_{\rho\sigma}$ has $6$ independent components, 
but the gauge symmetry 
\begin{equation}
\delta B_{\rho\sigma} = \partial_{[\rho}\Lambda_{\sigma]} = \frac{1}{2}\left(\partial_{\rho}\Lambda_{\sigma}-
\partial_{\sigma}\Lambda_{\rho} \right)~,
\end{equation}
with the extra redundancy $\Lambda^{\prime}_{\sigma} = \Lambda_{\sigma} +\partial_{\sigma}\lambda$ correctly 
reduces to one
the number of dynamical degrees of freedom.

\subsection{From Minkowski to Euclidean space}\label{app:Minkowski2Euclidean}

Let us consider the action in curved space-time, with flat metric $\eta_{\mu\nu} = {\rm diag}(+1,-1,-1,-1)$
\begin{equation}\label{eq:AppA1}
\mathcal{S} = \int d^4x\sqrt{-g}\left[
\frac{M_{\rm Pl}^2}{16\pi}\,\mathcal{R}  +  \frac{f_a^2}{2}(\partial_{\rho}\theta)(\partial^{\rho}\theta)
\right]~,
\end{equation}
describing a massless Goldstone boson field $\theta$ minimally coupled to Einstein gravity.
As discussed in section~\ref{sec:BulkAction}, the field $\theta$ admits a dual description 
in terms of an antisymmetric two-form with field strength $H_{\mu\nu\rho}$ given by 
$H_{\mu\nu\rho} = f_a^2\epsilon_{\mu\nu\rho\sigma}\left(
\partial^{\sigma}\theta
\right)$. In Lorentzian space-time, we have the following identity  
\begin{equation}\label{eq:DualMinkowski}
\frac{f_a^2}{2}(\partial_{\rho}\theta)(\partial^{\rho}\theta)  = -\frac{\mathcal{F}}{2}H_{\mu\nu\rho}H^{\mu\nu\rho}~,
\end{equation} 
with $\mathcal{F}$ defined right below eq.~(\ref{eq:WormholeAction}). The minus sign follows from the Levi-Civita contraction 
$\epsilon_{\mu\nu\rho\sigma}\epsilon^{\mu\nu\rho\lambda} = (-1)^t\,3!\,\delta_{\sigma}^{\lambda}$, with $t=1$ in Minkowski space-time.
The action in eq.~(\ref{eq:AppA1}) takes the form 
\begin{equation}\label{eq:AppA2}
\mathcal{S} = \int d^4x\sqrt{-g}\left[
\frac{M_{\rm Pl}^2}{16\pi}\,\mathcal{R}  -\frac{\mathcal{F}}{2}H_{\mu\nu\rho}H^{\mu\nu\rho}
\right]~.
\end{equation}
We are now ready to go from Minkowski -- with coordinates $x\equiv (t,\vec{x})$, and metric $g_{\mu\nu}(x)$ -- to Euclidean space-time -- with 
coordinates $\tilde{x}\equiv (t_{\rm E},\vec{x})$, and metric $\tilde{g}_{\mu\nu}(\tilde{x})$ --  by means of the Wick rotation $t_{\rm E} \equiv \imath t$.
The Euclidean action is related to eq.~(\ref{eq:AppA1}) via $S_{\rm E} \equiv -\imath\left.\mathcal{S}\right|_{t = -\imath t_{\rm E}}$. 
The analytical continuation can be seen as the coordinate transformation $x\equiv (t,\vec{x}) \to \tilde{x}\equiv (t_{\rm E},\vec{x}) = (\lambda t,\vec{x})$, with $\lambda = \imath$.
We can therefore write the general coordinate transformation 
\begin{equation}
g_{\alpha\beta}(x) = \tilde{g}_{\mu\nu}(\tilde{x})\frac{\partial \tilde{x}^{\mu}}{\partial x^{\alpha}}
\frac{\partial \tilde{x}^{\nu}}{\partial x^{\beta}} = 
\left(
\begin{array}{cccc}
 \lambda^2 \tilde{g}_{00} &  \lambda \tilde{g}_{01}  &  \lambda \tilde{g}_{02} &  \lambda \tilde{g}_{03} \\
 \lambda \tilde{g}_{01} & \tilde{g}_{11}  & \tilde{g}_{12} & \tilde{g}_{13} \\
\lambda \tilde{g}_{02}  &\tilde{g}_{12}  & \tilde{g}_{22} & \tilde{g}_{23} \\
\lambda \tilde{g}_{03}  & \tilde{g}_{13}  & \tilde{g}_{23}  & \tilde{g}_{33}
\end{array}
\right)~,
\end{equation}
from which it follows that $g = \lambda^2 \tilde{g} = -\tilde{g}$, 
and $d^4x\sqrt{-g} = d^4\tilde{x}\sqrt{\tilde{g}}(-\imath)$. The Euclidean version of eq.~(\ref{eq:AppA2}) is
\begin{equation}
\mathcal{S}_{\rm E} = \int d^4x\sqrt{\tilde{g}}\left[
-\frac{M_{\rm Pl}^2}{16\pi}\,\mathcal{R}   + \frac{\mathcal{F}}{2}H_{\mu\nu\rho}H^{\mu\nu\rho}
\right]~,
\end{equation}  
which coincides, after renaming $\tilde{g}$, with eq.~(\ref{eq:WormholeAction}).
In Euclidean space, the duality relation in eq.~(\ref{eq:DualMinkowski}) reads 
\begin{equation}\label{eq:DualEuclidean}
\frac{f_a^2}{2}(\partial_{\rho}\theta)(\partial^{\rho}\theta)  = \frac{\mathcal{F}}{2}H_{\mu\nu\rho}H^{\mu\nu\rho}~,
\end{equation} 
where we now used $\epsilon_{\mu\nu\rho\sigma}\epsilon^{\mu\nu\rho\lambda} = (-1)^t\,3!\,\delta_{\sigma}^{\lambda}$, with $t=0$ in Euclidean 
space-time. Consequently, we reconstruct the Euclidean action in eq.~(\ref{eq:AxionAction}).

Notice that it is crucial to define the Euclidean action 
by analytical continuation of eq.~(\ref{eq:AppA2}) rather then eq.~(\ref{eq:AppA1}).
Indeed, the analytical continuation of eq.~(\ref{eq:AppA1}) would generate the Euclidean action 
\begin{equation}\label{eq:WrongAxionAction}
\mathcal{S}_{\rm E} = \int d^4x\sqrt{g}\left[
-\frac{M_{\rm Pl}^2}{16\pi}\,\mathcal{R} - \frac{f_a^2}{2}(\partial_{\rho}\theta)(\partial^{\rho}\theta)
\right]~,
\end{equation}
that features the `wrong' minus sign in front of the Goldstone kinetic term.

Finally, we stress again that the construction we put forward in this appendix, 
as well as the consequent wormhole solutions, does not rely on the transformation properties under parity 
of the Goldstone boson.
This suggests that wormhole solutions can be defined for both free massless scalar and pseudo-scalar fields.

\subsection{Axion charge quantization}\label{app:AxionChargeQuantization}

In this appendix we provide a more quantitative picture of the quantization condition discussed at the end of section~\ref{sec:BulkAction}.
To this end, we exploit the analogy with the quantization of the electric charge in the presence of a magnetic monopole.

The magnetic field of a putative magnetic monopole placed at the origin of  $\mathbb{R}^3$ is
\begin{equation}\label{eq:MagneticMonopole}
\vec{\mathcal{B}}(\vec{r}) = \frac{g_{\rm M}}{4\pi}\frac{\vec{r}}{r^3}~,~~~~~~~g_{\rm M} = \int_{S_2}\vec{\mathcal{B}}\cdot d\vec{S}~,
\end{equation}
where in the second equation the magnetic charge $g_{\rm M}$ corresponds to the magnetic flux through the sphere $S_2$.
Bearing in mind the relation with the electromagnetic field strength, $F_{ij} = -\epsilon_{ijk}\mathcal{B}_k$, one can already 
identify the analogy with the three-form field $H_{ijk} = \epsilon_{ijk}\mathcal{H}_0 = \epsilon_{ijk}n/2\pi^2 r^3$ and 
the computation of the flux in eq.~(\ref{eq:Charge}). Of course, in the electromagnetic case we are dealing with a two-form 
field strength $F_{\mu\nu} = \partial_{\mu}A_{\nu} - \partial_{\nu}A_{\mu}$
while in the axion case the field strength is the three-form $H_{\mu\nu\rho} = 
\partial_{\mu}B_{\nu\rho} + \partial_{\nu}B_{\rho\mu} + \partial_{\rho}B_{\mu\nu}$ but the comparison is evident.
The r\^ole of the electromagnetic potential $A_{\mu}$ is played by the antisymmetric tensor 
 $B_{\mu\nu}$.
 The obstacle is that it is not possible to find a vector potential such that $\vec{\nabla}\times \vec{A} = \vec{\mathcal{B}}$
 since the monopole field is not divergenceless. The best that one can do is to define (in ordinary spherical coordinates) the vector potentials
 \begin{equation}
 \vec{A}_{1} \equiv \frac{g_{\rm M}(1-\cos\theta)}{4\pi r \sin\theta}\hat{\phi}~,~~~~~~~
 \vec{A}_{2} \equiv - \frac{g_{\rm M}(1+\cos\theta)}{4\pi r\sin\theta}\hat{\phi}~,
 \end{equation}
 defined, respectively, in $R_1:\theta \in [0,\pi/2 +\delta)$ and $R_2:\theta \in (\pi/2 -\delta,\pi]$ with an overlap region
 $\pi/2 -\delta < \theta < \pi/2 +\delta$. In their domains, both potentials satisfy $\vec{\nabla}\times \vec{A}_{1,2} = \vec{\mathcal{B}}$ 
but they are singular in the complementary region. This is the same ambiguity one encounters 
for the  two-form potential $B$ in the axion case. The only way to obtain a consistent  picture 
is to show that the two vector potentials describe the same physics in the overlap region. Said differently, $\vec{A}_{1}$ and $\vec{A}_{2}$ must be 
related by a gauge transformation $\vec{A}\to \vec{A}^{\prime} = \vec{A} + \vec{\nabla}\chi$. 
From $\vec{A}_1 - \vec{A}_2 = \vec{\nabla}\chi$ one immediately finds  $\chi \equiv g_{\rm M}\phi/2\pi$.
Notice that the function $\chi(\phi)$ is not continuous since the azimuthal angle is defined modulo $2\pi$.
This is actually crucial for the existence of a non-zero flux in eq.~(\ref{eq:MagneticMonopole}) since 
\begin{equation}
g_{\rm M} = \int_{S_2}\vec{\mathcal{B}}\cdot d\vec{S} = \int_{R_1}(\vec{\nabla}\times \vec{A}_{1})\cdot d\vec{S} + 
\int_{R_2}(\vec{\nabla}\times \vec{A}_{2})\cdot d\vec{S} = 
\int_0^{2\pi}(\partial\chi/\partial\phi)d\phi =
\chi(2\pi) - \chi(0)~.
\end{equation}  
Dirac quantization proceeds as follows. 
We consider a test particle of mass $m$ and charge $q$ in the field of the magnetic monopole.
Its wavefunction satisfies the Schr\"odinger equation $-\slashed{h}^2/2m(\vec{\nabla} + \imath e\vec{A})^2\psi  = 
\imath\slashed{h}\partial\psi/\partial t$, with $e= q/\slashed{h}$, and the gauge transformation $\vec{A}\to \vec{A}^{\prime} = \vec{A} + \vec{\nabla}\chi$ 
transforms $\psi$ into $\psi^{\prime} = e^{-\imath e\chi}\psi = e^{-\imath e g_{\rm M}\phi/2\pi}\psi$.
Since $\phi = 0$ and $\phi = 2\pi$ are the same physical point, consistency of Quantum Mechanics 
requires the wavefunction $\psi^{\prime}$ to be single valued. 
The quantization condition $e g_{\rm M} = 2\pi k$, with $k \in \mathbb{Z}$, then follows.

Dirac quantization involves the product $e g_{\rm M}$ of electric and magnetic charge. 
The electric charge of a particle is given by the surface integral at spatial infinity over a $S_2$ sphere of the 
Hodge dual of the electromagnetic field strength two-form $F_{\mu\nu}$. 
Conversely, the magnetic charge of a particle is given by the surface integral at spatial infinity over a sphere $S_2$ of the electromagnetic 
field strength two-form $F_{\mu\nu}$, as already discussed  in eq.~(\ref{eq:MagneticMonopole}).

The quantization in the axion case is based on the fact that potentials of higher rank admit the following generalization.
Formally, in $D$ dimensions, 
the `electric' charge $Q_{\rm E}$ is the integral over the sphere $S_{D-(p+2)}$ of the Hodge dual of the $(p+2)$-form field strength.
For the axion three-form $H$, $p=1$, and in $D=4$ dimensions $Q_{\rm E} = \int_{S_1} {^*H}$. 
The `magnetic' charge, as already noticed,  is $Q_{\rm M} = \int_{S_3} H = n$ (see eq.~(\ref{eq:Charge})), that in our case corresponds to
 the axion charge flowing  through the wormhole throat.
 Dirac quantization therefore generalizes to $Q_{\rm E} n= 2\pi k$.
 If $Q_{\rm E}\neq 0$, the quantization of $n$ trivially follows.
 In terms of the Goldstone  field, we have $Q_{\rm E} = \int_{S_1} {^*H} = \int_{S_1}d\theta$.
 The key observation is that this integral encompasses a closed path. 
 This means that $Q_{\rm E}$  can be different from zero only if $\theta$ is multi-valued, and this is exactly the case 
 for the Goldstone field since it is a phase, and  possesses the periodicity $\theta = \theta + 2\pi$. 
 This concludes the argument, 
 and shows the fundamental importance of the periodicity of the Goldstone boson field for the quantization of the axion 
 charge along the wormhole throat.

\section{The P\"oschl-Teller potential}\label{App:PT}

We want to solve the spectral problem for the differential operator 
\begin{equation}
\left[
-\frac{d^2}{d\tau^2} + 1 - \frac{15}{\cosh^2(2\tau)}
\right]u(\tau) = \lambda u(\tau)~,
\end{equation}
or, equivalently
\begin{equation}\label{eq:PT}
\left[-\frac{d^2}{dx^2} - \frac{15/4}{\cosh^2x}\right]u(x) = 2E u(x)~,~~~~{\rm with}~~2E=\left(
\frac{\lambda - 1}{4}
\right)~.
\end{equation}
Eq.~(\ref{eq:PT}) is describes the one-dimensional motion of a particle in the P\"oschl-Teller potential~\cite{Flugge}.
In full generality, the dynamics in the P\"oschl-Teller potential is described by the differential equation 
$-u^{\prime\prime}(x) - \frac{\alpha^2 \xi(\xi -1)}{\cosh^2(\alpha x)} u(x) = 2Eu(x)$. 
Eq.~(\ref{eq:PT}) corresponds to $\alpha = 1$, $\xi = 5/2$.
By changing variable $y\equiv \cosh^2x$ we find
\begin{equation}
y(1-y)u^{\prime\prime}(y) + \left(
\frac{1}{2} - y
\right)u^{\prime}(y) - \left(
\frac{15}{16 y} + \frac{E}{2}
\right)u(y) = 0~.
\end{equation}
After the rescaling $u(y) = y^{5/4}v(y)$ the problem reduces to the hypergeometric differential equation
\begin{equation}
y(1-y)v^{\prime\prime}(y) + v^{\prime}(y)\left[
\xi + \frac{1}{2} - y(\xi +1)
\right] - \frac{v(y)}{4}\left(
\xi^2 + 2E
\right)= 0~,
\end{equation}
with $\xi = 5/2$. We have the general solution
\begin{equation}
u(x) = y^{\xi/2}\left[
B\sqrt{1-y}~_2F_1\left(
a+\frac{1}{2}, b+\frac{1}{2};\frac{3}{2};1-y
\right) + A~_2F_1\left(
a,b;\frac{1}{2};1-y
\right)
\right]~,
\end{equation}
where $_2F_1(a,b;c;z) $ 
is the Gaussian or ordinary hypergeometric function, and we introduced the short-hand notation
\begin{equation}
a \equiv \frac{1}{2}\left(
\xi + \imath\sqrt{2E}
\right)~,~~~~~b\equiv \frac{1}{2}\left(
\xi - \imath\sqrt{2E}
\right)~.
\end{equation}
We can separate the two independent solutions. 
If $B=0$, $A=1$, we have the even solution
\begin{equation}\label{eq:EvenSolutions}
u_{\rm even}(x) = \cosh^{\xi}x~_2F_1\left(
a,b;\frac{1}{2};-\sinh^2x
\right)~.
\end{equation}
If $A=0$, $B = \imath$ we have the odd solution
\begin{equation}\label{eq:OddSolutions}
u_{\rm odd}(x) = \cosh^{\xi}x\sinh x~_2F_1\left(
a+\frac{1}{2},b+\frac{1}{2};\frac{3}{2};-\sinh^2x
\right)~.
\end{equation}
In order to identify the bound states with negative energy solutions, we need to study the asymptotic behavior.
Using the trigonometric definitions $\sinh x = (e^x - e^{-x})/2$, $\cosh x = (e^x + e^{-x})/2$ we find 
\begin{eqnarray}
u_{\rm even}(x) &\overset{x \to \pm \infty}{\rightarrow}& \sqrt{\pi}\left[
\frac{2^{2a-\xi}e^{-\imath\sqrt{2E}|x|}\Gamma(-a+b)}{\Gamma(1/2 - a)\Gamma(b)}  +
\frac{2^{2b-\xi}e^{\imath\sqrt{2E}|x|}\Gamma(a-b)}{\Gamma(a)\Gamma(1/2 - b)}
\right]~, \label{eq:Even} \\
u_{\rm odd}(x) &\overset{x \to \pm \infty}{\rightarrow}& \pm\sqrt{\pi}\left[
\frac{2^{2a-(\xi+1)}e^{-\imath\sqrt{2E}|x|}\Gamma(-a+b)}{\Gamma(1 - a)\Gamma(1/2 + b)}  +
\frac{2^{2b-(\xi+1)}e^{\imath\sqrt{2E}|x|}\Gamma(a-b)}{\Gamma(1/2  + a)\Gamma(1 - b)}
\right]~.\label{eq:Odd}
\end{eqnarray} 
In the presence of negative energy states, 
we have
\begin{equation}
e^{-\imath\sqrt{2E}} = e^{+\sqrt{2|E|}}~,~~~~~e^{\imath\sqrt{2E}} = e^{-\sqrt{2|E|}}~.
\end{equation}
The coefficient of $e^{-\imath\sqrt{2E}}$ must therefore vanish in order to ensure the correct asymptotic behavior.
This condition leads to quantization of energy.
The Gamma function diverges for negative integer numbers, and  we can impose this condition 
to cancel the coefficient of $e^{-\imath\sqrt{2E}}$ in both $u_{\rm even}(x) $ and $u_{\rm odd}(x)$.
For the even functions we find
\begin{equation}
\frac{1}{2}  - a = -m~~~~\Longrightarrow~~~~2E_m^{\rm (even)} = - (-2m -1 +\xi)^2~,
\end{equation}
whereas for the even functions we find
\begin{equation}
1 - a = -m~~~~\Longrightarrow~~~~2E_m^{\rm (odd)} = - (-2m -2 +\xi)^2~,
\end{equation}
with $m=0,1,2,\dots$. 

With $\xi = 5/2$ we only have one negative (even) state with energy $2E_0^{\rm (even)}  = -9/4$ or, equivalently, $\lambda = -8$.
Next, we have two states with $\lambda = 0$ corresponding to 
 $2E_1^{\rm (even)}  = -1/4$ and $2E_0^{\rm (odd)}  = -1/4$.
 Using eqs.~(\ref{eq:EvenSolutions},\ref{eq:OddSolutions}), we find
\begin{align}
{\rm Odd\,eigenfunction:}~~&\left\{u_{\rm odd}^{\lambda = 0}(\tau) = \frac{\sinh(2\tau)}{\cosh^{3/2}(2\tau)}\right\}~,\\
{\rm Even\,eigenfunctions:}~~&\left\{u_{\rm even}^{\lambda = -8}(\tau) = \frac{1}{\cosh^{3/2}(2\tau)}~,~~
u_{\rm even}^{\lambda = 0}(\tau) =  \frac{3 - \cosh(4\tau)}{2\cosh^{3/2}(2\tau)}\right\}~.
\end{align}
The eigenfunction $u_{\rm even}^{\lambda = 0}(\tau) $ is not square-integrable.

 \section{Spectrum of tensor and inhomogeneous scalar perturbations}\label{app:Perturbations}
 
 In this appendix we compute the spectrum of tensor and inhomogeneous scalar perturbations of the quadratic action expanded around the wormhole background.
 
 \subsection{Tensor perturbations}
 
 We  exploit the results of~\cite{Hertog:1999kg}  in which tensor fluctuations around instanton background were studied. 
 The perturbed line element, in Euclidean space, is $ds^2 = a^2(\tau)\left[d\tau^2 + (\gamma_{ij} + t_{ij})dx^i dx^j\right]$, where $\gamma_{ij}$
  is the background metric on the three-sphere
 and $a(\tau)$ the background conformal factor. As discussed in section~\ref{sec:Fluctuations},  $t_{ij}$ is a transverse trace-free symmetric tensor.
  The Euclidean quadratic action takes the form~\cite{Hertog:1999kg}
 \begin{equation}
 \left.\delta \mathcal{S}_{\rm E}^{(2)}\right|_{\rm tensor}  = \frac{1}{8\kappa}\int d\tau d^3\vec{x}\sqrt{\gamma}a^2\left(
 t^{\prime\,ij}t^{\prime}_{ij} +
 \nabla^{i}t^{jk}\nabla_{i}t_{jk} + 2t^{ij}t_{ij}
 \right)~,
 \end{equation}
 where covariant derivatives and contractions are defined w.r.t. the metric $\gamma$. 
 By introducing the rescaling $\tilde{t}_{ij} \equiv at_{ij}$, and integrating by parts, one finds~\cite{Hertog:1999kg}
 \begin{equation}\label{eq:TensorQuadratic}
  \left.\delta \mathcal{S}_{\rm E}^{(2)}\right|_{\rm tensor}  = \frac{1}{8\kappa}
  \int d\tau d^3\vec{x}\sqrt{\gamma}\left[
  \tilde{t}_{ij}\left(
  \hat{K} + 3 -\Delta
  \right)\tilde{t}^{ij}
  - \frac{d}{d\tau}\left(
 \frac{a^{\prime}}{a}\tilde{t}_{ij}\tilde{t}^{ij}
  \right)
  \right]~,
 \end{equation}
 with the Schr\"odinger-type operator
 \begin{equation}\label{eq:TensroSch}
 \hat{K} \equiv -\frac{d^2}{d\tau^2}  + \frac{a^{\prime\prime}}{a} -1 =  -\frac{d^2}{d\tau^2} + \frac{1}{\cosh^2(2\tau)}~,
 \end{equation}
 where in the last step we substituted the wormhole background solution. Notice that the quadratic action in eq.~(\ref{eq:TensorQuadratic})
 does not present the conformal factor problem. This is indeed correct since, as reviewed in section~\ref{sec:Fluctuations}, the quadratic 
 action for the trace-free part of the tensor
 fluctuations is not unbounded from below.
 The differential operator $ \hat{K}$ acts on the $\tau$ variable while the Laplacian operator $\Delta$ acts on the three-dimensional 
 space. We can therefore diagonalize them independently and add their spectra.
 The differential operator $\hat{K}$  does not possess negative eigenvalues since the potential is a positive function and there are no bound states.
 The tensor harmonics $(G_{ij})^{n}_{lm}(\psi,\phi,\varphi)$ are tensor eigenfucntions of the Laplacian operator on $S_3$. They satisfy the
 eigenvalue equation $\Delta(G_{ij})^{n}_{lm} = -(n^2 - 3)(G_{ij})^{n}_{lm}$, with $n=3,4,5,\dots$, supported by the 
 transverse and traceless conditions $\nabla^{i}(G_{ij})^{n}_{lm} = 0$ and $(G_{i}^{i})^{n}_{lm} = 0$.
 Because of the degeneracy in $l$ and $m$, the most general solution of the eigenvalue equation is 
 $G_{ij}^{(n)}(\psi,\phi,\varphi) = \sum_{l=2}^{n-1}\sum_{m=-l}^l C_{lm}^{(n)}(G_{ij})^{n}_{lm}(\psi,\phi,\varphi)$,
 with $C_{lm}^{(n)}$ arbitrary constants.  From this general discussion it follows that 
 the tensor Laplacian $-\Delta$ in eq.~(\ref{eq:TensorQuadratic}) has positive eigenvalues, starting from $\lambda = 6$.
 We conclude that tensor fluctuations around the wormhole background have a positive spectrum.
 
 
 \subsection{Inhomogeneous scalar perturbations}
 
 We now move to discuss inhomogeneous scalar perturbations. In section~\ref{sec:Fluctuations} we 
 focused on scalar homogeneous perturbations. 
 A separated analysis is required since, as well known~\cite{Gratton:2000fj,Gratton:1999ya}, inhomogeneous perturbations 
 have positive Euclidean action and do not suffer from the conformal factor problem.
 We exploit the results of~\cite{Gratton:1999ya}. In the Euclidean space, the quadratic action takes in form
 \begin{equation}\label{eq:InhomogeneousScalar}
 \left.\delta\mathcal{S}_{\rm E}^{(2)}\right|_{\Delta \neq 0} =
 \frac{1}{2}\int d\tau d^{3}\vec{x}\sqrt{\gamma}\left\{\left(
 -\Delta + 3\right)q\left(
 \hat{O} - \Delta + 3
 \right)q + \left(
 -\Delta + 3
 \right)\frac{d}{d\tau}\left[
 qq^{\prime}  + \left(
 -\frac{\kappa \phi^{\prime\,2}}{4a^{\prime}/a} + \frac{\phi^{\prime\prime}}{\phi^{\prime}}
 \right)q^2
 \right]
 \right\}~,
 \end{equation}
 with the Schr\"odinger-type operator
 \begin{equation}\label{eq:DifferentialInhom}
 \hat{O} \equiv -\frac{d^2}{d\tau^2} - \frac{\kappa\phi^{\prime\,2}}{2} + \phi^{\prime}\left(\frac{1}{\phi^{\prime}}\right)^{\prime\prime}  
 = -\frac{d^2}{d\tau^2}  + 4 - \frac{3}{\cosh^2(2\tau)}~.
 \end{equation}
As before,  the differential operator $ \hat{O}$ acts on the $\tau$ variable while the Laplacian operator $\Delta$ acts on the three-dimensional 
 space. We can therefore diagonalize them independently and add their spectra. Furthermore, we already know that $-\Delta$ has positive eigenvalues 
 starting from $\lambda = 3$ (ignoring the homogeneous mode).  In eq.~(\ref{eq:InhomogeneousScalar}) 
 $q$ is the only physical dynamical variable describing scalar perturbations 
 after gauge degrees of freedom are removed~\cite{Gratton:1999ya}.
 By defining $2\tau \equiv x$, the differential equation $\hat{O}u(\tau) = \lambda u(\tau)$ in eq.~(\ref{eq:DifferentialInhom}) takes the form
\begin{equation}\label{eq:PT2}
\left[-\frac{d^2}{dx^2} - \frac{3/4}{\cosh^2x}\right]u(x) = 2E u(x)~,~~~~{\rm with}~~2E=\left(
\frac{\lambda}{4} - 1
\right)~.
\end{equation}
This is again a P\"oschl-Teller potential with $\alpha = 1$, $\xi = 3/2$. We can repeat the analysis presented in appendix~\ref{App:PT}.
With $\xi = 3/2$, we have a double-degenerate negative eigenvalue $E_{0}^{\rm even/odd} = -1/4$
but it corresponds to the positive value  $\lambda = 3$, that is the starting point of the spectrum of the differential operator $\hat{O}$.
We therefore conclude that the spectrum of the bulk operator in eq.~(\ref{eq:InhomogeneousScalar}) is fully positive.

\def\hhref#1{\href{http://arxiv.org/abs/#1}{#1}} 

\end{document}